\documentclass[3p,times]{elsarticle}

\usepackage{amsmath}      

\usepackage{mathtools, cuted, cases}
\usepackage[hidelinks]{hyperref}
\usepackage{amsfonts}
\usepackage{tikz}
\usepackage{caption}
\usepackage[labelformat=simple]{subcaption}    
\usepackage{algorithm}
\usepackage{algpseudocode}

\renewcommand{\algorithmiccomment}[1]{\bgroup\textcolor{teal}{\#~#1}\egroup}
\algnewcommand{\NoLineNumber}[1]{\Statex \hskip #1}

\biboptions{sort&compress}

\DeclarePairedDelimiterXPP\myln[1]{\operatorname{ln}}{(}{)}{}{#1}

\usepackage{xfrac}
\usepackage{siunitx}
\usepackage{multirow}

\usepackage{amssymb}
\usepackage{mathrsfs}
\usepackage{tcolorbox}

\newtcolorbox[auto counter]{MyBox}[2][]{%
colback=white, boxrule=.3mm,title=Box~\thetcbcounter: #2,#1}

\usepackage[makeroom]{cancel}

\DeclareMathOperator*{\argmin}{arg\,min}
\DeclareMathAlphabet\mathbfcal{OMS}{cmsy}{b}{n}

\journal{Computers in Biology and Medicine}

\begin{document}
\begin{frontmatter}

\title{A probabilistic reduced-order modeling framework for patient-specific cardio-mechanical analysis}

\author[1,2]{Robin Willems\corref{cor1}}
\cortext[cor1]{Corresponding author}
\ead{R.Willems@tue.nl}
\author[3,5]{Peter Förster}
\author[5]{Sebastian Schöps}
\author[4]{Olaf van der Sluis}
\author[1]{Clemens V. Verhoosel}
\address[1]{Department of Mechanical Engineering, Energy Technology and Fluid Dynamics, Eindhoven University of Technology, The Netherlands}
\address[2]{Department of Biomedical Engineering, Cardiovascular Biomechanics, Eindhoven University of Technology, The Netherlands}
\address[3]{Department of Mathematics and Computer Science, Computational Science, Eindhoven University of Technology, The Netherlands}
\address[4]{Department of Mechanical Engineering, Mechanics of Materials, Eindhoven University of Technology, The Netherlands}
\address[5]{Department of Electrical Engineering and Information Technology, Computational Electromagnetics, Technical University of Darmstadt, Germany}

\begin{abstract}
Cardio-mechanical models can be used to support clinical decision-making. Unfortunately, the substantial computational effort involved in many cardiac models hinders their application in the clinic, despite the fact that they may provide valuable information. In this work, we present a probabilistic reduced-order modeling (ROM) framework to dramatically reduce the computational effort of such models while providing a credibility interval. In the online stage, a fast-to-evaluate generalized one-fiber model is considered. This generalized one-fiber model incorporates correction factors to emulate patient-specific attributes, such as local geometry variations. In the offline stage, Bayesian inference is used to calibrate these correction factors on training data generated using a full-order isogeometric cardiac model (FOM). A Gaussian process is used in the online stage to predict the correction factors for geometries that are not in the training data. The proposed framework is demonstrated using two examples. The first example considers idealized left-ventricle geometries, for which the behavior of the ROM framework can be studied in detail. In the second example, the ROM framework is applied to scan-based geometries, based on which the application of the ROM framework in the clinical setting is discussed. The results for the two examples convey that the ROM framework can provide accurate online predictions, provided that adequate FOM training data is available. The uncertainty bands provided by the ROM framework give insight into the trustworthiness of its results. Large uncertainty bands can be considered as an indicator for the further population of the training data set.
\end{abstract}


\end{frontmatter}

\section{Introduction}\label{sec:Intro}

% Background
Over the past decades, the use of patient-specific computational models in biomedical research and clinical practice has become abundant \cite{trayanova_cardiac_2011, sun_computational_2014, niederer_computational_2019, sung_whole-heart_2021, de_lepper_evidence-based_2022}. In cardiology, a large variety of patient-specific computational models has been developed in the context of specific diseases and phenomena \cite{aguado-sierra_patient-specific_2011, kuijpers_modeling_2012, arevalo_arrhythmia_2016, niederer_computational_2019, sung_whole-heart_2021}. In our previous work \cite{willems_isogeometric_2024,willems_echocardiogram-based_2024,willems_isogeometric-mechanics-driven_2023} we have, for example, developed a computational model to study cardiac mechanics in the context of Ventricular Tachycardias (VTs), taking into consideration echocardiogram data. The objective of patient-specific models varies from gaining a fundamental understanding to supporting clinical decision making.

% Models in research vs models in clinical practice
The requirements of a computational model strongly depend on its intended purpose. A computational model used in biomedical research to enhance fundamental understanding typically requires detailed anatomical and physiological representations. The evaluation of such a model often involves substantial computational effort. Although this is generally not problematic in view of the modeling purpose, the computational demand can make calibration procedures impractical, warranting the accurate determination/measurement of the model parameters. In contrast, a computational model used for clinical decision support is typically required to have limited computational effort, as to not delay the decision-making process. Limiting the computational effort is typically achieved by reducing the level of detail of the anatomical and physiological representations to what is needed to support the clinician. In this setting, it is very important that the model can be tailored easily to a patient through model calibration and that it can accommodate the large uncertainties inherent to the clinical setting.

% Reduced-order modeling
The results obtained from a biomedical research model are often highly informative for clinical decision making, but its application in clinical practice can be hindered by the substantial computational effort. This creates a need for \emph{reduced-order modeling (ROM)} approaches~\cite{manzoni_reduced_2018}, which involve an \emph{offline stage} in which a computationally demanding \emph{full-order model (FOM)} is simplified to a computationally affordable \emph{reduced-order model}. This reduced-order model -- which should balance model details and accuracy with computational effort -- is then used to support the clinical decision-making process in the \emph{online stage}.

\begin{figure}[!t]
\centering
\includegraphics[width=0.7\textwidth]{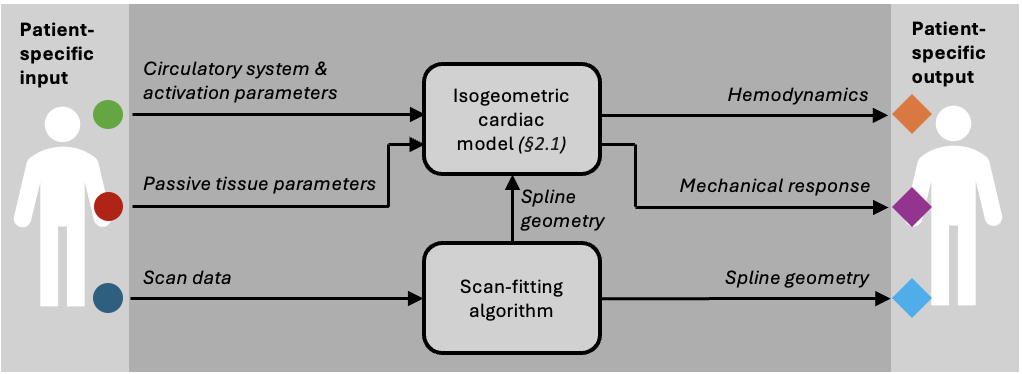}
\caption{Schematic of the considered full-order cardiac model. Patient-specific input (circles) is mapped on patient-specific output (diamonds) by combining a NURBS-based fitting algorithm with an isogeometric cardiac solver.} \label{fig:HFmodel}
\end{figure}

% Specific setting of the current work
In this work we intend to develop a ROM approach for a patient-specific cardiac mechanics model. We specifically consider the echocardiogram-based isogeometric analysis model developed by Willems et al.~\cite{willems_isogeometric_2024} (Figure~\ref{fig:HFmodel}) as the FOM. This model represents the anatomy of the ventricles by means of non-uniform rational B-splines (NURBS), which can be fitted to echocardiogram data~\cite{willems_echocardiogram-based_2024}. The time-dependent mechanical behavior is modeled by a tissue-scale nonlinear continuum mechanics description, which is coupled to a lumped parameter model for the arterial system and a phenomenological activation potential relation. The model parameters are selected based on expertise from clinical research. This model computes the time-dependent mechanical response in the form of displacement fields, from which many other properties, such as stress fields and pressure-volume curves, are derived. When solving the model for multiple cardiac cycles, as required to reach cyclic steady-state conditions, typical simulation times surmount to multiple hours. Although the model settings can be optimized for computational efficiency, usage of this model for clinical decision making is impractical, especially because calibration procedures in that setting would require many evaluations of the model. The model would, however, be very suitable for application in the offline stage of a ROM approach, where it is used to generate training data.

% Requirements of the reduced-order model
Besides the requirement that the ROM should be computationally affordable, it should also be able to incorporate as much of the information about a patient as possible. We therefore require the reduced-order model to use the same input as the FOM, \emph{i.e.}, the echocardiograms, tissue stiffness parameters and parameters for the circulatory system and activation potential law. We assume that for clinical decision making the pressure-volume curves are most important, and that displacement and stress fields as obtained from the full-order model are not essential in this setting. Optionally, such detailed output should, however, be obtainable in a subsequent offline evaluation of the full-order model.

% Strategies to reduce model complexity
The choice for a specific ROM approach depends on the characteristics of the FOM on which it is based, as well as on functionality, accuracy and computational complexity requirements of the ROM. A myriad of methods to construct a ROM is available. Inspired by Peherstorfer \emph{et al.}\footnote{The authors present their classification in a multi-fidelity modeling context.}~\cite{peherstorfer_survey_2018}, we classify the ROM models resulting from these methods as:
\begin{itemize}
    \item A \emph{simplified reduced-order model} is constructed from a FOM by making additional modeling assumptions, for example regarding the geometry or physics, to allow for mathematical simplification of the problem formulation. In essence, this follows a traditional modeling approach, in which domain-specific expertise is used to obtain a simplified ROM that adequately balances model functionality and accuracy on the one hand with computational complexity on the other. Consequently, in many research domains there is a wealth of literature on simplified ROMs. Due to the additional modeling assumptions, the simplified reduced-order model parameters and results in general do not coincide with those of the FOM on which it is based. For the model parameters, this means that available data for the FOM needs to be cast into a form suitable for the ROM, a process that typically involves loss of information. The results of simplified ROMs are typically less detailed compared to their FOM counterparts, restricting the objectives for which they can be used.
    \item A \emph{projection-based reduced-order model} retains the formulation of the FOM but reduces its complexity by identifying an adequate solution subspace in which this formulation is solved. Projection-based methods, of which proper orthogonal decomposition \cite{liang_proper_2002} and the reduced basis method \cite{quarteroni_reduced_2016} are prominent examples, exploit the mathematical structure of the FOM in a generic way and hence do not require (extensive) domain-specific expertise. A projection-based ROM typically has the same parameters and outcomes as the full-order model. The accuracy and computational complexity of the ROM can be balanced through the selection of the subspace on which the FOM is projected. For complex problems, for example incorporating time-dependent and nonlinear phenomena, finding an adequate balance can be complicated~\cite{boulakia_reduced-order_2012, rama_towards_2020}.
    \item A \emph{data-fit reduced-order model} -- also commonly referred to as a surrogate model -- provides an abstract mapping between the input and output data of the FOM based on a set of FOM results. Such methods, of which neural networks \cite{gurney_introduction_2017} and Gaussian processes \cite{williams_gaussian_2006} are prominent examples, in principle consider the FOM as a black box. Although standard data-fit ROMs are physics-agnostic, they need to be tailored to the specific input and output under consideration. The amount of training data needed to attain a data-fit ROM with sufficient accuracy depends on the input and output spaces, which can make the generation of a FOM data set computationally demanding in the case of large parameter spaces.
\end{itemize}
We note that, as with any classification, not all ROM approaches fit perfectly, and that methods have been developed that combine ingredients of these classes. A typical example of such hybrid methods are physics-informed neural networks \cite{raissi_physics-informed_2019}.

\begin{figure}[!t]
\centering
\includegraphics[width=0.7\textwidth]{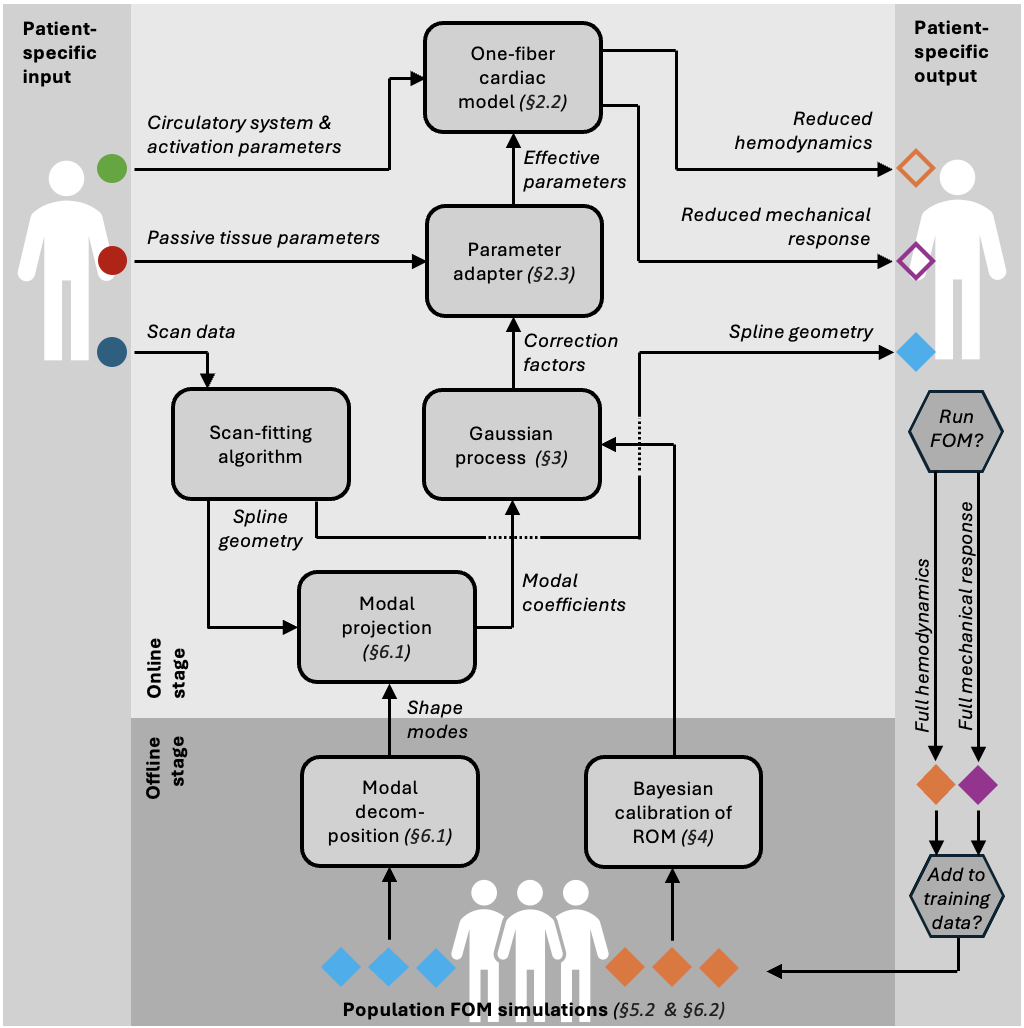}
\caption{Schematic of the developed ROM framework. The framework considers the same input (circles) as the FOM (Figure~\ref{fig:HFmodel}) on which it is based. Since the ROM framework incorporates the scan-fitting algorithm in the online stage, it generates the same spline anatomy output as the FOM (light blue filled diamond). In the online stage, it also generates hemodynamical and mechanical output in the same form as that of the FOM, albeit based on a simplified ROM (open diamonds). If required, these output quantities can also be evaluated using the FOM in a subsequent offline stage, after which they can be used as additional training data.} \label{fig:LFmodel}
\end{figure}

The ROM framework developed in this work, which is schematically illustrated in Figure~\ref{fig:LFmodel}, leverages functionality from different classes of ROMs. The core of the framework is the consideration of a \emph{simplified reduced-order cardiac model} in the form of an organ-scale one-fiber cardiac model, similar to that of Arts~\emph{et al.}~\cite{arts_relation_1991}. This simplified reduced-order modeling approach is motivated by the fact that the one-fiber reduced-order model is established in clinical practice as part of the CircAdapt model~\cite{arts_adaptation_2005}. Projection-based methods are expected to be computationally demanding on account of the nonlinear and time-dependent character of the considered FOM, and data-fit ROMs are impractical on account of the number of FOM simulations that would be required for training. While the physiological model parameters of the FOM can, to a large extent, be used directly in the employed ROM, the scan data cannot be incorporated directly. In order to introduce this input into the reduced-order model, we consider a \emph{projection-based reduced-order model} for the anatomy, in which we use a proper orthogonal decomposition to represent patient-specific geometries by a relatively small number of modal coefficients. A \emph{data-fit reduced-order model} in the form of a Gaussian process is then used to map this geometric input onto the effective parameters of the cardiac ROM.

We acknowledge that the choices made in our ROM framework are strongly driven by our own research experience and that alternative choices are possible and may even be better depending on the considered setting and goals. In light of this, our goal is to develop a versatile ROM framework, in the sense that it is not specific to our choice of models. Alternative cardiac models should be usable without making fundamental changes to the framework, and ideally different types of problems should also be possible to consider, provided that their structure is similar.

This paper is organized as follows. In Section~\ref{sec:mfidelity} we commence with the introduction of the two cardiac models that serve as the FOM and the ROM in our framework. In Section~\ref{sec:GP} we then introduce the Gaussian process that maps geometric inputs onto the ROM parameter space. Subsequently, Section~\ref{sec:BI} introduces the Bayesian calibration framework that is used to generate training data for this Gaussian process. With all elements of the ROM framework in place, in Section~\ref{sec:ApplicationS} we then first extensively test it in the context of an idealized NURBS ventricle parametrized by two geometric quantities. We then extend the application of the framework to a scan-based setting in Section~\ref{sec:ApplicationC}, in which the geometric input quantities are replaced by a modal decomposition of the NURBS ventricle. Finally, conclusions and recommendations are presented in Section~\ref{sec:ConcRec}.

\section{Cardio-mechanical modeling}\label{sec:mfidelity}
The developed reduced-order modeling framework (Figure~\ref{fig:LFmodel}) integrates two established cardio-mechanical models. The isogeometric cardiac analysis approach proposed by Willems \emph{et al.} \cite{willems_isogeometric_2024,willems_echocardiogram-based_2024} serves as the full-order model (FOM). This model will be briefly reviewed in Section~\ref{sec:fommodel}, addressing the most important physical assumptions, model parameters and output quantities. The reader is referred to Refs.~\cite{willems_isogeometric_2024,willems_echocardiogram-based_2024} for details regarding the model and the solution procedure. The one-fiber model (\emph{e.g.}, Refs.~\cite{arts_model_1979,arts_relation_1991}) introduced in Section~\ref{sec:lowfidelity} is considered as the simplified-physics reduced-order model (ROM). The similarities and differences in terms of model parameters will be discussed in Section~\ref{sec:fomromrelations}.

\subsection{Full-order model: isogeometric cardiac analysis}\label{sec:fommodel}
\begin{figure*}[!t]
     \centering
     \begin{subfigure}[b]{0.5\textwidth}
         \centering
         \includegraphics[width=\textwidth]{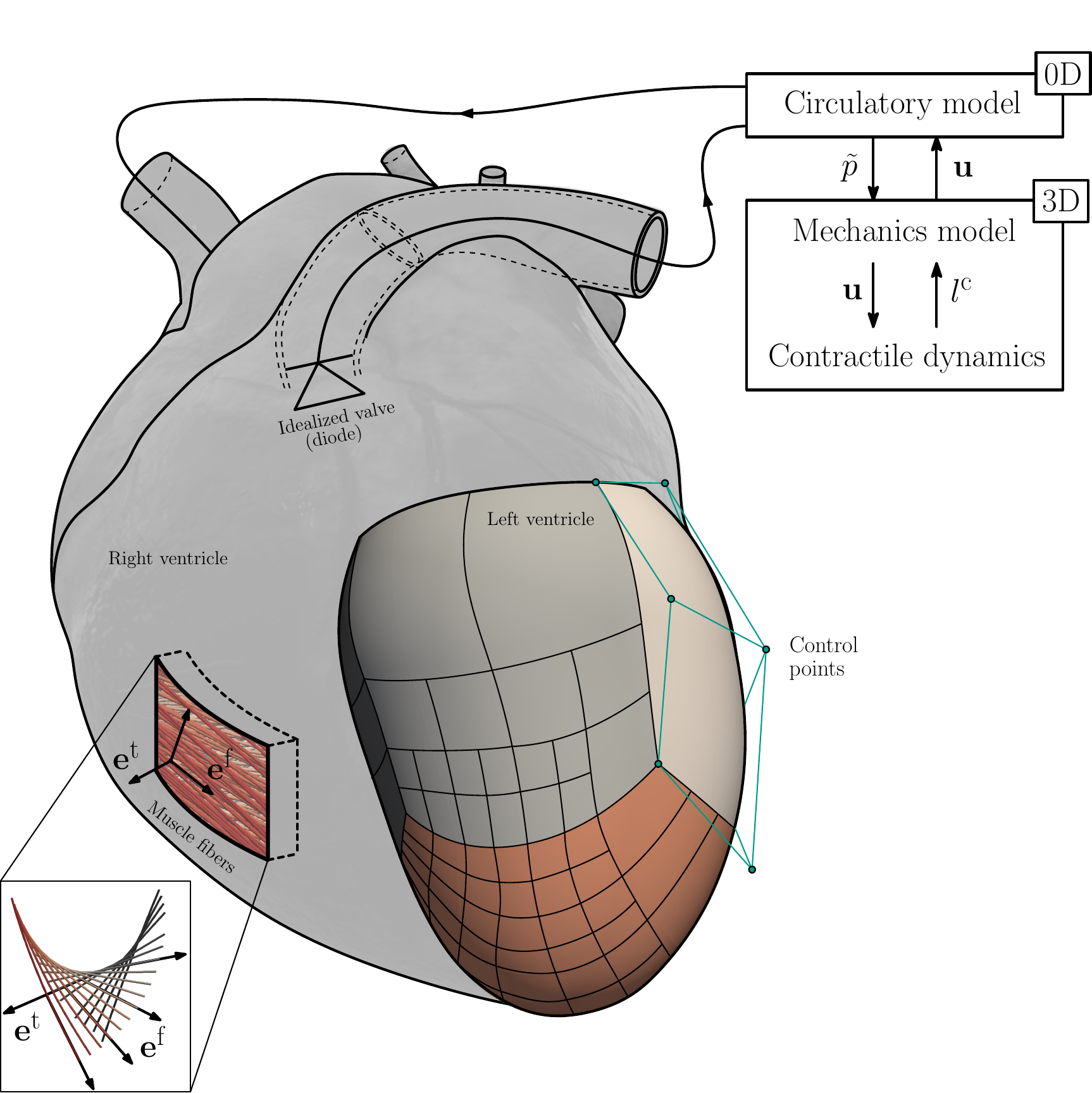}
                  \caption{}
                  \label{fig:igafoma}
     \end{subfigure}
     \hfill
     \begin{subfigure}[b]{0.45\textwidth}
         \centering
         \includegraphics[width=\textwidth]{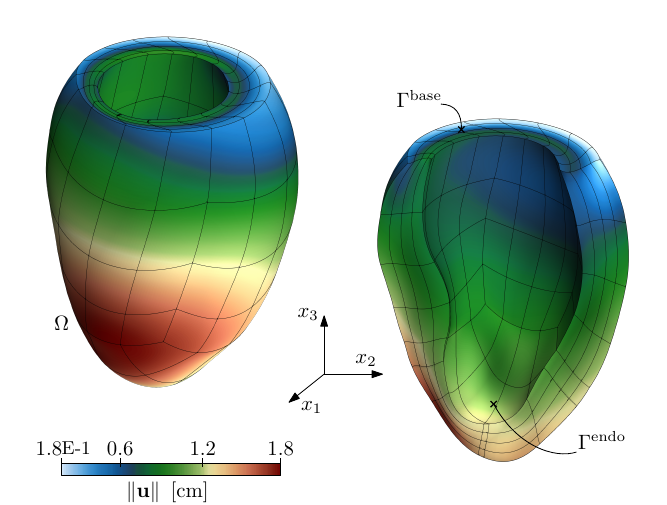}
                  \caption{}
                  \label{fig:igafomb}
     \end{subfigure}
        \caption{(a) Schematic representation of the isogeometric cardiac model, illustrating the coupling between the 0D and 3D model components. (b) Typical isogeometric cardiac analysis result (see Ref.~\cite{willems_echocardiogram-based_2024} for details), showing the displacement magnitude at end-systole relative to end-diastole.}
        \label{fig:igafom}
\end{figure*}

The isogeometric cardiac model of Willems \emph{et al.}~\cite{willems_isogeometric_2024} solves the temporal cardiac mechanics on a NURBS-based left-ventricle domain, which may be patient-specific~\cite{willems_echocardiogram-based_2024} and possibly be diseased~\cite{willems_isogeometric-mechanics-driven_2023}. The model, illustrated in Figure~\ref{fig:igafom}, is subdivided into three coupled components, \emph{viz.} the mechanical passive response of the tissue, the active tissue contraction, and the circulatory model.

The time-dependent mechanical response of the left ventricle is described by the displacement field, $u_i$, which maps a material point in the undeformed configuration, $\Omega_0$, onto the deformed configuration, $\Omega$. The left ventricle is loaded on the endocardium, $\Gamma^{\rm endo}$, by the time-dependent cavity pressure, $p$, which is coupled to the circulatory system model. Rigid body translations and rotations are constraint on the basal plane, $\Gamma^{\rm base}$. The initial-boundary value problem for the mechanical response is summarized in Box~\ref{box:fom}.1.

The mechanical equilibrium equation~\eqref{eq:B11a} is complemented by the constitutive relations in Box~\ref{box:fom}.3, which provide a relation between the stress tensor $\sigma_{ij}$ and the displacement field $u_i$ through the (quasi-)invariants of the deformation gradient and Green-Lagrange strain vector in Box~\ref{box:fom}.2. The quasi-invariant $I_4$ depends on the fiber orientation $\mathrm{e}^{\rm f0}_i$, for which the rule-based method outlined in Ref.~\cite{willems_isogeometric_2024} is used. The stress tensor \eqref{eq:B13a} is composed of a passive part \eqref{eq:B13b} and an active part \eqref{eq:B13c}. The constitutive model parameters associated with the nearly incompressible transversely isotropic passive stress contribution are the elastic modulus of the myocardium, $a_0$, the bulk modulus, $\kappa$, and the coefficients $a_1$, $a_2$ and $a_3$ of the quadratic strain form $Q$. The constitutive model parameters associated with the active part are the initial sarcomere length, $l^{\rm s0}$, and the sarcomere elastic stiffness, $E^{\rm a}$. 

The isometric function $f^{\rm iso}$ in \ref{eq:B21a} governs the influence of the contractile length, $l^{\rm c}$, on the active stress. This function depends on the $T^0$ and $a^l$ parameters and on the initial contractile length $l^{\rm c0}$. The contractile length follows the phenomenological model in Box~\ref{box:fomrom}.2 \cite{hill_heat_1997}, which depends on the parameter $v^{\mathrm{0}}$. The function $f^{\rm twitch}$ in \ref{eq:B21b} models the rise and decay of the myocyte contraction. This function depends on the time since activation, $t^{\rm a}=t-t^{\rm act}$, where muscle activation is considered to be instantaneous and homogeneous at a predefined time, $t^{\mathrm{act}}$, within a single cardiac cycle $0 \leq t^{\mathrm{act}} \leq t^{\mathrm{cycle}}$. The twitch function is parametrized by $b$, $l^{\rm d}$ and the time scales $\tau^{\rm r}$ and $\tau^{\rm d}$.

The circulatory system is represented by a two-element 0D Windkessel model \eqref{eq:B23}. This model provides a differential relation between the pressure in the ventricle, $p$, the ventricle volume, $V$, and a set of auxiliary pressures, $\tilde{p}=\{ p^{\rm art}, p^{\rm ven}\}$, representing the arterial and venous pressures. The circulatory model is parametrized by the reference volumes $V^{\rm art0}$ and $V^{\rm ven0}$, the compliances $C^{\rm art}$ and $C^{\rm ven}$, and the resistances $R^{\rm art}$, $R^{\rm ven}$ and $R^{\rm per}$.

The system of equations in Box 1 and Box 2 is solved monolithically for each time step. Use is made of the total-Lagrange formulation for the mechanical equilibrium, where the reference configuration is assumed to be stress free. For patient-specific geometries -- which are inherently subject to a cavity pressure at the moment of image registration -- an extension is proposed in Ref.~\cite{willems_echocardiogram-based_2024} to include the effects of prestress. The mechanical equilibrium equations are discretized using a Bubnov-Galerkin formulation with Lagrange multipliers for the enforcement of the in-plane boundary conditions. Maximum-regularity cubic B-splines -- which are also used to parametrize the geometry of the left ventricle -- are employed for both the displacement field and the contractile length field. This isogeometric analysis approach has been demonstrated to closely resemble a finite element reference simulation using far fewer degrees of freedom \cite{willems_isogeometric_2024}. Detailed convergence studies have been presented in Ref.~\cite{willems_isogeometric_2024}. 

\begin{MyBox}[label={box:fom},nameref={FOM overview}]{Overview of the full-order model (FOM) based on Ref.~\cite{willems_isogeometric_2024}}
Mechanical equilibrium:
\begin{align}
  \frac{\partial}{\partial x_j} \sigma^{\phantom{}}_{ij}( l^{\rm{c}},u_k, t )  &= 0_i & \text{in } &\Omega & \times & (0, t^{\mathrm{cycle}}] \tag{Box1.1a} \label{eq:B11a}\\
  \sigma^{\phantom{}}_{ij} n_j &= -p\ n_i & \text{on } & \Gamma^{\mathrm{endo}} & \times & (0, t^{\mathrm{cycle}}] \tag{Box1.1b} \label{eq:B11b}\\
  n_i u_i = u_3 &= 0  & \text{on } &\Gamma^{\mathrm{base}} & & \tag{Box1.1c} \label{eq:B11c}\\
   \int u_1 \ \mathrm{d}\Gamma^{\mathrm{base}} &= 0 & \text{on } &\Gamma^{\mathrm{base}} & & \tag{Box1.1d} \label{eq:B11d}\\
   \int u_2 \ \mathrm{d}\Gamma^{\mathrm{base}} &= 0 & \text{on } &\Gamma^{\mathrm{base}} & & \tag{Box1.1e} \label{eq:B11e}\\
   \int \left( \frac{\partial u_2}{\partial x_1} - \frac{\partial u_1}{\partial x_2} \right) \ \mathrm{d}\Gamma^{\mathrm{base}} &= 0 & \text{on } &\Gamma^{\mathrm{base}} & & \tag{Box1.1f} \label{eq:B11f}\\
   u_i & = 0_i & \text{in } &\Omega & \times & 0 \tag{Box1.1g} \label{eq:B11g}
\end{align}
Invariants of the deformation gradient  $\mathbf{F} = \mathbf{I} + \nabla \mathbf{u}$ and Green-Lagrange strain $\mathbf{E}=\frac{1}{2}(\mathbf{F}^{\mathrm{T}} \mathbf{F} - \mathbf{I})$:
\begin{align}
  I_1 &= E^{\phantom{}}_{ii} & \text{in } &\Omega & & \tag{Box1.2a} \label{eq:B12a}\\
  I_2 & = \frac{1}{2} \left( E^{\phantom{}}_{ii} E^{\phantom{}}_{jj} -  E^{\phantom{}}_{ji} E^{\phantom{}}_{ij}\right) & \text{in } &\Omega & & \tag{Box1.2b} \label{eq:B14b}\\
  I_3 &= J = \text{det}\left( \mathbf{F} \right) & \text{in } &\Omega & & \tag{Box1.2c} \label{eq:B12c}\\
  I_4 &= {\mathrm{e}}^{\mathrm{f0}}_i E^{\phantom{}}_{ij}  {\mathrm{e}}^{\mathrm{f0}}_j & \text{in } &\Omega & & \tag{Box1.2d} \label{eq:B12d}
\end{align}
Constitutive relation:
\begin{align}
  \sigma^{\phantom{}}_{ij} &= \frac{1}{J} F^{\phantom{}}_{ik} \left( S^{\mathrm{pas}}_{kl} + S^{\mathrm{act}}_{kl} \right) F^{\phantom{}}_{jl} & \text{in } &\Omega  & \times & (0, t^{\mathrm{cycle}}] \tag{Box1.3a} \label{eq:B13a}\\
  S^{\mathrm{pas}}_{ij} &= \frac{\partial }{\partial E^{\phantom{}}_{ij}} \left( a_0 [ \mathrm{exp}\left( Q \right) - 1    ] + \frac{1}{2} \kappa [ J^2 - 1    ]^2 \right) & \text{in } &\Omega_0 & & \tag{Box1.3b} \label{eq:B13b}\\
  S^{\mathrm{act}}_{ij} &= \frac{l^{\mathrm{s0}}}{l^{\mathrm{s}}}f^{\mathrm{iso}}\left( l^{\mathrm{c}} \right) \ f^{\mathrm{twitch}}\left(t^{\mathrm{a}}, l^{\mathrm{s}}  \right) \ E^{\mathrm{a}} \left( l^{\mathrm{s}} - l^{\mathrm{c}} \right) \ {\mathrm{e}}^{\mathrm{f0}}_i {\mathrm{e}}^{\mathrm{f0}}_j & \text{in } &\Omega_0  & \times & (0, t^{\mathrm{cycle}}]\tag{Box1.3c} \label{eq:B13c}\\
  Q &=  a_1 I_1^2 - a_2 I_2 + a_3  I_4^2  & \text{in } &\Omega_0 & & \tag{Box1.3d} \label{eq:B13d} \\
  l^{\mathrm{s}} &= l^{\mathrm{s0}} \sqrt{2 I_4 + 1}  & \text{in } &\Omega_0 & & \tag{Box1.3e} \label{eq:B13e}
\end{align}
\end{MyBox}

\begin{MyBox}[label={box:fomrom},nameref={Overview of contraction and circulatory model}]{Overview of the contraction and circulatory model for both the FOM and the ROM (based on Ref.~\cite{willems_isogeometric_2024})}
Active contraction functions:
\begin{align}
    f^{\mathrm{iso}} \left( l^{\mathrm{c}} \right ) &= \left\{\begin{matrix} T^0 \text{tanh}^2 \left[ a^{l} \left( l^{\mathrm{c}} - l^{\mathrm{c0}}\right) \right] & l^{\mathrm{c}} \geq l^{\mathrm{c0}} \\   0 & l^{\mathrm{c}} < l^{\mathrm{c0}}  \end{matrix}\right. & \text{in } & \Omega_0 &\times & (t^{\mathrm{act}}, t^{\mathrm{cycle}}] \tag{Box2.1a} \label{eq:B21a}\\    
    f^{\mathrm{twitch}} \left( t^{\mathrm{a}}, l^{\mathrm{s}} \right ) &= \left\{\begin{matrix}
    0  &  t^{\mathrm{a}} < 0 \\
    \mathrm{tanh}^2 \left( \frac{t^{\mathrm{a}}}{\tau^{\mathrm{r}}}  \right ) \mathrm{tanh}^2 \left( \frac{t^{\mathrm{max}}-t^{\mathrm{a}}}{\tau^{\mathrm{d}}}\right) & 0 \leq t^{\mathrm{a}} \leq t^{\mathrm{max}} \\
    0  & t^{\mathrm{a}} > t^{\mathrm{max}} 
    \end{matrix}\right. & \text{in } & \Omega_0 &\times & (t^{\mathrm{act}},t^{\mathrm{cycle}}] \tag{Box2.1b} \label{eq:B21b} \\
    t^{\mathrm{max}} &= b \left(  l^{\mathrm{s}} - l^{\mathrm{d}} \right) & \text{in } & \Omega_0 &\times & (t^{\mathrm{act}},t^{\mathrm{cycle}}] \tag{Box2.1c} \label{eq:B21c} 
\end{align}
Contractile dynamics:
\begin{align}
  \frac{d l^{\mathrm{c}}}{dt} &= \left[ E^{\mathrm{a}} \left( l^{\mathrm{s}} - l^{\mathrm{c}} \right) - 1\right] v^{\mathrm{0}} & \text{in } & \Omega_0 &\times & (t^{\mathrm{act}}, t^{\mathrm{cycle}}] \tag{Box2.2a} \label{eq:B22a}\\
  l^{\mathrm{c}} &= l^s & \text{in } & \Omega_0 &\times & (0, t^{\mathrm{act}}] \tag{Box2.2b} \label{eq:B22b}
\end{align}  
Circulatory model:
\begin{align}
  \mathcal{M}_{\mathrm{circ}} \left( p, V , \tilde{p}\right) &= 0 & \text{in } & & & (0, t^{\mathrm{cycle}}] \tag{Box2.3} \label{eq:B23}
\end{align}
\end{MyBox}

\subsection{Reduced-order model: one-fiber cardiac modeling}\label{sec:lowfidelity}
While the isogeometric FOM is relatively fast compared to standard finite element models, the time-dependent nonlinear continuum characteristics make it computationally still too demanding for direct usage in a clinical setting. The computational complexity of the FOM resides mainly in the continuum mechanics component, as simplified-physics descriptions are used for the active tissue contraction and the circulatory system model. Therefore, in the developed ROM, the continuum mechanics component is substituted by a simplified-physics relation. The active tissue contraction model and circulatory system model as outlined in Box~\ref{box:fomrom} are used in unaltered form, as these are not critical for the computational costs.

From an abstract perspective, the continuum mechanics component of the FOM provides a relation between the pressure in the ventricle and its cavity volume. By making additional assumptions regarding the geometry and mechanical behavior of the ventricle, the continuum equations can be simplified up to a point that an analytical approximation of this pressure-volume relation can be found. A variety of such approximations has been discussed in the literature \cite{arts_adaptation_2005,lumens_three-wall_2009,bovendeerd_modeling_2012,huntjens_influence_2014}, where the variations are a result of differences in assumptions. A common characteristic of these approximations is their insensitivity to local geometrical changes, as often the only geometric parameters involved are the cavity and wall volumes. In our ROM framework, we aim to introduce local geometric sensitivity by calibrating a sufficiently versatile simplified-physics ROM to the FOM. For this ROM we consider a generalization of the one-fiber model originally proposed by Arts \emph{et al.}~\cite{arts_relation_1991}. We introduce this model in Section~\ref{sec:genonefiber}, after which we discuss its relation to models established in the literature in Section~\ref{sec:genrelations}. 

\subsubsection{The generalized one-fiber model}\label{sec:genonefiber}
A one-fiber model provides an analytical relation between the global left-ventricular pump function, \emph{i.e.}, the cavity pressure, $p$, and volume, $V$, and the local wall tissue function, \emph{i.e.}, the fiber stress, $\tau_{\mathrm{fiber}}$, of the form 
\begin{equation}\label{eq:fmechanicalequilibrium}
    \frac{\tau_{\mathrm{fiber}}}{p} =  f\left(  \frac{V}{V_{\mathrm{w}}} \right),
\end{equation}
where $V_{\mathrm{w}}$ is the ventricle wall volume and $f(\cdot)$ is an arbitrary function. By equating the global pumping work to the fiber-generated work, \emph{i.e.},
\begin{equation}\label{eq:work}
    p\ \mathrm{d}V = V_{\mathrm{w}} \tau_{\mathrm{fiber}}\ \mathrm{d}\epsilon_{\mathrm{f}},
\end{equation}
the fiber strain is expressed in terms of the cavity and wall volumes as
\begin{equation}
    \epsilon_{\mathrm{fiber}} =  \int_{V_0}^{V} \frac{1}{f\left(  \sfrac{V}{V_{\mathrm{w}}} \right)} \ \mathrm{d}\left( \frac{V}{V_{\mathrm{w}}} \right).
    \label{eq:ffiberstrain}
\end{equation}
To obtain a relation between the cavity pressure and volume, these relations are complemented with a constitutive law that relates the fiber stress to the fiber strain.

For the generalized one-fiber model -- which is summarized in Box~\ref{box:rom} -- we consider the linear function
\begin{equation}
    f\left(  \frac{V}{V_{\mathrm{w}}} \right) = 2 \alpha + 3 \beta \frac{V}{V_{\mathrm{w}}},
    \label{eq:fgenonefiber}
\end{equation}
where -- as will be discussed in Section~\ref{sec:genrelations} -- the parameters $\alpha$ and $\beta$ are interpreted as correction factors that take into consideration aspects such as the shape of the ventricle (other than the cavity and wall volume), restraints and fiber orientation. The mechanical equilibrium \eqref{eq:B31} and fiber strain \eqref{eq:B32} follow from substitution of Equation~\eqref{eq:fgenonefiber} into Equations \eqref{eq:fmechanicalequilibrium} and \eqref{eq:ffiberstrain}, respectively. The constitutive relation in Box~\ref{box:rom}.3 is adopted from Bovendeerd \emph{et al.}~\cite{bovendeerd_dependence_2006}. Similar to the FOM model discussed above, this constitutive relation decomposes the fiber stress in an active and a passive part \eqref{eq:B33a}. Both stress components depend on the sarcomere length \eqref{eq:B33d}, the extension of which relative to the undeformed length, $l^{\rm s0}$, is directly related to the fiber strain \eqref{eq:B32}. The passive stress component \eqref{eq:B33b} is parametrized by the stiffnesses $T^{\rm p0}$ and $c^{\rm p}$. The active stress component \eqref{eq:B33c} is the one-dimensional version of that of the FOM as discussed in Section~\ref{sec:fommodel}.

The generalized model \eqref{eq:fgenonefiber} is sufficiently versatile to capture geometric effects, but there are limitations to what the model can represent. For example, we recognize that the model does not correct for low cavity volume ratios~\cite{bovendeerd_dependence_2006}, an aspect relevant for patients suffering from heart failure. Incorporation of such effects into our ROM framework is conceptually possible through further generalization of the simplified-physics ROM, but is considered beyond the scope of this work.

\begin{MyBox}[label={box:rom},nameref={ROM overview}]{Overview of the generalized one-fiber ROM}
Mechanical equilibrium:
\begin{align}
  \frac{\tau_{\mathrm{fiber}}}{p} &= 2 \alpha + 3 \beta \frac{V}{V_{\mathrm{w}}}\tag{Box3.1} \label{eq:B31}
\end{align}
Fiber strain:
\begin{align}
  \epsilon_{\mathrm{fiber}} &= \frac{1}{3 \beta} \ln \left(\frac{2 \alpha + 3\beta \frac{V}{V_{\mathrm{w}}}}{2 \alpha + 3 \beta \frac{V_{0}}{V_{\mathrm{w}}}}\right) \tag{Box3.2} \label{eq:B32}
\end{align}
Constitutive relation:
\begin{align}
  \tau_{\mathrm{fiber}} &= \frac{l^{\mathrm{s}}}{l^{\mathrm{s0}}} \left( T^{\mathrm{pas}}_{\mathrm{fiber}} + T^{\mathrm{act}}_{\mathrm{fiber}}  \right)  \tag{Box3.3a} \label{eq:B33a}\\ 
  T^{\mathrm{pas}}_{\mathrm{fiber}} &=  \left\{\begin{matrix}
0 & l^{\mathrm{s}} \leq l^{\mathrm{s0}} \\
T^{\mathrm{p0}}\left[ \mathrm{exp}\left( c^{\mathrm{p}} (l^{\mathrm{s}} - l^{\mathrm{s0}}) \right) - 1\right] &  l^{\mathrm{s}} > l^{\mathrm{s0}}\\
\end{matrix}\right.  \tag{Box3.3b} \label{eq:B33b}\\ 
  T^{\mathrm{act}}_{\mathrm{fiber}} &= f^{\mathrm{iso}}\left( l^{\mathrm{c}} \right) \ f^{\mathrm{twitch}}\left(t^{\mathrm{a}}, l^{\mathrm{s}}  \right) \ E^{\mathrm{a}} \left( l^{\mathrm{s}} - l^{\mathrm{c}} \right)\tag{Box3.3c} \label{eq:B33c}\\ 
  l^{\mathrm{s}} &= l^{\mathrm{s0}} \, \mathrm{exp}( \epsilon_{\mathrm{fiber}} ) \tag{Box3.3d} \label{eq:B33d}
\end{align}
\end{MyBox}

\subsubsection{Relation to existing models}\label{sec:genrelations}
The frequently used one-fiber model as proposed by Arts \emph{et al.}~\cite{arts_model_1979,arts_epicardial_1982} corresponds to the function
\begin{equation}
  f_{\rm cyl.}\left(  \frac{V}{V_{\mathrm{w}}} \right) = 1 + 3 \frac{V}{V_{\mathrm{w}}}.
  \label{eq:fempirical}
\end{equation}
This model was empirically derived from a cylindrical-shell model, assuming the fiber stress to be far greater than the passive tissue stress (making it more representative of the tissue response at systole). This empirical model \eqref{eq:fempirical} evidently is a special case of the generalized model \eqref{eq:fgenonefiber}, with the factors $\alpha=\frac{1}{2}$ and $\beta=1$. These factors are physiologically motivated \cite{arts_relation_1991,arts_dynamics_1989}, but the underlying assumptions leading to these factors are not evident. Since the factors are fixed in this model, its response is invariant to local geometry effects.

The continuum mechanics derivation of the one-fiber model for arbitrary axisymmetric geometries presented by Arts \emph{et al.}~\cite{arts_relation_1991} enhances its interpretability. Assuming homogeneous fiber stress and alignment of the fibers with the tissue pressure isobars, the nonlinear relation 
\begin{equation}\label{eq:of_log}
    f_{\mathrm{rsym.}}\left(  \frac{V}{V_{\mathrm{w}}} \right) = 3 \myln*{1 + \frac{V_{\mathrm{w}}}{V}}^{-1}
\end{equation}
is obtained. An alternative derivation, assuming a spherical geometry, was later presented by Bovendeerd \emph{et al.}~\cite{bovendeerd_dependence_2006}. The invariance of this nonlinear model to local geometry effects is a result of the homogeneity and rotational symmetry assumptions. The empirical relation \eqref{eq:fempirical} is within $8.4\%$ of the nonlinear model \eqref{eq:of_log} over the physiological range, $\frac{V}{V_{\rm w}} \in [0.15-0.70]$, which corroborates its mechanical soundness. As detailed in Box~\ref{box:linearization}, the empirical model \eqref{eq:fempirical} is not a Taylor series expansion of the nonlinear model \eqref{eq:of_log} around a point in the physiological regime. The Taylor series expansion that matches the empirical relation at zero cavity volume yields a generalized one-fiber model with factors $\alpha=\frac{1}{2}$ and $\beta=1\frac{1}{6}$. This linearization results in a maximum error of less than $3.4\%$ in the physiological range, stipulating that the generalized model \eqref{eq:fgenonefiber} can closely approximate the nonlinear model \eqref{eq:of_log}. The reduced compliance of the empirical model (with $\beta=1$) has been argued to be favorable from a physiological point of view, as this may be explained by the fiber orientation, shape constraints, right-ventricular asymmetry, and the presence of the mitral valve \cite{arts_relation_1991,arts_dynamics_1989}. In our ROM framework, such effects are systematically introduced into the generalized one-fiber model through calibration with the FOM.

\begin{MyBox}[label={box:linearization},nameref={Linearization}]{Linearization of the nonlinear one-fiber model \eqref{eq:of_log}}
The empirical model \eqref{eq:fempirical} can be considered as a physiologically motivated linearization of the mechanics-based model \eqref{eq:of_log}; see Ref.~\cite{arts_relation_1991}. The mathematical relation between these models has, to the best of our knowledge, not been studied in the literature. We here aim to provide additional understanding of this relation by studying the linear Taylor series expansion of the nonlinear model.
Let us consider the relation between the global pump function and fiber stress in the general form \eqref{eq:fmechanicalequilibrium},  where $f(\cdot)$ can either be a linear or a nonlinear function of $\frac{V}{V_{\mathrm{w}}}$. The linear Taylor series expansion of $f(\cdot)$ around an arbitrary point $\eta$ equals
\begin{equation}
    f_{\mathrm{lin}} = f(\eta) + f'(\eta) \left(\frac{V}{V_{\mathrm{w}}} - \eta \right),
\end{equation}
which results in the linear approximation
\begin{equation}\label{eq:linearized}
     \frac{\tau_{\mathrm{fiber}}}{p} \approx \underbrace{[f(\eta) - \eta f'(\eta)  ]}_{\alpha^*} + \underbrace{f'(\eta)}_{\beta^*} \frac{V}{V_{\mathrm{w}}}.
\end{equation}
Application to the nonlinear model \eqref{eq:of_log} proposed by Arts \emph{et al.}~\cite{arts_relation_1991} results in the  linearization coefficients
\begin{equation}\label{eq:lin_req}
    \begin{aligned}
        \alpha^*  &= 3 \myln*{ 1 +  \frac{1}{\eta} }^{-1} - \eta \beta^*,\\
        \beta^* &= 3 \left[ \eta \myln*{1 +  \frac{1}{\eta}} \right]^{-2} \left( 1 +  \frac{1}{\eta} \right)^{-1}. 
    \end{aligned} 
\end{equation}

For the empirical relation \eqref{eq:fempirical} to be a Taylor approximation of the nonlinear model \eqref{eq:of_log}, the coefficient should equate to $\alpha^*=1$ and $\beta^*=3$ for a given $\eta$. However, the condition that $\beta^*=3$ can only be satisfied when $\eta \rightarrow \infty$, which is not a sensible expansion point and does not result in $\alpha^*=1$ (instead $\alpha^*=\frac{3}{2}$ is found). This conveys that the empirical model \eqref{eq:fempirical} cannot be interpreted as a linear Taylor series expansion of \eqref{eq:of_log}. By making a Taylor series expansion around a point in the the physiological regime, \emph{i.e.}, taking $\eta \in [0.15-0.70]$, the mathematical approximation error of the empirical model can be drastically reduced. This is illustrated in Figure~\ref{fig:OFcomparison}, where the empirical model is compared to a Taylor series expansion around $\eta=0.345$. This expansion yields $\alpha^*=1$ and $\beta^*=3\frac{1}{2}$and results in a maximum error of less than $3.4\%$ over the physiological range compared to the nonlinear model, which substantially improves upon the empirical model error of approximately $8.5\%$.
\par\medskip
\begin{minipage}{\linewidth}
\centering
\begin{tabular}{c@{\hspace{2em}}c}
  \includegraphics[width=0.45\textwidth]{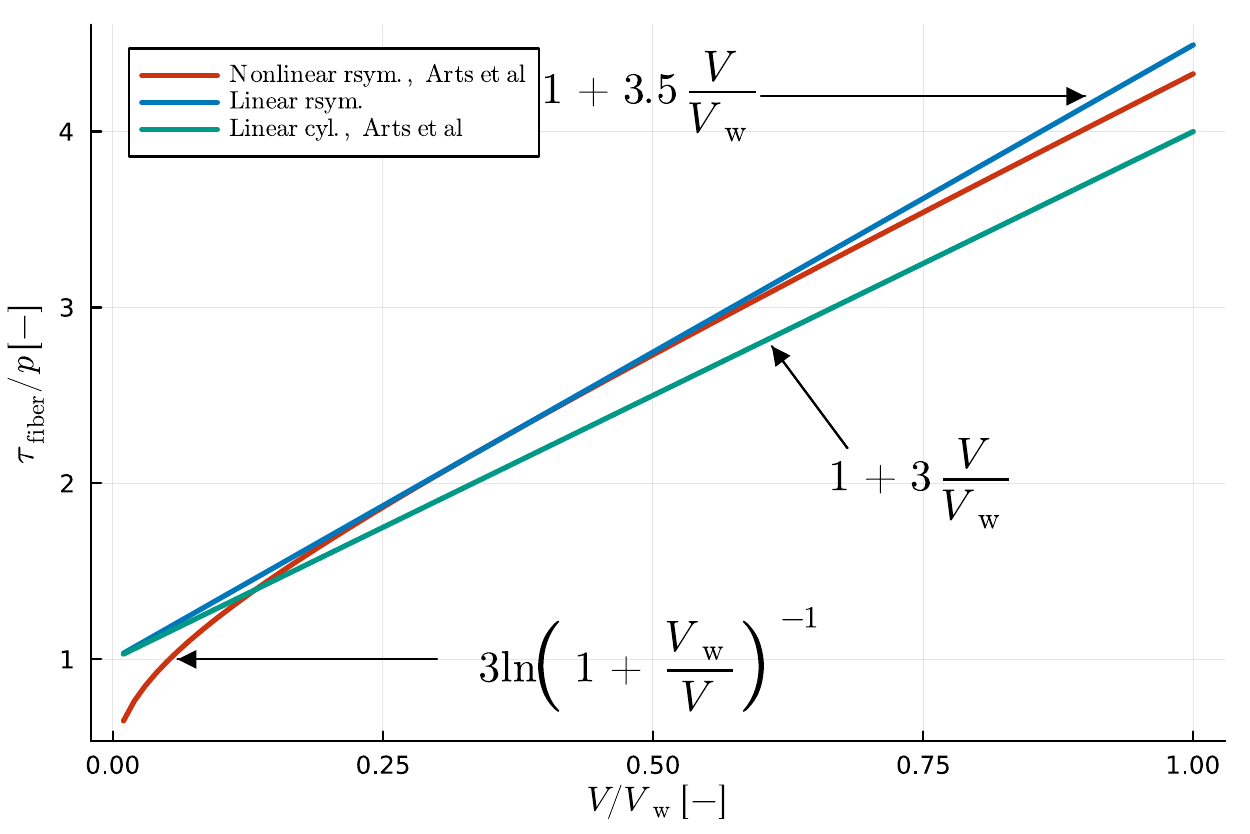} &
  \includegraphics[width=0.45\textwidth]{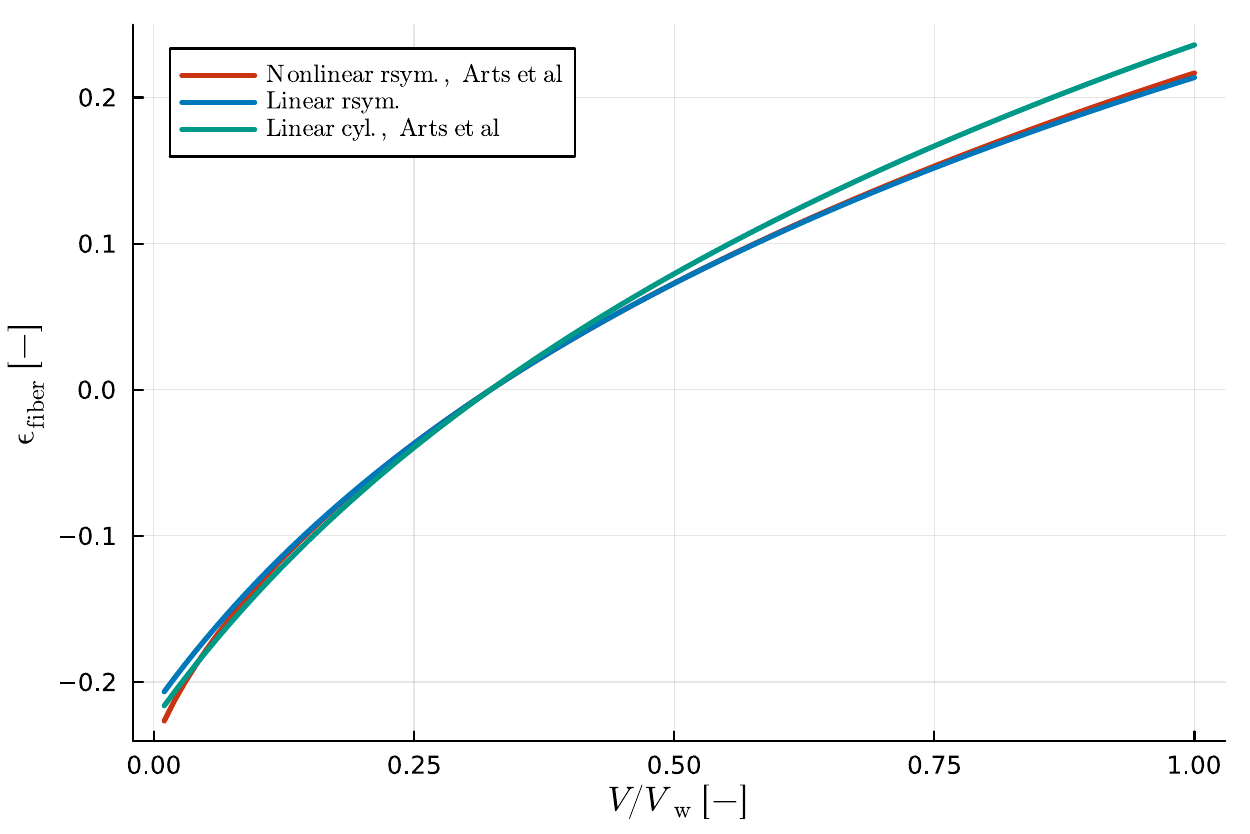} \\
\parbox{0.cm}{(a)} &
\parbox{0.cm}{(b)}
\end{tabular}
\captionof{figure}{Comparison between the nonlinear rotational-symmetric (nonlinear rsym.) function proposed by Arts \emph{et al.}~\cite{arts_relation_1991}, a linear Taylor series expansion of this function around $\eta=0.345$ (linear rsym.), and the linear empirical relation (linear cyl.) proposed by Arts \emph{et al.}~\cite{arts_epicardial_1982}. (a) Fiber stress-pressure ratio for varying $\sfrac{V}{V_{\mathrm{w}}}$ values; (b)  Fiber strain comparison for varying $\sfrac{V}{V_{\mathrm{w}}}$ values.}\label{fig:OFcomparison}
\end{minipage}
\end{MyBox}

\subsection{Coupling the FOM and ROM: calibration of correction factors}
\label{sec:fomromrelations}
Our reduced-order modeling framework (Figure~\ref{fig:LFmodel}) calibrates the ROM to FOM training data. While this calibration procedure will be discussed in detail in Section~\ref{sec:BI}, we here first pave the way for this discussion by 
categorizing the various model parameters. The goal of this categorization is to identify which ROM parameters are to be calibrated.

\setlength{\tabcolsep}{15pt} % Default value: 6pt
\renewcommand{\arraystretch}{1.5} % Default value: 1
\begin{table}[!h]
\caption{Categorized overview of the parameters of the full-order model (FOM) and reduced-order model (ROM). Parameters shared between the models are centered in the table.}\label{tab:relation}
\centering
 \begin{tabular}{c|lc|lc}
 & \multicolumn{2}{c|}{Full-order model} & \multicolumn{2}{c}{Reduced-order model}\\
 \hline
 \parbox[t]{2mm}{\multirow{3}{*}{\rotatebox[origin=c]{90}{Geometry}}} & NURBS control points & $\{\mathbf{x}_1,  \mathbf{x}_2, \mathbf{x}_3,...\}$ & Cavity volume (initial) & $V$\\
  & Fiber orientation & $\mathbf{e}^{\rm f0}$ & Wall volume & $V_{\mathrm{w}}$\\
  & & & Correction factors & $\alpha$, $\beta$\\
  \hline
\parbox[t]{2mm}{\multirow{3}{*}{\rotatebox[origin=c]{90}{Passive}}} & Stiffnesses & $a_0$, $a_1$, $a_2$, $a_3$ & Stiffnesses & $T^{\rm p0}= \gamma a_0$\\
& Bulk modulus & $\kappa$ & & $c^{\rm p}= 2 \lambda \left( a_1 -0.5a_2 + a_3\right)$ \\ \cline{2-5}
& \multicolumn{4}{c}{
\begin{tabular}{lc}
Sarcomere length (initial) & $l^{\rm s0}$
\end{tabular}
}\\
 \hline
 \parbox[t]{2mm}{\multirow{1}{*}{\rotatebox[origin=c]{90}{Active}}} & \multicolumn{4}{c}{
 \begin{tabular}{lc}
Contractile length (initial) & $l^{\rm c0}$\\
Contractile parameters & $E^{\rm a}$, $v^{\mathrm{0}}$\\
Cycle/activation time & $t^{\rm cycle}$, $t^{\rm act}$\\
Isometric parameters & $T^0$, $a^l$ \\
Twitch parameters & $\tau^{\rm r}$, $\tau^{\rm d}$, $b$, $l^{\rm d}$ \end{tabular}
 } \\
 \hline
 \parbox[t]{2mm}{\multirow{2}{*}{\rotatebox[origin=c]{90}{Circulation}}} & \multicolumn{4}{c}{
 \begin{tabular}{lc}
Reference volumes & $V^{\rm art0}$, $V^{\rm ven0}$\\
Compliances & $C^{\rm art}$, $C^{\rm ven}$\\
Resistances &  $R^{\rm art}$, $R^{\rm ven}$, $R^{\rm per}$\\
  \end{tabular}
 }\\
 & \multicolumn{4}{c}{}\\
 \end{tabular}
 \end{table}

The parameters introduced for both the FOM (Section~\ref{sec:fommodel}) and the ROM (Section~\ref{sec:lowfidelity}) are listed in Table~\ref{tab:relation}. The parameters are subdivided in four categories, \emph{viz.} geometry parameters, passive tissue parameters, activation parameters, and circulatory system parameters. Each of these categories corresponds to the patient-specific model inputs in Figure~\ref{fig:LFmodel}. The FOM and ROM incorporate the same activation model and circulatory system model (Box~\ref{box:fomrom}), as a result of which the parameters related to these categories are shared between the models. However, the geometry parameters and passive tissue parameters, with the exception of the sarcomere length, do differ between the FOM and the ROM. 

The FOM and ROM geometry parameters are fundamentally different. The FOM employs a complex 3D geometry encoded in the NURBS control net, including an inhomogeneous fiber field orientation. The generalized one-fiber ROM is parametrized by the cavity volume and wall volume. In addition, the FOM incorporates the correction factors, in which shape effects, \emph{a.o.}, are lumped. Through integration, the volume parameters can be directly derived from the FOM. The correction factors are to be calibrated.

The passive tissue model also differs between the FOM and ROM. Although both models have a similar exponential form, their stiffness parameters are different and cannot be formally related to one another. In our ROM framework we incorporate a parameter adapter (Figure~\ref{fig:LFmodel}) to accommodate a physically sensible relation between these stiffnesses. We postulate that the stiffness parameters are linearly related by
\begin{align}\label{eq:Stiffparams}
    T^{\mathrm{p0}} &= \gamma a_0  & &\mbox{and} & c^{\mathrm{p}} &= 2 \lambda \left( a_1 -0.5a_2 + a_3\right).
\end{align}
In these relations, $\gamma$ and $\lambda$ are correction factors in which effects from additional assumptions made on the ROM, such as its one-dimensional character, are lumped. The proposed relations should not be considered as the ground-truth, but merely as a starting point to calibrate the two models. More sophisticated and possibly fine-tuned relations may be found, but this is not considered in the current contribution.

The parameters of the ROM, $\boldsymbol{\theta}^{\rm ROM}$, can be divided into two classes, \emph{viz.} parameters that are shared with or can directly be derived from the FOM parameters, $\boldsymbol{\theta}^{\rm FOM}$, and parameters that are related to the FOM through the correction factors $\boldsymbol{\theta}^{\rm corr}=[\alpha, \beta, \gamma, \lambda]$. Through Bayesian calibration of these correction factors (Section~\ref{sec:BI}), the ROM that matches (in some sense) the FOM can be determined. The correction factors are defined such that they are expected to remain relatively close to unity. Values that deviate significantly from unity may indicate an incorrect relation between the two models, warranting reconsideration of the postulated relations in Equation~\eqref{eq:Stiffparams}.

\section{Gaussian process geometry parameter mapping}\label{sec:GP}
To introduce patient-specific geometric effects into the simplified-physics ROM, in our framework (Figure~\ref{fig:LFmodel}) a set of geometry parameters, $\mathbf{c} \in \mathbb{R}^{n^{\rm geom}}$, which provides a low-dimensional parametrization of the NURBS geometry, is mapped onto the correction factors, $\boldsymbol{\theta}^{\rm corr}=[\alpha, \beta, \gamma, \lambda]\in \mathbb{R}^{n^{\rm corr}}$ (with $n^{\rm corr}=4$), as
\begin{equation}
    \boldsymbol{\theta}^{\rm corr} = \vartheta( \mathbf{c} ).
    \label{eq:abstractmapping}
\end{equation}
This parametrization can either pertain to the geometric quantities on which the NURBS template is based (Section~\ref{sec:ApplicationS}) or to a modal decomposition based on scan data (Section~\ref{sec:ApplicationC}). To accommodate uncertainties associated with the geometry and the generalized one-fiber model, the mapping \eqref{eq:abstractmapping} is considered to be probabilistic. That is, both the geometric parameters, $\mathbf{c}$, and the correction factors, $\boldsymbol{\theta}^{\rm corr}$, are interpreted as random variables. In the proposed modeling framework, the mapping \eqref{eq:abstractmapping} is constructed based on training data of the form
\begin{equation}
    \mathbf{d}^{\rm corr} = [
    \begin{array}{cccc}
        (\mathbf{c}_1, \boldsymbol{\mu}_1^{\rm corr}, \boldsymbol{\Sigma}_1^{\rm corr}) & (\mathbf{c}_2, \boldsymbol{\mu}_2^{\rm corr}, \boldsymbol{\Sigma}_2^{\rm corr}) & \cdots & (\mathbf{c}_{n^{\rm obs}}, \boldsymbol{\mu}_{n^{\rm obs}}^{\rm corr}, \boldsymbol{\Sigma}_{n^{\rm obs}}^{\rm corr})
    \end{array} ],
    \label{eq:trainingdata}
\end{equation}
where $n^{\rm obs}$ are the number of FOM simulations that serve as observations. In the construction of this training data, the geometry parameters, $\mathbf{c}_i$, serve as the input to the FOM. Through Bayesian calibration, which will be discussed in Section~\ref{sec:BI}, the corresponding ROM correction factors, $\boldsymbol{\theta}_i^{\rm corr}$, are determined. In our training data these probabilistic factors are represented by their mean, $\boldsymbol{\mu}_i^{\rm corr}$, and covariance, $\boldsymbol{\Sigma}_i^{\rm corr}$. On account of the computational effort involved in the FOM, the number of observations at our disposal for training the mapping \eqref{eq:abstractmapping} is limited.

The mapping \eqref{eq:abstractmapping} can be modeled by a Gaussian process (GP) \cite{williams_gaussian_2006}, which is a common surrogate model choice when the amount of available data is limited and the number of parameters is moderate. This sets them apart from neural networks for example, which perform well for large numbers of parameters, but also require a lot of training data \cite{golestaneh_how_2024}. The advantages of GPs are mainly due to two factors: their close connection to radial basis function interpolation \cite{wendland_scattered_2004} and their inherently probabilistic nature. The first point intuitively explains their ability to provide good approximations even for limited amounts of data, of course depending on the smoothness of the considered mapping. The second point is what allows them to provide an uncertainty estimate along with their predictions, which is essential in the considered setting.

In Section~\ref{sec:GPscalar} we introduce the GP concept by considering its application to a scalar-valued function. The application of the concept to the vector-valued correction factor mapping \eqref{eq:abstractmapping} is then discussed in Section~\ref{sec:GPcorrection}.

\subsection{Gaussian processes for scalar-valued functions}\label{sec:GPscalar}
Let us consider the prediction of the scalar values $\theta \in \mathbb{R}$ of an unknown function $\vartheta(c)$, with variable $c \in \mathbb{R}$, based on (noisy) observations $(c_i, \mu_i, \sigma_i)$, where we assume the observations to be corrupted by Gaussian noise, such that $\mu_i$ is the mean of all observations of $\theta_i$ at $c_i$ and $\sigma_i$ is the associated standard deviation. Note that we maintain notational consistency with the mapping \eqref{eq:abstractmapping} in this scalar-valued case, but that we drop the superscript for notational convenience. A suitable GP is defined via a mean function $m: \mathbb{R} \to \mathbb{R}$, alongside a covariance (or kernel) function $k: \mathbb{R} \times \mathbb{R} \to \mathbb{R}$. Numerous covariance functions exist, but for the remainder of this article we will consider the RBF kernel
\begin{equation}\label{eq:RBFkernel}
    k(c_i, c_j) = \mathrm{exp}\left( -\frac{|c_i - c_j|^2}{2 l^2} \right),
\end{equation}
with scaling factor $l$. This kernel is one of the most common choices for predicting smooth functions, like the ones we will consider in Sections~\ref{sec:ApplicationS} and \ref{sec:ApplicationC}. Note that Equation~\eqref{eq:RBFkernel} extends to functions of multiple variables, by replacing the absolute value with a norm (and possibly scaling each variable by a different factor).

Independent of the choice of covariance function, GPs make their predictions based on Bayesian inference. Assuming that the observations $\boldsymbol{\mu} = [\mu_1, \mu_2, \cdots, \mu_{\rm n^{\rm obs}}]^{\mathrm{T}}$ and predictions $\hat{\boldsymbol{\theta}}$ at $\hat{\mathbf{c}}$, \emph{cf.}~$\hat{\theta}_i \approx \vartheta( \hat{c}_i )$, are jointly normally distributed according to
\begin{align*}
    \begin{bmatrix}
        \boldsymbol{\mu}\\
        \hat{\boldsymbol{\theta}}
    \end{bmatrix} \sim \mathcal{N} \left( \begin{bmatrix}
        \mathbf{m}\\
        \hat{\mathbf{m}}
    \end{bmatrix},\ \begin{bmatrix}
        \mathbf{K}(\mathbf{c}, \mathbf{c}, \boldsymbol{\sigma}) & \mathbf{K}(\mathbf{c}, \hat{\mathbf{c}})\\
        \mathbf{K}(\mathbf{c}, \hat{\mathbf{c}})^{\mathrm{T}} & \mathbf{K}(\hat{\mathbf{c}}, \hat{\mathbf{c}})
    \end{bmatrix} \right),
\end{align*}
where $\mathbf{m} = [m(c_1), m(c_2), \dotsc, m(c_{n^\mathrm{obs}})]^{\mathrm{T}}$ and $\hat{\mathbf{m}}$ is defined analogously, we can derive the posterior distribution (\emph{i.e.}, the prediction given the observational data) analytically \cite{williams_gaussian_2006} as
\begin{align}\label{eq:gp_posterior}
    \hat{\boldsymbol{\theta}} | \boldsymbol{\mu} \sim \mathcal{N} \left( \hat{\mathbf{m}} + \mathbf{K}(\mathbf{c}, \hat{\mathbf{c}})^{\mathrm{T}} \left( \mathbf{K}(\mathbf{c}, \mathbf{c}, \boldsymbol{\sigma}) \right)^{-1} \boldsymbol{\mu},\ \mathbf{K}(\hat{\mathbf{c}}, \hat{\mathbf{c}}) - \mathbf{K}(\mathbf{c}, \hat{\mathbf{c}})^{\mathrm{T}} \left( \mathbf{K}(\mathbf{c}, \mathbf{c}, \boldsymbol{\sigma}) \right)^{-1} \mathbf{K}(\mathbf{c}, \hat{\mathbf{c}}) \right).
\end{align}
Here we define $K_{ij}(\mathbf{c}, \mathbf{c}, \boldsymbol{\sigma}) \coloneqq k(c_i, c_j) + \sigma_i^2 \delta_{ij}$, where $\delta_{ij}$ is the Kronecker delta. The remaining covariance matrix blocks are defined analogously.

The length scale $l$, appearing in the definition of the kernel in \eqref{eq:RBFkernel}, is treated as a hyperparameter that may be optimized based on the data at hand. We herein employ the standard approach to maximize the log marginal likelihood
\begin{align}\label{eq:gp_lml}
    L(l) = -\frac{1}{2} \boldsymbol{\mu}^{\mathrm{T}} \left( \mathbf{K}(\mathbf{c}, \mathbf{c}, \boldsymbol{\sigma}) \right)^{-1} \boldsymbol{\mu} - \frac{1}{2} \log \left( \left| \mathbf{K}(\mathbf{c}, \mathbf{c}, \boldsymbol{\sigma}) \right| \right) - \frac{n^{\rm obs}}{2} \log (2 \pi),
\end{align}
in which the first term can be seen as measuring the goodness of fit, where the second term estimates the complexity of the model, and where the third term is a normalization constant. Maximizing this marginal likelihood automatically balances the goodness of fit with the model complexity.

\subsection{Application to the corrections factors}\label{sec:GPcorrection}
Conceptually it is possible to extend the scalar-valued GP concept introduced above to vector-valued functions \cite{bonilla_multi-task_2007}, such as the mapping \eqref{eq:abstractmapping}. Such an extension is not standard, however, in the sense that existing implementations are not versatile enough to directly deal with the problem at hand. Therefore, as an alternative approach for applying the Gaussian process concept to the mapping \eqref{eq:abstractmapping}, we consider the correction factors to be modeled by the multi-variate Gaussian distribution
\begin{align}
   \boldsymbol{\theta}^{\rm corr} &\sim \mathcal{N}(\boldsymbol{\mu}^{\rm corr}, \boldsymbol{\Sigma}^{\rm corr})    &   &\mbox{with}    &  \boldsymbol{\Sigma}^{\rm corr} &= {\rm diag}( \boldsymbol{\sigma}^{\rm corr} ) \, \boldsymbol{\rho}^{\rm corr} \, {\rm diag}( \boldsymbol{\sigma}^{\rm corr} ),
\end{align}
where $\boldsymbol{\mu}^{\rm corr}$ and $\boldsymbol{\sigma}^{\rm corr}$ are the mean and standard deviation of $\boldsymbol{\theta}^{\rm corr}$, respectively, and where $\boldsymbol{\rho}^{\rm corr}$ is its correlation matrix. We then approximate each of the components of $\boldsymbol{\theta}^{\rm corr}$ by a scalar-valued Gaussian process with mean $\mu_i^{\rm corr}$ and standard deviation $\sigma_i^{\rm corr}$, as well as the off-diagonal elements of the (symmetric) correlation matrix, $\rho_{ij}^{\rm corr}$, with zero noise. This alternative approach to construct a vector-valued GP is detailed in Algorithm~\ref{alg:GP}. Note that training data for both the correction factor components and the elements of the correlation matrix can be obtained from Equation~\eqref{eq:trainingdata}, \emph{i.e.}, the input parameters of this algorithm.

\begin{algorithm}
\caption{Gaussian process-based prediction of the correction factors.}\label{alg:GP}
    \begin{algorithmic}[1]
        \Require $\hat{\mathbf{c}}$, $\mathbf{d}^{\rm corr} = [(\mathbf{c}_1, \boldsymbol{\mu}_1^{\rm corr}, \boldsymbol{\Sigma}_1^{\rm corr}) ~ \cdots ~ (\mathbf{c}_{n^{\rm obs}}, \boldsymbol{\mu}_{n^{\rm obs}}^{\rm corr}, \boldsymbol{\Sigma}_{n^{\rm obs}}^{\rm corr})]$  \Comment{Geometry parameters for prediction; Training data}
        \Statex
        \For{$i \in \{1, \ldots,n^{\rm geom}\}$} \Comment{Iterate over the parameter directions}
            \State $\mathbf{c}^i = [c_{k,i}^{\rm corr}~\mathbf{for}~k \in \{1, \ldots,n^{\rm obs}\}]$ \Comment{Coordinates in the $i^{\rm th}$ parameter direction}
            \State $\boldsymbol{\mu}^i = [\mu_{k,i}^{\rm corr}~\mathbf{for}~k \in \{1, \ldots,n^{\rm obs}\}]$ \Comment{Means in the $i^{\rm th}$ parameter direction}
            \State $\boldsymbol{\sigma}^i = [\sqrt{\Sigma_{k,ii}^{\rm corr}}~\mathbf{for}~k \in \{1, \ldots,n^{\rm obs}\}]$ \Comment{Standard deviations in the $i^{\rm th}$ parameter direction}
            \Statex
            \State $\vartheta^i = \mathcal{GP}(\mathbf{c}^i, \boldsymbol{\mu}^i, \boldsymbol{\sigma}^i)$ \Comment{Gaussian process for the $i^{\rm th}$ parameter}
            \Statex
            \State $\hat{\mu}^i, \hat{\sigma}^i = \vartheta^i(\hat{c}_i)$ \Comment{Prediction for the $i^{\rm th}$ parameter}    
            \Statex
            \State $\hat{\Sigma}^{ii} = (\hat{\sigma}^i)^2$ \Comment{Diagonal terms of the prediction covariance matrix} 
            \For{$j \in \{i+1, \ldots,n^{\rm geom}\}$} \Comment{Iterate over the off-diagonal terms of the covariance matrix}
                \State $\boldsymbol{\rho}^{ij} = \boldsymbol{\rho}^{ji} = [\frac{\Sigma_{k,ij}^{\rm corr}}{\sqrt{\Sigma_{k,ij}^{\rm corr} \Sigma_{k,jj}^{\rm corr}}}~\mathbf{for}~k \in \{1, \ldots,n^{\rm obs}\}]$ \Comment{Correlation coefficients}
                \Statex
                \State $\vartheta^{ij} = \mathcal{GP}(\mathbf{c}^i, \boldsymbol{\rho}^{ij}, \mathbf{0})$ \Comment{Zero-noise Gaussian process for the correlation coefficients}
                \Statex
                \State $\hat{\rho}_{ij} = \vartheta^i(\hat{c}_{ij})$ \Comment{Prediction correlation coefficients}
                \State $\hat{\Sigma}^{ij} = \hat{\rho}_{ij} \hat{\sigma}^i \hat{\sigma}^j$ \Comment{Off-diagonal terms of the prediction covariance matrix}
            \EndFor 
        \EndFor
        \Statex 
        \State $\lambda_{\rm min} = {\rm min}( \ {\rm eig}( \hat{\boldsymbol{\Sigma}} ), \ 0\ )$ \Comment{Smallest eigenvalue of the predicted covariance matrix}
        \State $\hat{\boldsymbol{\Sigma}} = \hat{\boldsymbol{\Sigma}} - \lambda_{\rm min} \mathbf{I}$ \Comment{Restore (semi-)positive definiteness if required}
        \Statex
        \State \Return $\mathcal{N}(\hat{\boldsymbol{\mu}}, \hat{\boldsymbol{\Sigma}})$
    \end{algorithmic}
\end{algorithm}

In our framework, the GP is used in a sequential learning setting, \emph{i.e.}, only a small number of observations is available at first and more are collected and added to the training data at a later point (Figure~\ref{fig:LFmodel}). Our strategy for introducing new observations is to consider whether a new observation is sufficiently far away from the ones already in the training set. That is, we assess whether the newly considered geometry parameters, $\mathbf{c}$, do not closely resemble any of those already in the training set. This criterion for introducing new points into the GP is \emph{ad hoc} and ideally should be replaced by a criterion based on an estimate of the approximation error of the GP. While the development of a rigorous error-based criterion is beyond the scope of this work, in Section~\ref{sec:ApplicationC} we discuss the possibility to use the widths of the uncertainty bands as an error indicator.

\begin{figure}
\begin{center}
    \begin{subfigure}[t]{0.33\textwidth}
    \begin{center}
        \includegraphics[width=\textwidth]{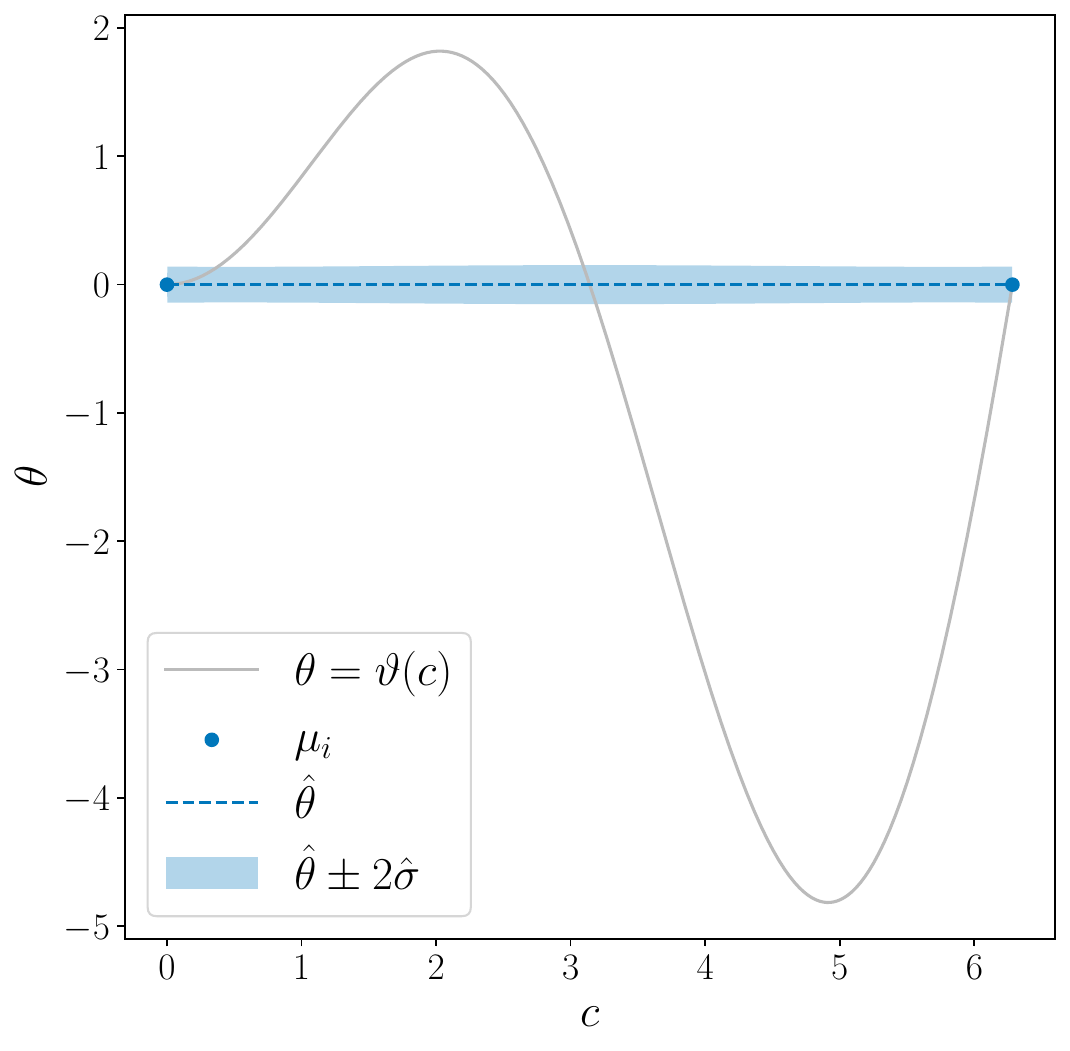}
        \caption{}
        \label{fig:gp_example_a}
    \end{center}
    \end{subfigure}
    \begin{subfigure}[t]{0.33\textwidth}
    \begin{center}
        \includegraphics[width=\textwidth]{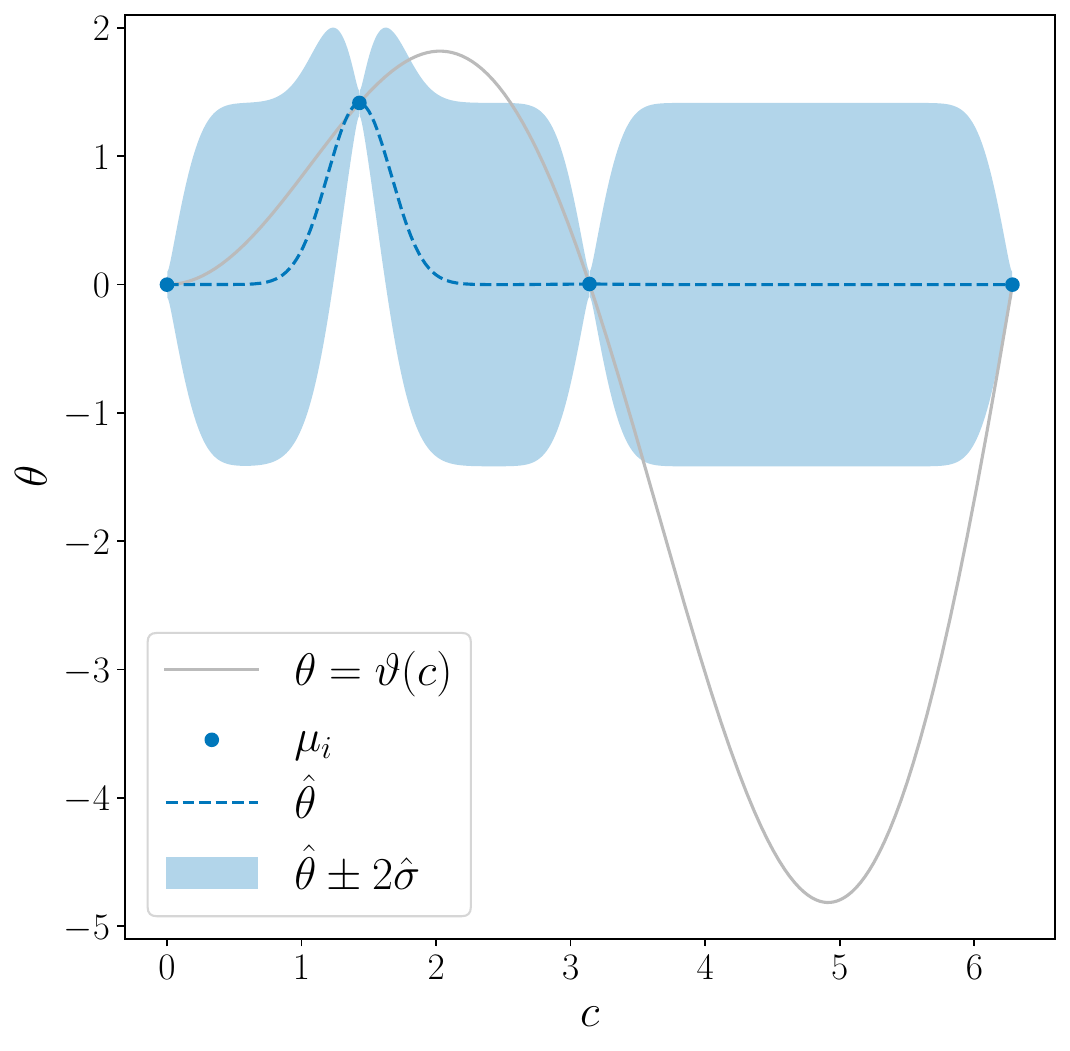}
        \caption{}
        \label{fig:gp_example_b}
    \end{center}
    \end{subfigure}
    \begin{subfigure}[t]{0.33\textwidth}
    \begin{center}
        \includegraphics[width=\textwidth]{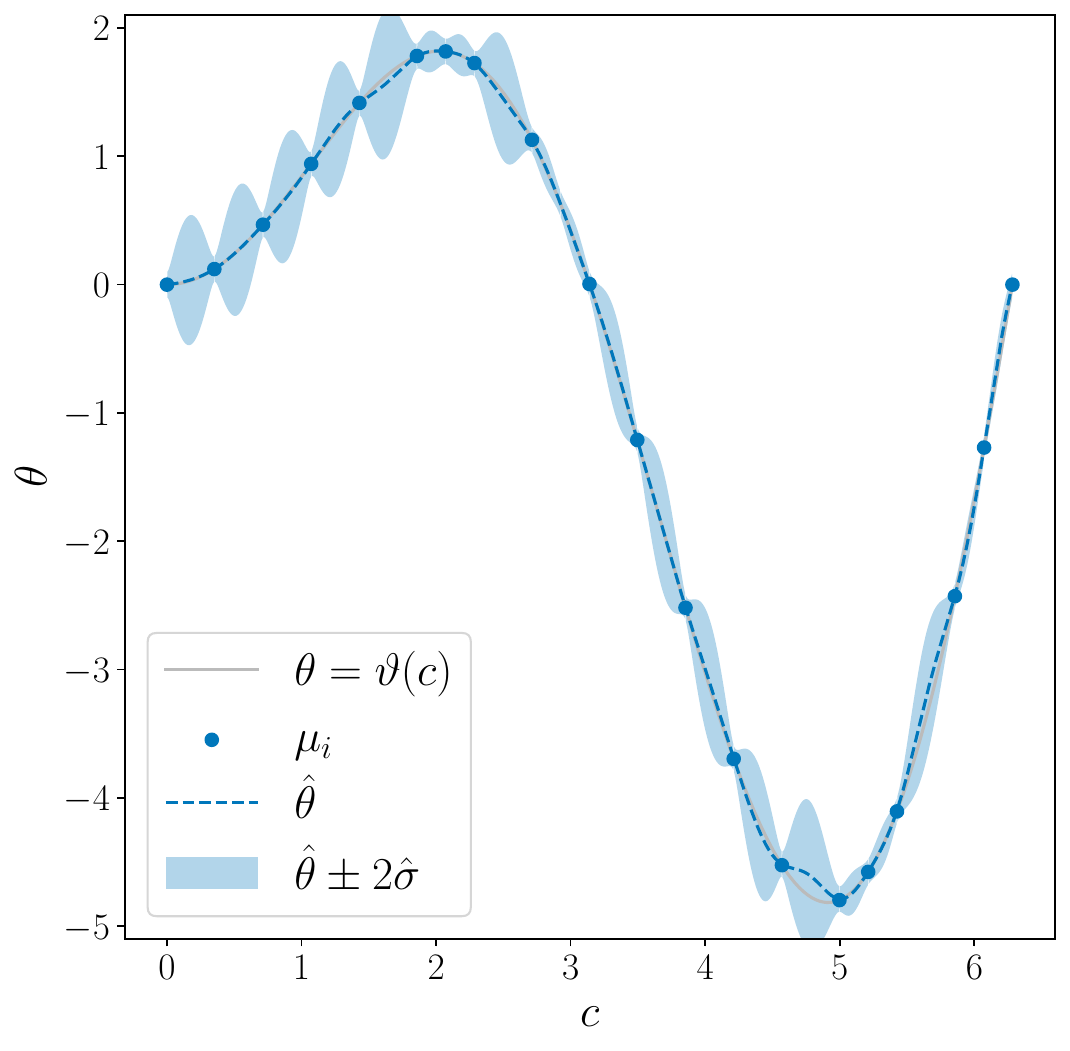}
        \caption{}
        \label{fig:gp_example_c}
    \end{center}
    \end{subfigure}
    \caption{Example illustrating sequential learning using a Gaussian process. (a) The initial fit between the GP prediction and only two observations or data points. (b) The fit after the addition of two data points. (c) The fit after adding more data points and maximizing the marginal likelihood.}
    \label{fig:gp_example}
\end{center}
\end{figure}

Figure~\ref{fig:gp_example} illustrates the sequential learning strategy on a simple example function, where Figure~\ref{fig:gp_example_a} shows the initial fit without maximizing the marginal likelihood, \emph{i.e.}, $l$ is set to twice the size of the input domain ($l = 12$). The prediction matches the observed data almost exactly. However, we observe that the GP is overconfident in this case and underestimates the complexity of the ground truth. To avoid this, an appropriate length scale should be selected by maximizing the marginal likelihood. Adding more data points also improves both the fit and the variance estimate, as Figure~\ref{fig:gp_example_b} shows. Finally, Figure~\ref{fig:gp_example_c} shows the fit after adding more data points and maximizing the marginal likelihood.

\section{Bayesian ROM calibration}\label{sec:BI}
The training of the Gaussian process that maps the geometry parameters, $\boldsymbol{c}$, onto the geometric correction factors, $\boldsymbol{\theta}^{\rm corr}=[\alpha, \beta, \gamma, \lambda]$, is based on data points obtained from FOM simulations \emph{cf.}~Equation \eqref{eq:trainingdata}. This training data for the Gaussian process takes into account uncertainties, in the sense that it comprises both the mean, $\boldsymbol{\mu}^{\rm corr}$, and the covariance, $\boldsymbol{\Sigma}^{\rm corr}$, of the geometric correction factors for each FOM simulation in the training set. To determine these statistical moments from a FOM simulation, Bayesian inference is employed as a probabilistic model calibration procedure \cite{lambert_students_2018, biehler_towards_2015}.

To calibrate the geometric correction factors, a quantity of interest is required which is sensitive to these ROM parameters, and which can be modeled by both the FOM and the ROM. The steady-state pressure-volume loop is a clinically relevant quantity of interest that meets these requirements and is therefore suitable for calibration. We represent this quantity of interest by the concatenated vector
\begin{equation}
\mathbf{q} =  \left[ p(t^{\mathrm{start}}), p(t^{\mathrm{start}} + \Delta t), \ldots, p(t^{\mathrm{end}}), V(t^{\mathrm{start}}), V(t^{\mathrm{start}}+\Delta t), \ldots, V(t^{\mathrm{end}}) \right] \in \mathbb{R}^{2n},
 \label{eq:pvdata}
\end{equation}
where $t^{\mathrm{start}}$ and $t^{\mathrm{end}}$ respectively correspond to the start and end time of the steady-state cycle, and $n$ is the number of time steps.

The calibration of the geometric correction factors is performed in a scenario, \emph{i.e.}, for a specific setting of the FOM parameters, $\boldsymbol{\theta}^{\rm FOM}$. Except for the calibration parameters, $\boldsymbol{\theta}^{\rm corr}$, all ROM parameters, $\boldsymbol{\theta}^{\rm ROM}$, are directly inherited from the FOM (see Section~\ref{sec:fomromrelations}). This means that, for a specific calibration scenario and selection of calibration parameters, the quantity of interest \eqref{eq:pvdata} can be evaluated using both models. In the setting considered here, the quantity of interest determined from the FOM is used to generate synthetic data of the form
\begin{equation}
    \mathbf{d}^{p-V} = \mathbf{q}^{\rm FOM}( \boldsymbol{\theta}^{\rm FOM}) + \boldsymbol{\nu}^{\rm FOM},
    \label{eq:FOMdata}
\end{equation}
where $\boldsymbol{\nu}^{\rm FOM}$ represents the data noise. This noise contribution, which will be discussed in detail below, is in this particular setting primarily associated with the bias of the FOM.

\subsection{Bayesian inference}\label{ssec:BIformulation}
\begin{figure}
    \centering
    \includegraphics[width=0.5\linewidth]{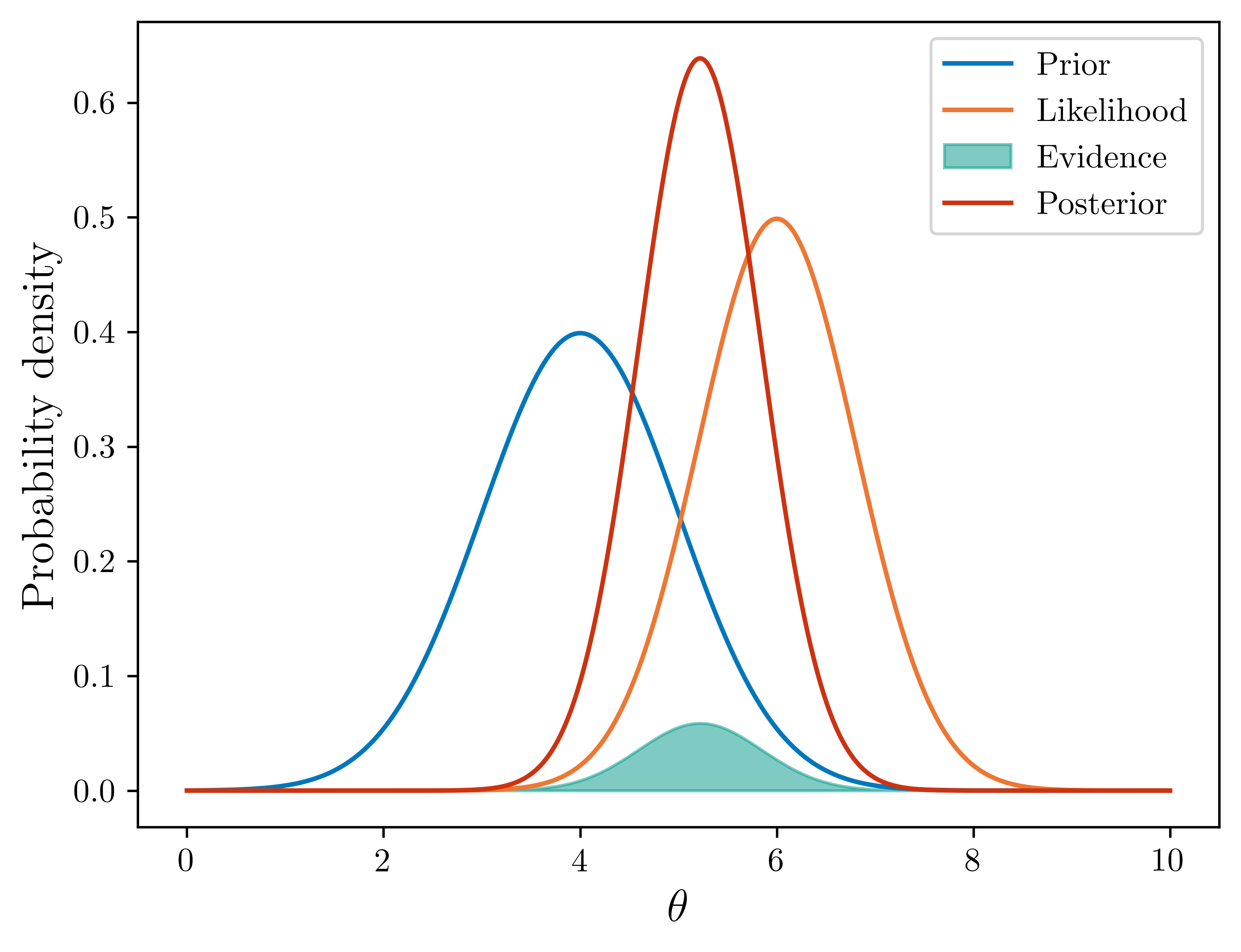}
    \caption{Illustration of Bayes' rule \eqref{eq:bayes} with a single parameter $\theta$ and a single data point at $d=6$. The prior is set to $\mathcal{N}(4,1)$ and the likelihood noise to $\Sigma_{\rm noise}=0.8^2$. For this case, the evidence (green area) is equal to 0.092.}
    \label{fig:bayes}
\end{figure}

In Bayesian inference, calibration pertains to determining the probability density function for the calibration parameters (\emph{i.e.}, the geometric correction factors), conditional to the data, \emph{i.e.}, $\rho( \boldsymbol{\theta}^{\rm corr} | \mathbf{d}^{p-V} )$. This conditional probability is referred to as the \emph{posterior}, which expresses the probability of the calibration parameters given the data. The statistical moments of the correction factors, required as training data for the Gaussian process, follow directly from the posterior distribution as
\begin{equation}
    \boldsymbol{\mu}^{\rm corr} = \int \boldsymbol{\theta}^{\rm corr} \, \rho( \boldsymbol{\theta}^{\rm corr} | \mathbf{d}^{p-V} ) \, {\rm d}\boldsymbol{\theta} \quad  \mbox{and} \quad \boldsymbol{\Sigma}^{\rm corr} = \int (\boldsymbol{\theta}^{\rm corr} - \boldsymbol{\mu}^{\rm corr}) \otimes (\boldsymbol{\theta}^{\rm corr} - \boldsymbol{\mu}^{\rm corr})\, \rho( \boldsymbol{\theta}^{\rm corr} | \mathbf{d}^{p-V} ) \, {\rm d}\boldsymbol{\theta}.
    \label{eq:moments}
\end{equation}
The posterior follows from Bayes' rule (illustrated in Figure~\ref{fig:bayes})
\begin{equation}
    \rho( \boldsymbol{\theta}^{\rm corr} | \mathbf{d} )= \frac{\mathcal{L}( \boldsymbol{\theta}^{\rm corr} | \mathbf{d}^{p-V} ) \rho(\boldsymbol{\theta}^{\rm corr})}{\rho(\mathbf{d}^{p-V})},
    \label{eq:bayes}
\end{equation}
where the probability density function $\rho(\boldsymbol{\theta}^{\rm corr})$ is the \emph{prior},  $\mathcal{L}( \boldsymbol{\theta}^{\rm corr} | \mathbf{d}^{p-V} ):= \rho( \mathbf{d}^{p-V} | \boldsymbol{\theta}^{\rm corr} )$ is the likelihood, and the normalization factor $\rho(\mathbf{d})$ is the model \emph{evidence}. Existing knowledge about the parameters, \emph{e.g.}, physical bounds, is encoded into the prior distribution. The likelihood function -- which expresses the probability that the data, $\mathbf{d}^{p-V}$, is observed given a particular choice of the geometric correction factors -- follows from equating the FOM data Equation~\eqref{eq:FOMdata} with its ROM equivalent to yield
\begin{equation}
    \mathbf{d}^{p-V} - \mathbf{q}^{\rm ROM}(\boldsymbol{\theta}^{\rm ROM}) = \boldsymbol{\nu}^{\rm FOM} + \boldsymbol{\nu}^{\rm ROM},
\end{equation}
where the noise contribution, $\boldsymbol{\nu}^{\rm ROM}$, is associated with the bias of the ROM. Assuming a multi-variate normal distribution with zero mean and covariance matrix $\boldsymbol{\Sigma}^{\rm noise} \in \mathbb{R}^{2n\times 2n}$ for the combined noise model,
\begin{equation}
    \boldsymbol{\nu} := \boldsymbol{\nu}^{\rm FOM} + \boldsymbol{\nu}^{\rm ROM} \sim \mathcal{N}(\boldsymbol{0}, \boldsymbol{\Sigma}_{\rm noise}),
    \label{eq:noisemodel}
\end{equation}
the FOM data follows the distribution
\begin{equation}
     \boldsymbol{d}^{p-V} \sim \mathcal{N}(\boldsymbol{q}^{\rm ROM}(\boldsymbol{\theta}^{\rm ROM}), \boldsymbol{\Sigma}_{\rm noise}).
\end{equation}
Expressing the quantity of interest as a function of the calibration parameters, \emph{i.e.}, $\boldsymbol{q}^{\rm corr}(\boldsymbol{\theta}^{\rm corr}):=\boldsymbol{q}^{\rm ROM}(\boldsymbol{\theta}^{\rm ROM})$, the likelihood function can then be expressed as
\begin{equation}
    \mathcal{L}( \boldsymbol{\theta}^{\rm corr} | \mathbf{d}^{p-V} ) = \frac{1}{\sqrt{(2\pi)^{2n} | \boldsymbol{\Sigma}_{\rm noise} |}}\exp{\left( -\frac{1}{2}(\mathbf{d}^{p-V}-\mathbf{q}^{\rm corr}(\boldsymbol{\theta}^{\rm corr}))^T \boldsymbol{\Sigma}_{\rm noise}^{-1} (\mathbf{d}^{p-V}-\mathbf{q}^{\rm corr}(\boldsymbol{\theta}^{\rm corr}))\right)}.
    \label{eq:likelihoodfunction}
\end{equation}

As the noise model \eqref{eq:noisemodel} is associated with the FOM and ROM biases, we assume the noise on the pressure and volume data to be temporally correlated, but uncorrelated from one another. For example, if the pressure is overestimated in a particular time step, it is likely that the pressure in the next time step is also overestimated. This pressure overestimation does, however, not say anything about the modeling error for the volume. These assumptions mean that we can express the covariance matrix as
\begin{equation}\label{eq:covtotal}
    \boldsymbol{\Sigma}_{\rm noise} = \left[ \begin{array}{cc} \boldsymbol{\Sigma}_{\rm noise}^{pp} & \boldsymbol{0} \\ \boldsymbol{0} & \boldsymbol{\Sigma}_{\rm noise}^{VV} \end{array} \right],
\end{equation}
where $\boldsymbol{\Sigma}_{\rm noise}^{pp} \in \mathbb{R}^{n \times n}$ and $\boldsymbol{\Sigma}_{\rm noise}^{VV} \in \mathbb{R}^{n \times n}$ are the pressure and volume covariance matrices, respectively. To accommodate the temporal correlation, we express the elements of these matrices as
\begin{equation}
    \Sigma_{{\rm noise},ij}^{pp}  = \sigma_i^p \sigma_j^p \rho^{pp}_{ij} \qquad \mbox{and} \qquad \Sigma_{{\rm noise},ij}^{VV}  = \sigma_i^V \sigma_j^V \rho^{VV}_{ij},
    \label{eq:covariancematrices}
\end{equation}
where the correlation matrices follow an exponentially decaying correlation function, \emph{i.e.},
\begin{equation}\label{eq:correlationmatrices}
 \rho_{ij}^{pp} = \exp{\left( -|i-j| \frac{\Delta t}{\tau^p} \right)}   \qquad \mbox{and} \qquad \rho_{ij}^{VV} = \exp{\left( -|i-j| \frac{\Delta t}{\tau^V} \right)},
\end{equation}
where $\tau^p$ and $\tau^V$ are correlation time scales for the pressure and volume signals, respectively. The standard deviations in the covariance matrices \eqref{eq:covariancematrices} are, in general, time-step dependent. This time-step dependence allows us to assign importance to different parts of the pressure-volume loop, achieving a similar effect to likelihood weighing \cite{agostinelli_weighted_2012}. Specifically, we assign relatively small standard deviations to the volume signal during the isovolumetric phases in order to correct for the fact that relatively few sample points are present in these short-duration phases. The specific choices for the standard deviations and correlation time scales will be discussed in the context of the numerical results in Section~\ref{ssec:BI2D}.

\subsection{Adaptive Metropolis-Hastings sampling}\label{sec:metropolis}
In theory, the mean and covariance of the correction factors,  Equation~\eqref{eq:moments}, can be computed using tensor-product integration quadrature rules for the parameter domain. However, this evaluation method becomes impractical when the number of calibration parameters increases, as this leads to an exorbitant number of integration points in high-dimensional spaces. Since we wish to develop a generic ROM framework, the ability to evaluate the statistical moments of the calibration parameters should not be impeded by the number of parameters. Therefore, instead of evaluating the integrals in Equation~\eqref{eq:moments} directly, we approximate these integrals using Markov Chain Monte Carlo (MCMC) sampling \cite{brooks_handbook_2011,lambert_students_2018} as
\begin{equation}
\boldsymbol{\mu}^{\rm corr} \approx \frac{1}{N}\sum \limits_{i=1}^N \boldsymbol{\theta}^{\rm corr}_i \quad \mbox{and} \quad  \boldsymbol{\Sigma}^{\rm corr} \approx \frac{1}{N-1}\sum \limits_{i=1}^N ( \boldsymbol{\theta}^{\rm corr}_i - \boldsymbol{\mu}^{\rm corr}) \otimes ( \boldsymbol{\theta}^{\rm corr}_i - \boldsymbol{\mu}^{\rm corr}),
\end{equation}
where the samples $\{ \boldsymbol{\theta}_i^{\rm corr} \}_{i=1}^N$ are drawn from the posterior distribution $\rho( \boldsymbol{\theta}^{\rm corr} | \mathbf{d}^{p-V} )$.

\begin{figure}
    \centering
    \includegraphics[width=0.65\linewidth]{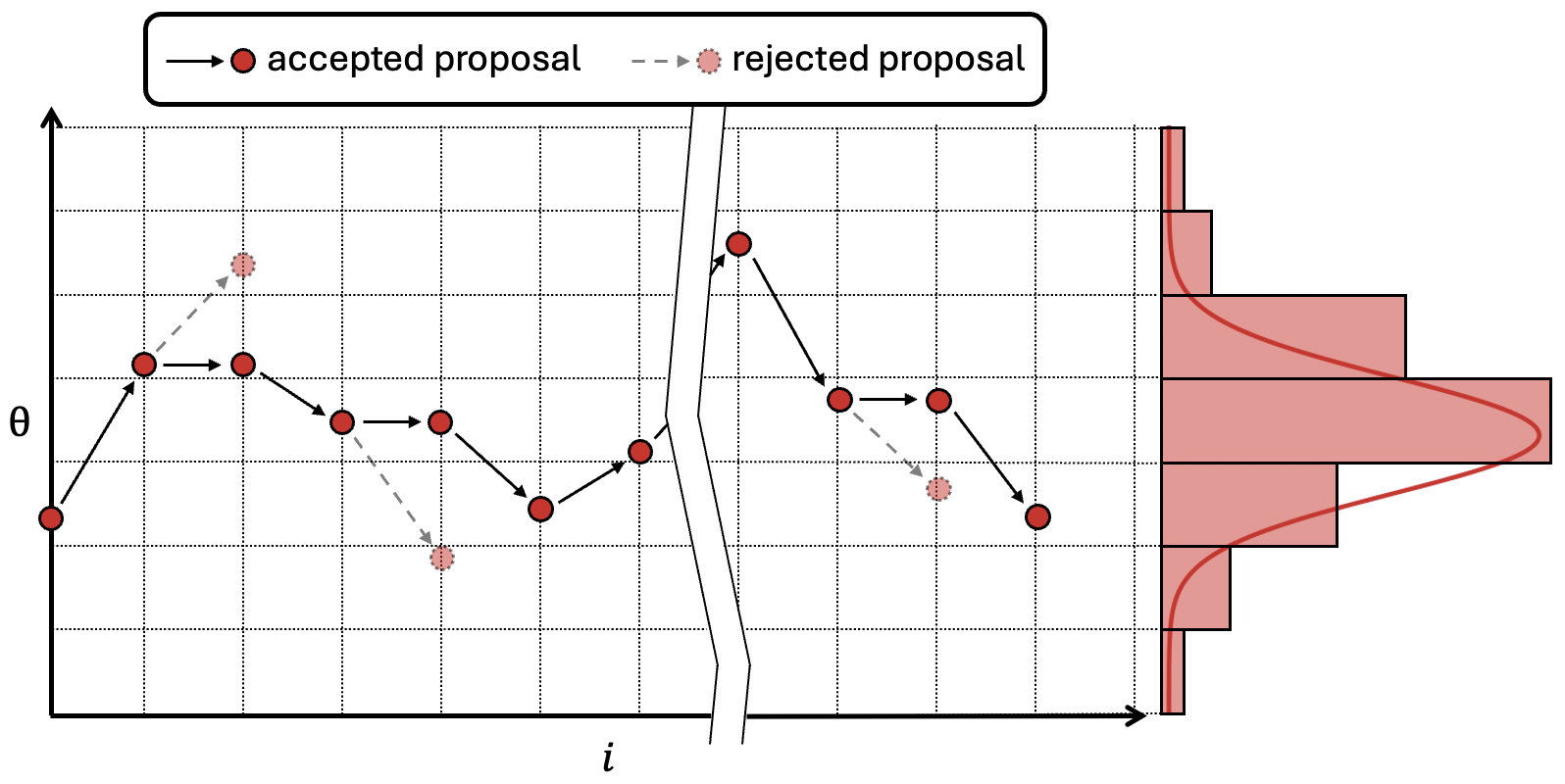}
    \caption{Illustration of the Metropolis-Hastings algorithm. Proposals in the Markov chain are accepted or rejected based on a probabilistic acceptance criterion.}
    \label{fig:metropolishastings}
\end{figure}

To draw samples from the posterior distribution we employ the Adaptive Metropolis-Hastings algorithm \cite{haario_adaptive_2001}. The standard Metropolis-Hastings algorithm is schematically visualized in Figure~\ref{fig:metropolishastings}. Starting from a randomly chosen setting of the calibration parameters, $\boldsymbol{\theta}^{\rm corr}$, a proposal for the next sample in the Markov-Chain is made by sampling from a proposal distribution, $\rho_{\rm proposal}(\boldsymbol{\theta}^{\rm corr}_{i+1}|\boldsymbol{\theta}^{\rm corr}_{i})$, conditioned on the head of the Markov Chain. An acceptance/rejection criterion then decides whether the newly proposed sample is added to the chain, or whether the current position is repeated. The Markov Chain that follows from the repetitive application of this procedure follows the limit distribution, \emph{i.c.}, the posterior.

For the proposal distribution, we employ a multivariate normal distribution centered at the previous sample and with a covariance matrix $\boldsymbol{\Sigma}_{\rm proposal}$. We optimize the choice for this covariance matrix using the adaptive MCMC algorithm proposed in Ref.~\cite{haario_adaptive_2001}. When starting a Markov chain from a random location, it typically requires a number of steps before the limit distribution is sampled. This \emph{burn-in period} is removed from the chain using visual inspection of the chain's trace. Multiple chains are generated to assess convergence.

\section{Framework demonstration for idealized ventricles}\label{sec:ApplicationS}
We demonstrate the developed reduced-order modeling framework (Figure~\ref{fig:LFmodel}) for a controlled test case, using idealized left ventricle geometries. This test case serves two purposes. First, it showcases the complete workflow, from input to output, also indicating its limitations. Second, it will be demonstrated that the reduced-order model based on the generalized one-fiber model can capture geometric changes other than the wall and cavity volume, this in contrast to the standard one-fiber models.

In Section~\ref{sec:2Dcasegeometry} we commence the discussion of the test case by introducing the idealized left ventricle geometry. The generation of the FOM training data is then discussed in Section~\ref{sec:2Dcasetrainingset}. Subsequently, the Bayesian calibration of the simplified-physics ROM on the FOM training data is discussed in Section~\ref{ssec:BI2D}. The Gaussian process to determine the correction factors is then considered in Section~\ref{ssec:GP2D}. With all components of the reduced-order modeling framework in place, the results for the idealized test case are then discussed in Section~\ref{ssec:2Dresults}.

\subsection{Idealized ventricle geometry parametrization}\label{sec:2Dcasegeometry}
\begin{figure*}[!t]
     \centering
     \begin{subfigure}[b]{0.4\textwidth}
         \centering
         \includegraphics[width=\textwidth]{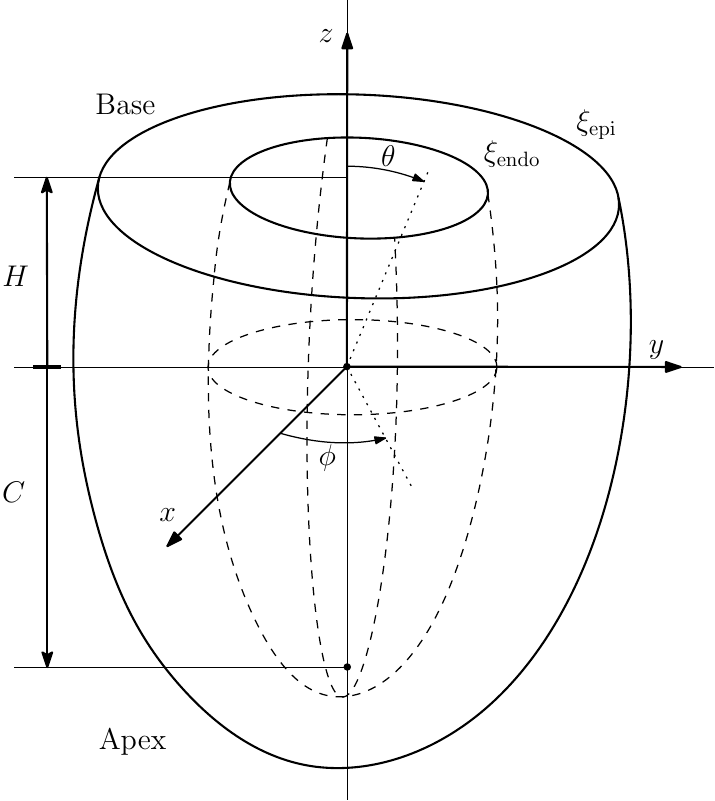}
                  \caption{}
                  \label{fig:LVideal3D}
     \end{subfigure}
     \hfill
     \begin{subfigure}[b]{0.48\textwidth}
         \centering
         \includegraphics[width=\textwidth]{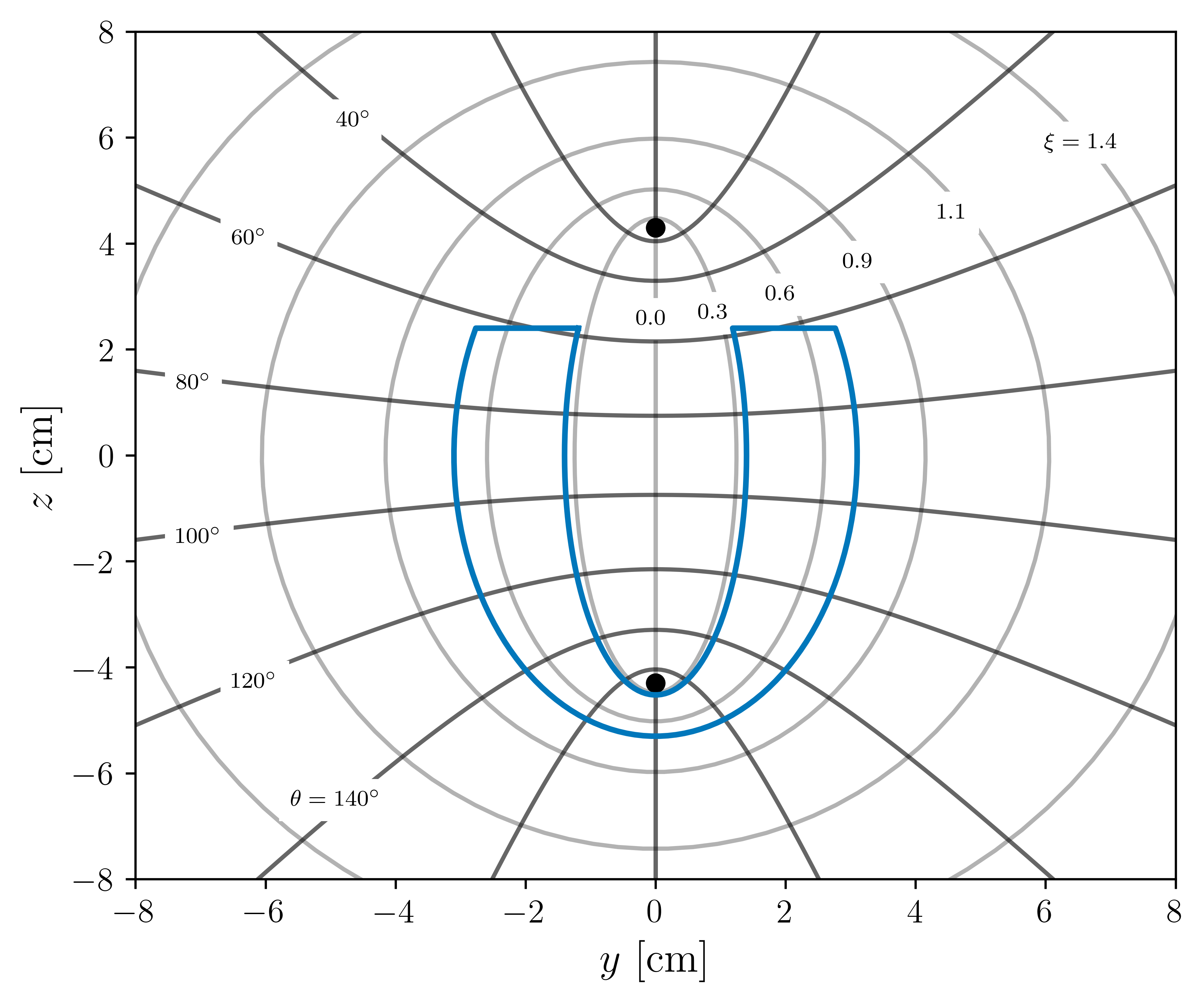}
                  \caption{}
                  \label{fig:LVideal2D}
     \end{subfigure}
        \caption{Parametrized left ventricle geometry defined in prolate coordinates $(\xi,\theta,\phi)$ (a) Three-dimensional representation of the left ventricle domain in the Cartesian coordinate system. The ventricle parameters are denoted by the truncation height, $H$, the focal length, $C$, and the endo- and epicardial transmural coordinates, $\xi_{\mathrm{endo}}$ and $\xi_{\mathrm{epi}}$. (b) Two-dimensional representation of the left ventricle, showcasing the isocontours of the $\theta$-angle and the transmural coordinate.}
        \label{fig:LVideal}
\end{figure*}

We consider a thick-walled truncated ellipsoid idealized left ventricle geometry~\cite{bovendeerd_determinants_2009}. This ellipsoid is defined in prolate coordinates $\{ \xi, \theta, \phi \}$ as
\begin{equation}\label{eq:prolate}
    \begin{aligned}\mathbf{x}^{\mathrm{ell}}(C, \xi, \theta, \phi )=
    C \left(\begin{array}{c}
          \mathrm{sinh}(\xi)\mathrm{sin}(\theta) \mathrm{cos}(\phi)\\ 
          \mathrm{sinh}(\xi)\mathrm{sin}(\theta) \mathrm{sin}(\phi)\\ 
          \mathrm{cosh}(\xi)\mathrm{cos}(\theta)
         \end{array}\right),
\end{aligned}
\end{equation}
where the focal length $C$ and the transmural coordinate $\xi$ control the radii. The truncated ellipsoid geometry is therefore defined by $4$ parameters: the focal length, $C$, the endo- and epicardial transmural values, $\xi_{\mathrm{endo}}$ and $\xi_{\mathrm{epi}}$, and the truncation height, $H$. The left ventricle domain is then defined as
\begin{equation}\label{eq:lvgeom}
\Omega(C, \xi_{\mathrm{endo}}, \xi_{\mathrm{epi}}, H) := \left\{  \mathbf{x}^{\mathrm{ell}}(C, \xi, \theta, \phi ) \mid 0 \leq \phi <  2\pi,~\xi_{\mathrm{endo}}\leq \xi \leq \xi_{\mathrm{epi}},~\mathrm{arccos}\left(\frac{H}{C \mathrm{cosh}(\xi)}\right) \leq  \theta  \leq \pi \right\}.
\end{equation}
Following Ref.~\cite{bovendeerd_determinants_2009}, we define a reference geometry (Figure~\ref{fig:LVideal3D}) according to the parameters in Table~\ref{tab:refparam}, which serves as the basis to generate new geometric variations. The reference geometry has a ventricle cavity volume of $V=44$~[ml] and a wall volume of $V_{\mathrm{w}}=136$~[ml].

\begin{table}[]
\caption{Parameter values for the reference left ventricle geometry.}\label{tab:refparam}
\centering
\begin{tabular}{cccc}
\hline
$C_{\mathrm{ref}}$          & $H_{\mathrm{ref}}$          & $\xi_{\mathrm{endo}}$ & $\xi_{\mathrm{epi}}$ \\ \hline
4.3 {[}cm{]} & 2.4 {[}cm{]} & 0.371 {[}-{]}         & 0.678 {[}-{]}       
\end{tabular}
\end{table}

\subsection{Training data generation}\label{sec:2Dcasetrainingset}
In Section~\ref{sec:genrelations} we showed that the standard one-fiber models are invariant to geometric changes other than the cavity and wall volumes. However, since the ellipsoidal ventricle is defined by four parameters, two additional constraints are required to generate a unique ventricle shape, \emph{cf.} Equation~\eqref{eq:lvgeom}. To investigate the effects of geometry variations under the assumption of fixed cavity and wall volume, we consider the truncation length, $H$, and the focal length, $C$, as our two-dimensional parameter input space to construct variations. We consider variations relative to the reference geometry following
\begin{subequations}
    \begin{align}
        H &= \tilde{H} H_{\mathrm{ref}}, \\
        C &= \tilde{C} C_{\mathrm{ref}},
    \end{align}
\end{subequations}
where the subscript $'\mathrm{ref}'$ refers to the reference geometry values listed in Table~\ref{tab:refparam}. This implies that the geometry parameters in the Gaussian process mapping \eqref{eq:abstractmapping} are equal to $\mathbf{c}=[\tilde{H}, \tilde{C}]$. The goal of this test case is to understand the influence of these parameters on the FOM results while keeping the cavity and wall volume constant, and eventually to infer the parameters of the generalized one-fiber model through the reduced-order modeling framework.

To generate the parameter space for the multiplication factors $\tilde{H}$ and $\tilde{C}$, we consider physiological data to restrict the geometrical shapes that can be created. We assume that the reference geometry represents the heart at end-systole. This geometry assumption is supported by the fact that the reference geometry's cavity volume is similar to the cavity volume at end-systole during the FOM simulations. While this may be true for the volumes, it should be emphasized that the internal and external forces at end-systole are significantly different when compared to the stress-free reference geometry. Nonetheless, this assumption is appropriate from a geometrical point of view. We therefore consider end-systolic population data to ensure that the generated geometries are physiologically sound.

\begin{table}[]
\caption{Physiological dimension values of the left ventricle at end-systole. The indicated ranges reflects the 95\% credible intervals.}\label{tab:ESphysvalues}
\centering
\begin{tabular}{llcc}
\hline
                         &Symbol     & Expression                                                                                                                                          & Range    \\ \hline
LV diameter \cite{lang_recommendations_2015,galderisi_standardization_2017}       & $D^{\mathrm{lv}}$ & $C \mathrm{sinh}(\xi_{\mathrm{endo}})$                                                                                                              & $2.8 - 4.0$ {[}cm{]}   \\
LV total length \cite{lang_recommendations_2005,nielsen_accuracy_2010}   & $L^{\mathrm{lv}}$ & $H + C \mathrm{cosh}(\xi_{\mathrm{endo}})$                                                                                                          & $4.2 - 8.6$ {[}cm{]} \\
LV basal diameter \cite{anwar_true_2007,de_groot-de_laat_how_2019} & $B^{\mathrm{lv}}$ & $D^{\mathrm{lv}} \mathrm{sin}\left(  \mathrm{arccos}\left(  \frac{H}{ C \mathrm{cosh}(\xi_{\mathrm{endo}}) } \right)\right)$ & $\geq 2.0$ {[}cm{]}       
\end{tabular}
\end{table}

\begin{figure*}[!t]
     \centering
     \begin{subfigure}[b]{0.48\textwidth}
         \centering
         \includegraphics[width=\textwidth]{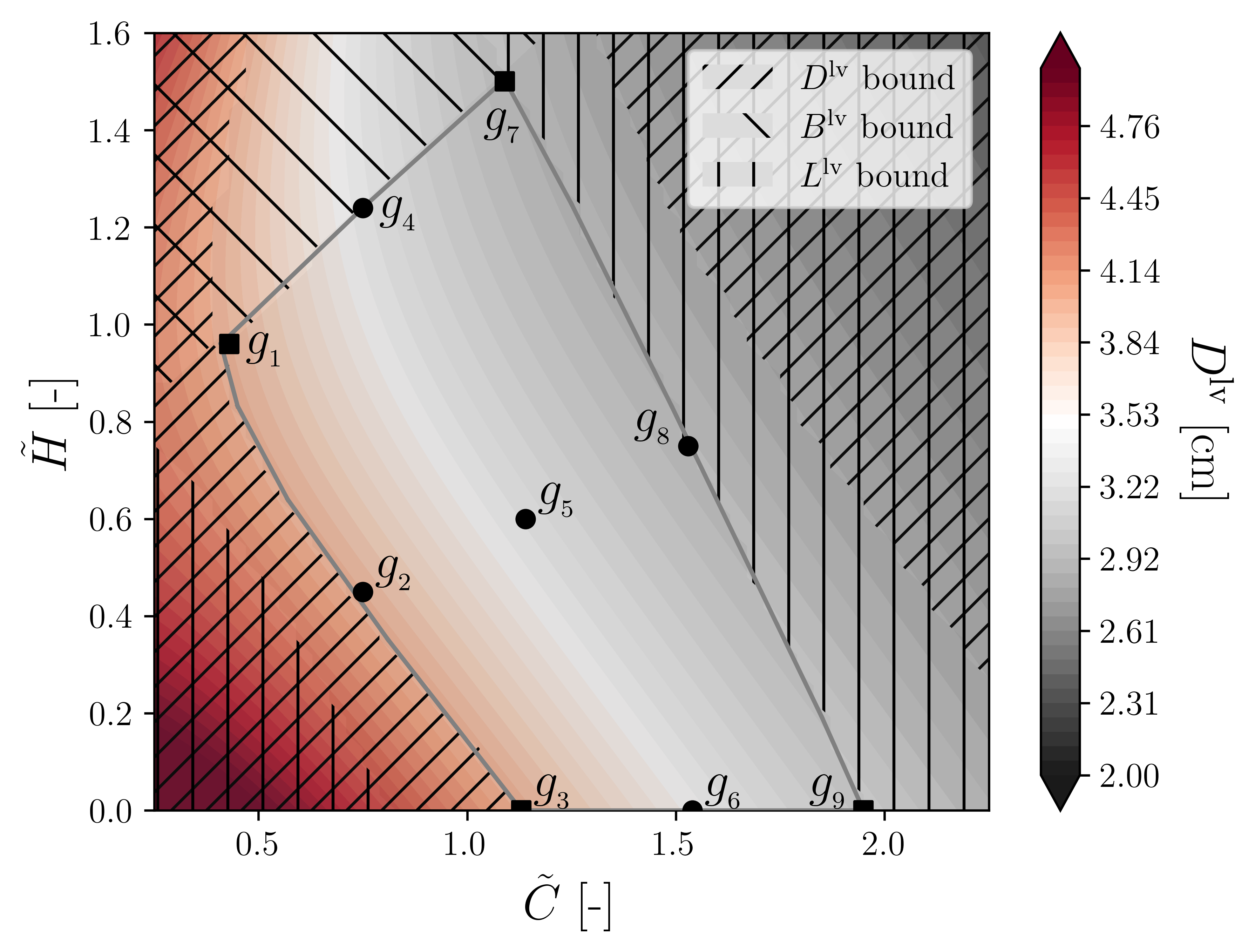}
                  \caption{}
                  \label{fig:paramspace2D}
     \end{subfigure}
     \hfill
     \begin{subfigure}[b]{0.35\textwidth}
         \centering
         \includegraphics[width=\textwidth]{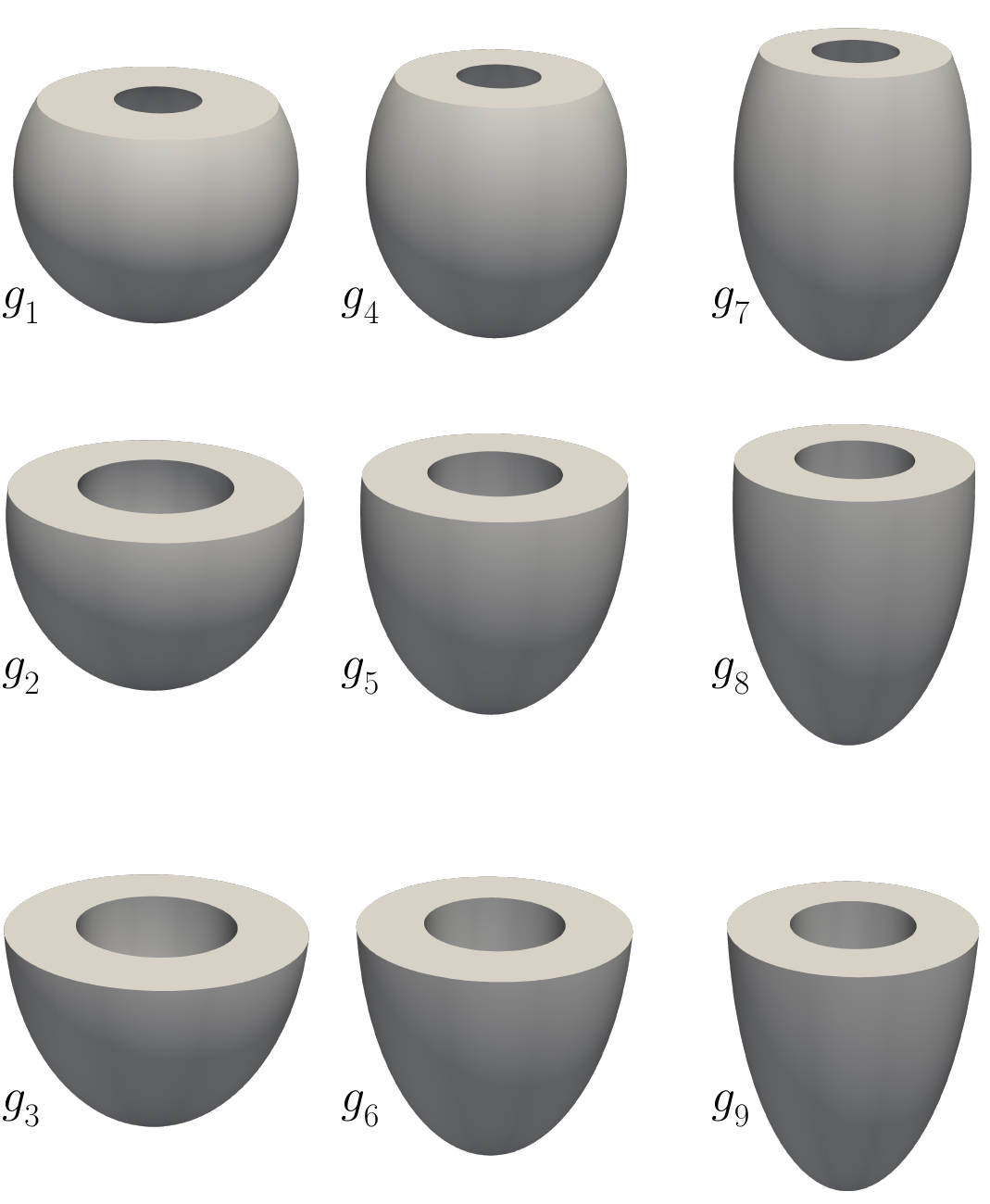}
                  \caption{}
                  \label{fig:variations2D}
     \end{subfigure}
        \caption{(a) Contour plot of the left ventricle diameter at $z=0$ for different combinations of $\tilde{H}$ and $\tilde{C}$ for a constant cavity volume of $44$~[ml] and constant wall volume of $136$~[ml]. The hatched areas indicate the physiological bounds representing non-physiological ventricle diameter values, basal diameter values, and total ventricle lengths. (b) The nine three-dimensional geometries defined in the physiological range.}
        \label{fig:LVvariations2D}
\end{figure*}

The physiological ranges for the ventricle geometry parameters are listed in Table~\ref{tab:ESphysvalues}.
The total length of the ventricle is based on the area-length method~\cite{lang_recommendations_2005}, such that $V=5/6D^{\mathrm{lv}}L^{\mathrm{lv}}$. The minimal value for the basal diameter is related to the observed mitral valve minimal diameter~\cite{de_groot-de_laat_how_2019}. The basal diameter has no explicit maximum, although it should be smaller than the ventricle diameter. The parameter space corresponding to these physiological constraints is visualized in Figure~\ref{fig:paramspace2D}.

\begin{figure*}[!t]
     \centering
     \begin{subfigure}[b]{0.45\textwidth}
         \centering
         \includegraphics[width=\textwidth]{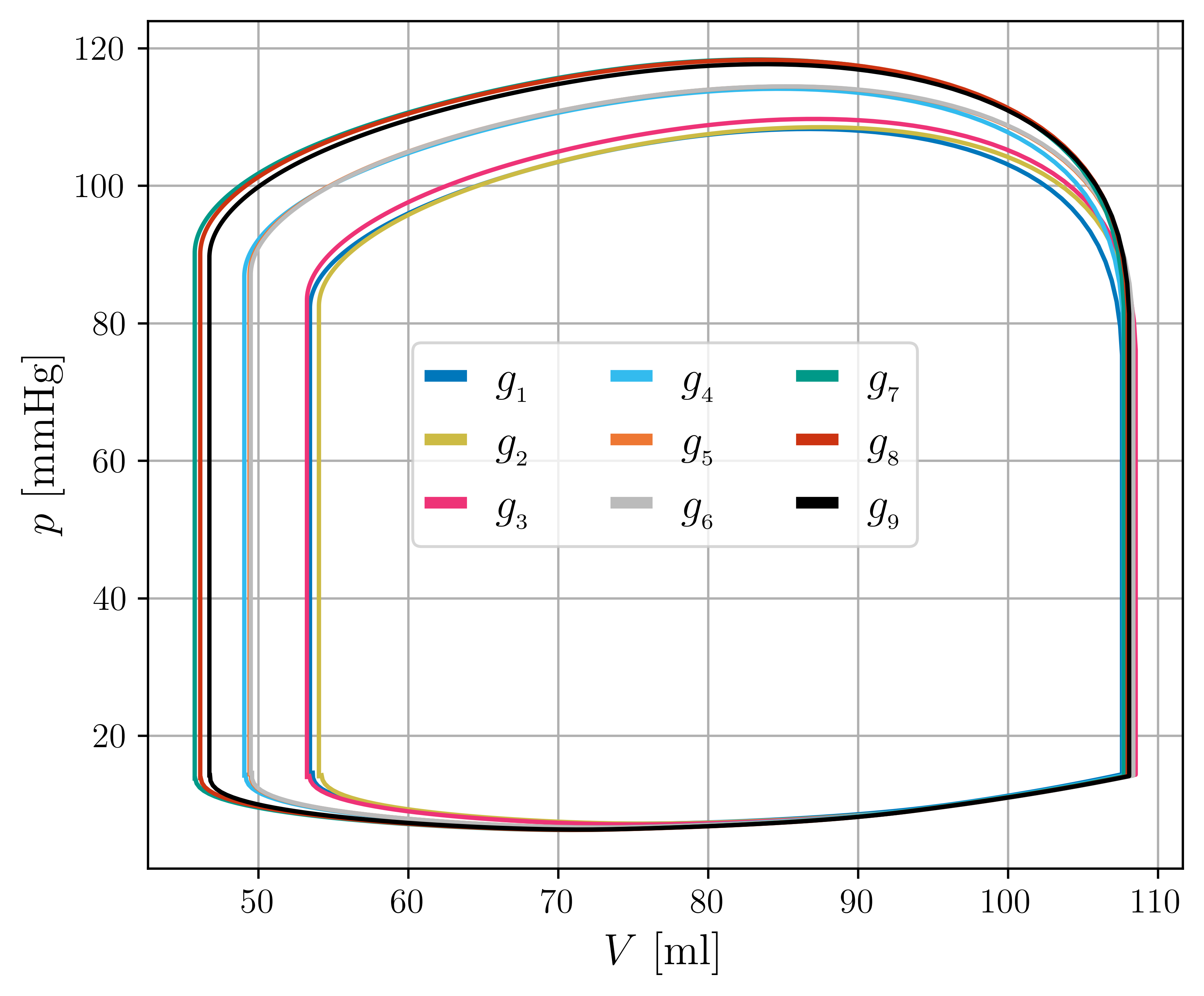}
                  \caption{}
                  \label{fig:Hemodynamics2Da}
     \end{subfigure}
     \hfill
     \begin{subfigure}[b]{0.45\textwidth}
         \centering
         \includegraphics[width=\textwidth]{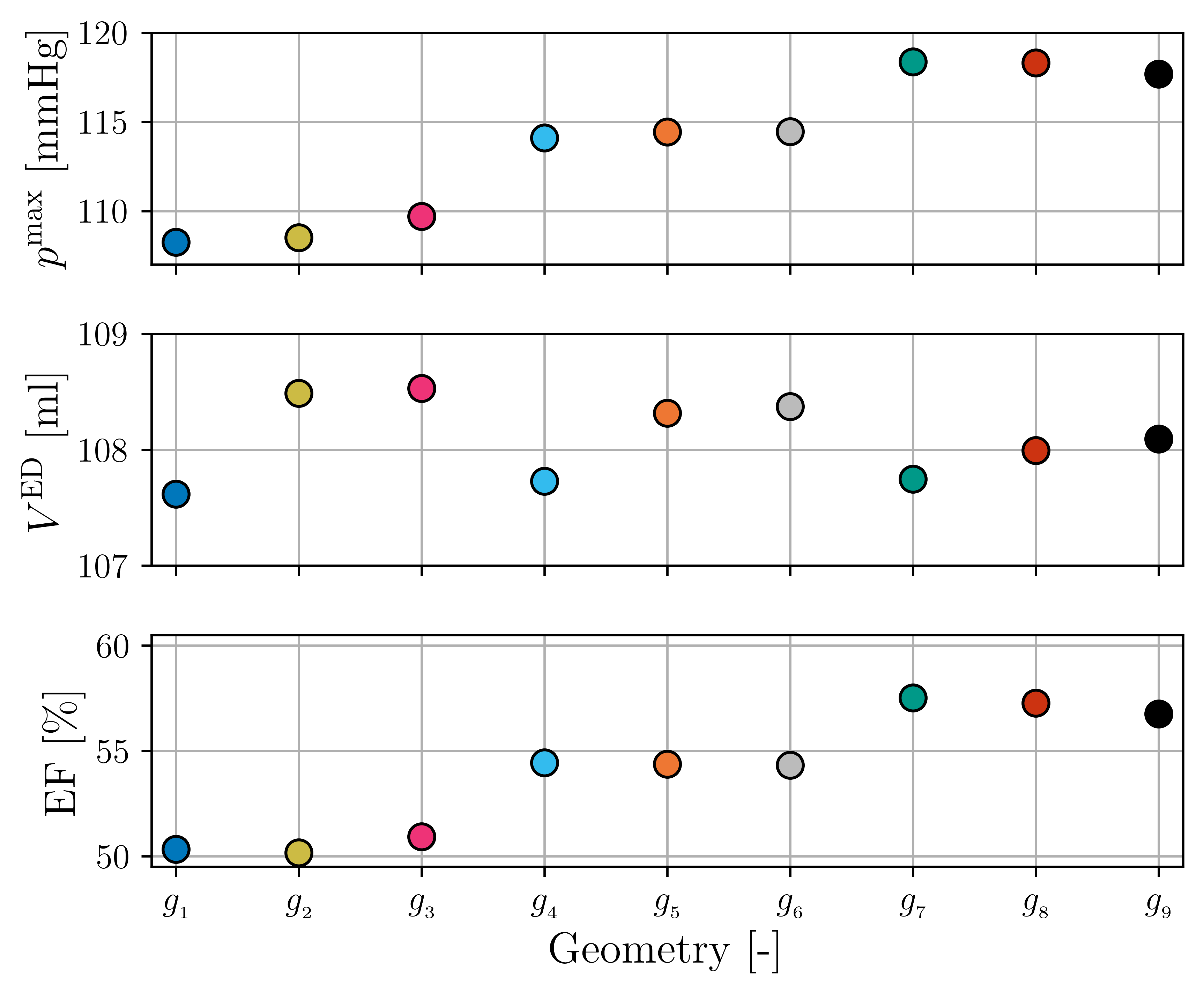}
                  \caption{}
                  \label{fig:Hemodynamics2Db}
     \end{subfigure}
        \caption{Hemodynamics results of the full-order IGA model for each ellipsoidal geometry $\{g_1, g_2, ..., g_9\}$. (a) The pressure-volume curve for each geometry. (b) Quantities of interest that follow from the pressure-volume curves: the maximum cavity pressure, $p^{\mathrm{max}}$, the end-diastolic volume, $V^{\mathrm{ED}}$, and the ejection fraction, $\mathrm{EF}$.}
        \label{fig:Hemodynamics2D}
\end{figure*}

Let us define a set of geometries, $\{g_1, g_2, ..., g_9 \}$, located in the constrained parameter space. The full-order IGA model results of these geometries are compared to investigate the effect of geometry changes while keeping the wall and cavity volumes fixed. The results of this analysis are visualized in Figure~\ref{fig:Hemodynamics2D}, from which it is apparent that the main influence on the results comes from the change in ventricle diameter, as the sets $\{g_1,g_2,g_3\}$, $\{g_4,g_5,g_6\}$, and $\{g_7,g_8,g_9\}$, are positioned at noticeably different isobars inside the $(\tilde{H},\tilde{C})$-space, \emph{cf.}~Figure~\ref{fig:paramspace2D}. We also observe minor differences in the end-diastolic volume.

\subsection{Bayesian calibration of the generalized one-fiber model}\label{ssec:BI2D}
The correction factors for the generalized one-fiber model are inferred following the Bayesian calibration procedure outlined in Section~\ref{sec:BI}. We aim to ensure that the inferred ROM pressure-volume curves, Figure~\ref{fig:Hemodynamics2Da}, closely match the FOM ejection fraction, $\mathrm{EF} = (V^{\mathrm{ED}} - V^{\mathrm{ES}})/V^{\mathrm{ED}}$, where $V^{\mathrm{ED}}$ and $V^{\mathrm{ES}}$ are the end-diastolic and end-systolic volumes, respectively. To accurately capture these isovolumetric phases, use is made of a time-dependent noise level for the volumetric part of the likelihood function \eqref{eq:likelihoodfunction}. This time-dependent noise level, specified in \ref{app:noise}, essentially assigns a higher weight to the isovolumetric phases compared to the other phases. The magnitude of the time-dependent noise level, Equation~\eqref{eq:timednoise}, is based on the variation observed in end-diastolic volume, Figure~\ref{fig:Hemodynamics2Db}, which ranges between approximately $107.5-108.5$~[ml]. Assuming the range to correspond to a 95\% credible interval, we define a minimum volumetric noise level of $\sigma^{V}_{\mathrm{min}}=0.25$~[ml]. The maximum volumetric noise level is set to $\sigma^{V}_{\mathrm{max}}=1.15$~[ml], which is approximately $1.5\%$ of the total stroke volume, $V^{\mathrm{ED}} - V^{\mathrm{ES}}$. The pressure deviation is a constant value related to the maximum pressures, $p^{\mathrm{max}}$ in Figure~\ref{fig:Hemodynamics2Db}, with a range of $10.0$~[mmHg]. This results in a standard deviation of $\sigma_i^{p}=2.5$~[ml] to be used in Equation~\eqref{eq:covariancematrices}. The time scales in Equation~\eqref{eq:correlationmatrices} are set to $\tau^{p}=\tau^{V}=t^{\mathrm{cycle}} - \Delta t$.

Since the correction factors, $\boldsymbol{\theta}^{\mathrm{corr}}=[\alpha, \beta, \gamma, \lambda]$, defined in Section~\ref{sec:lowfidelity}, are considered to be variations around unity, their prior distribution is assumed Gaussian and uncorrelated as
\begin{equation}\label{eq:2Dprior}
\boldsymbol{\theta}^{\mathrm{prior}} \sim \mathcal{N}(\boldsymbol{\mu}^{\mathrm{prior}}, \boldsymbol{\Sigma}^{\mathrm{prior}}),
\end{equation}
with prior means $\mu_{\alpha}^{\mathrm{prior}}=\mu_{\beta}^{\mathrm{prior}}=\mu_{\gamma}^{\mathrm{prior}}=\mu_{\lambda}^{\mathrm{prior}}=1$ and standard deviations $\sigma^{\mathrm{prior}}_{\alpha}=\sigma_{\beta}^{\mathrm{prior}}=\sigma_{\gamma}^{\mathrm{prior}}=\sigma_{\lambda}^{\mathrm{prior}}=0.1$. The correction factors are inferred using the $6^{\rm th}$ cardiac cycle, which represents the hemodynamic steady-state for both models. After inference, we obtain the correlated posterior distribution, $\boldsymbol{\theta}^{\mathrm{post}}$, for each geometry, $g(\tilde{H},\tilde{C})$. It is important to note that the resulting posterior covariance matrix, $\boldsymbol{\Sigma}^{\mathrm{post}}$, contains non-zero off-diagonal entries, in contrast to the prior distribution \eqref{eq:2Dprior}. Furthermore, it is stressed that the posterior covariance matrix is not related to the covariance matrix of the proposal constructed by the adaptive MCMC algorithm in Section~\ref{sec:metropolis}. A sufficient number of MCMC samples should be considered to achieve posterior convergence.

\subsection{Gaussian process construction}\label{ssec:GP2D}
In this controlled test case, we consider a given set of $9$ training geometries, $\{g_1, g_2, ..., g_9\}$, with corresponding posterior distributions, $\{ \boldsymbol{\theta}^{\mathrm{post}}_1, \boldsymbol{\theta}^{\mathrm{post}}_2, \ldots, \boldsymbol{\theta}^{\mathrm{post}}_9 \}$. We employ the Gaussian Process (GP) outlined in Section~\ref{sec:GP} to interpolate between these known geometries (patients). To initialize the GP, it is desired to consider the vertices of the bounding box of the constrained parameter space (Figure~\ref{fig:paramspace2D}), such that predictions are interpolated rather than extrapolated. Since the training geometries considered here are evenly distributed in the constrained parameter space, these vertices can directly be identified. We note, however, that, in the case of population data, there is limited control over the training data distribution inside the constrained parameter space. In this general case, the determination of the bounding box vertices is not trivial, as will be discussed in Section~\ref{sec:ApplicationC}. 

The GP length scale bounds for each geometric parameter direction, $\tilde{H}$ and $\tilde{C}$, are specified as
\begin{equation}\label{eq:lengscales}
    l \in \left[ 0.02 \ l_{\mathrm{max}}, 2l_{\mathrm{max}} \right], \quad \text{with} \quad l_{\mathrm{max}}=\mathrm{max}(x)-\mathrm{min}(x), \quad \text{and} \quad x \in \{ \tilde{H}, \tilde{C} \}.
\end{equation}
The $l_{\mathrm{max}}$ length scale represents the maximum distance encountered between the data points. By setting the upper bound to $2l_{\mathrm{max}}$, we imply that points remain moderately correlated throughout the entire constrained parameter space, with a minimum correlation of $\exp(-2)\approx0.135$, following the definition of the RBF kernel \eqref{eq:RBFkernel}. The minimum length scale is taken sufficiently small, such that points within this minimum length scale distance from the already available points in the training set, \emph{i.e.}, with $l\leq 0.02l_{\mathrm{max}}$, can be assumed to be fully correlated.

Given the initial GP using only the vertex geometry points and length-scale bounds, we sequentially insert the remaining five points in the training set, starting with the one farthest away from the points already in the training data. We note that all additional points are outside of the minimum length scale distance of the points already in the training set, meaning that all of them will be considered as new training data. Once finalized, predictions can be made by evaluating the vector-valued GP, \emph{cf}.~Algorithm~\ref{alg:GP}.

\subsection{Idealized ventricle results}\label{ssec:2Dresults}
\begin{figure*}[!t]
     \centering
     \begin{subfigure}[b]{0.48\textwidth}
         \centering
         \includegraphics[width=\textwidth]{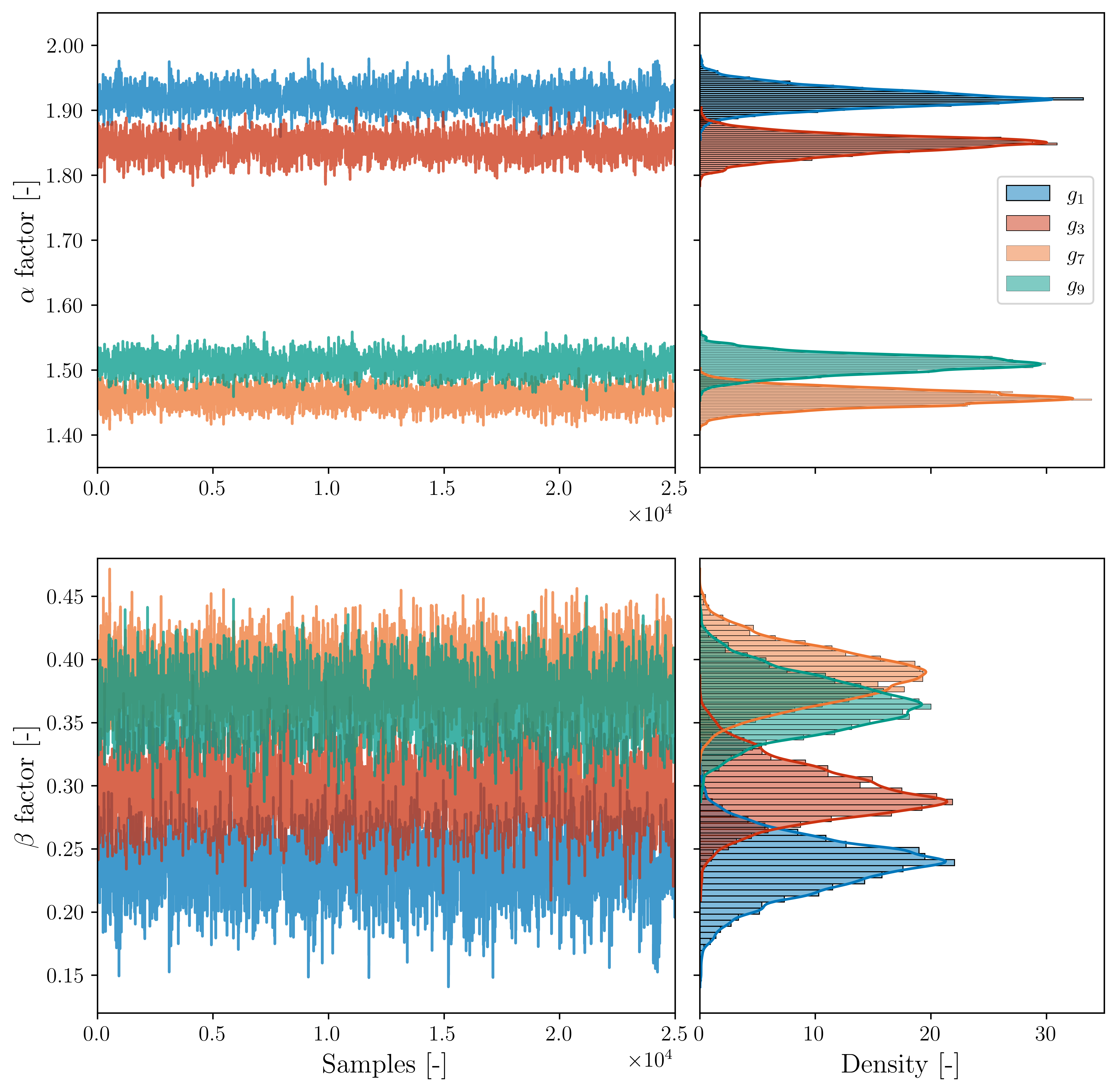}
     \end{subfigure}
     \hfill
     \begin{subfigure}[b]{0.48\textwidth}
         \centering
         \includegraphics[width=\textwidth]{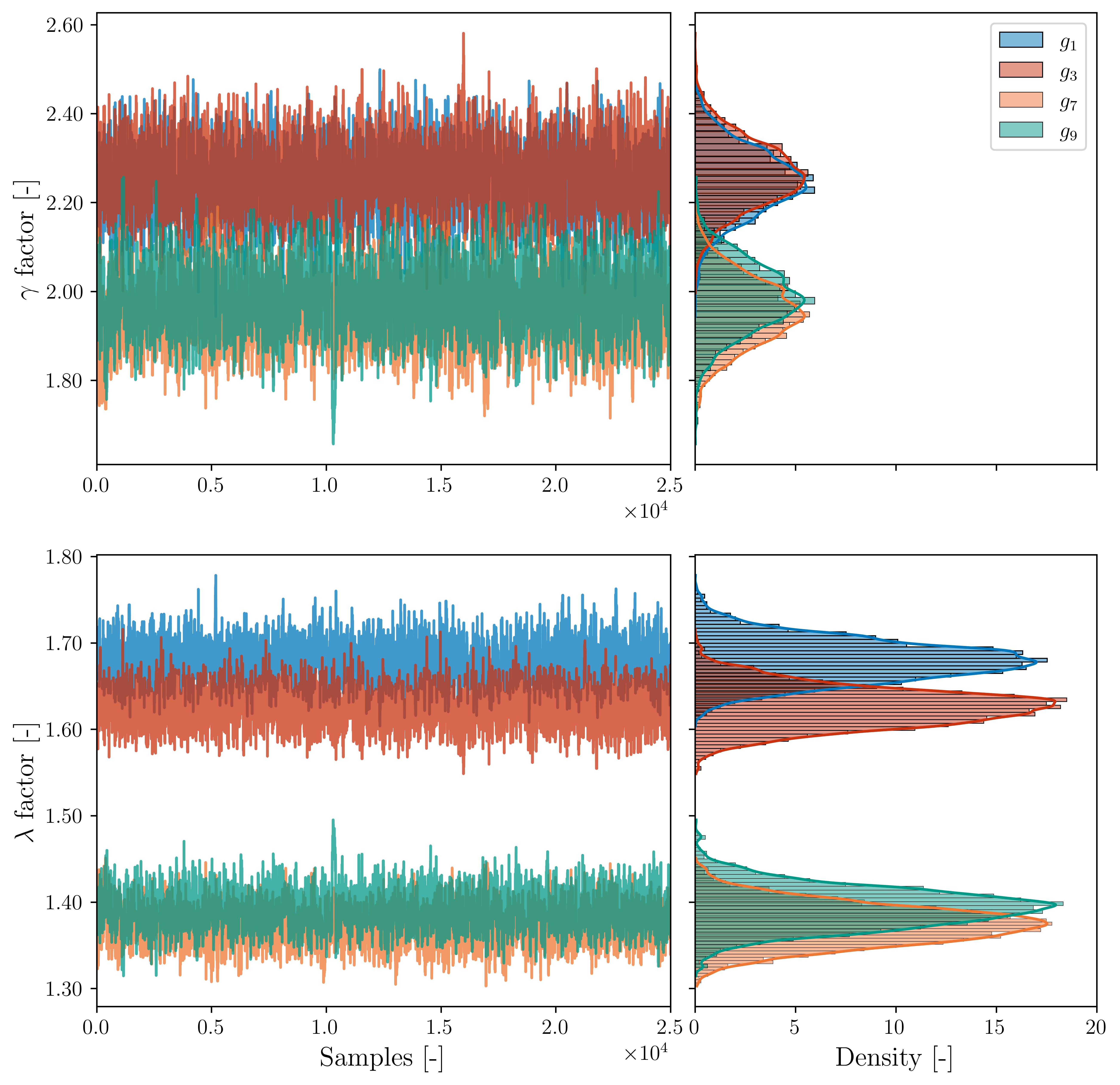}
     \end{subfigure}
        \caption{Markov chains of the four correction factors, $\boldsymbol{\theta}^{\rm corr}=[\alpha,\beta,\gamma,\lambda]$, resulting from the adaptive Metropolis-Hastings algorithm applied to the geometries $g_1$, $g_3$, $g_7$, and $g_9$. The chains are visualized analogously to the schematic in Figure~\ref{fig:metropolishastings}, showing the random walk on the left and the corresponding posterior distribution on the right. Note that Gaussian-like distributions are observed for all factors.}
        \label{fig:BI2Dchains}
\end{figure*}

\begin{figure*}[!t]
     \centering
     \begin{subfigure}[b]{0.48\textwidth}
         \centering
         \includegraphics[width=\textwidth]{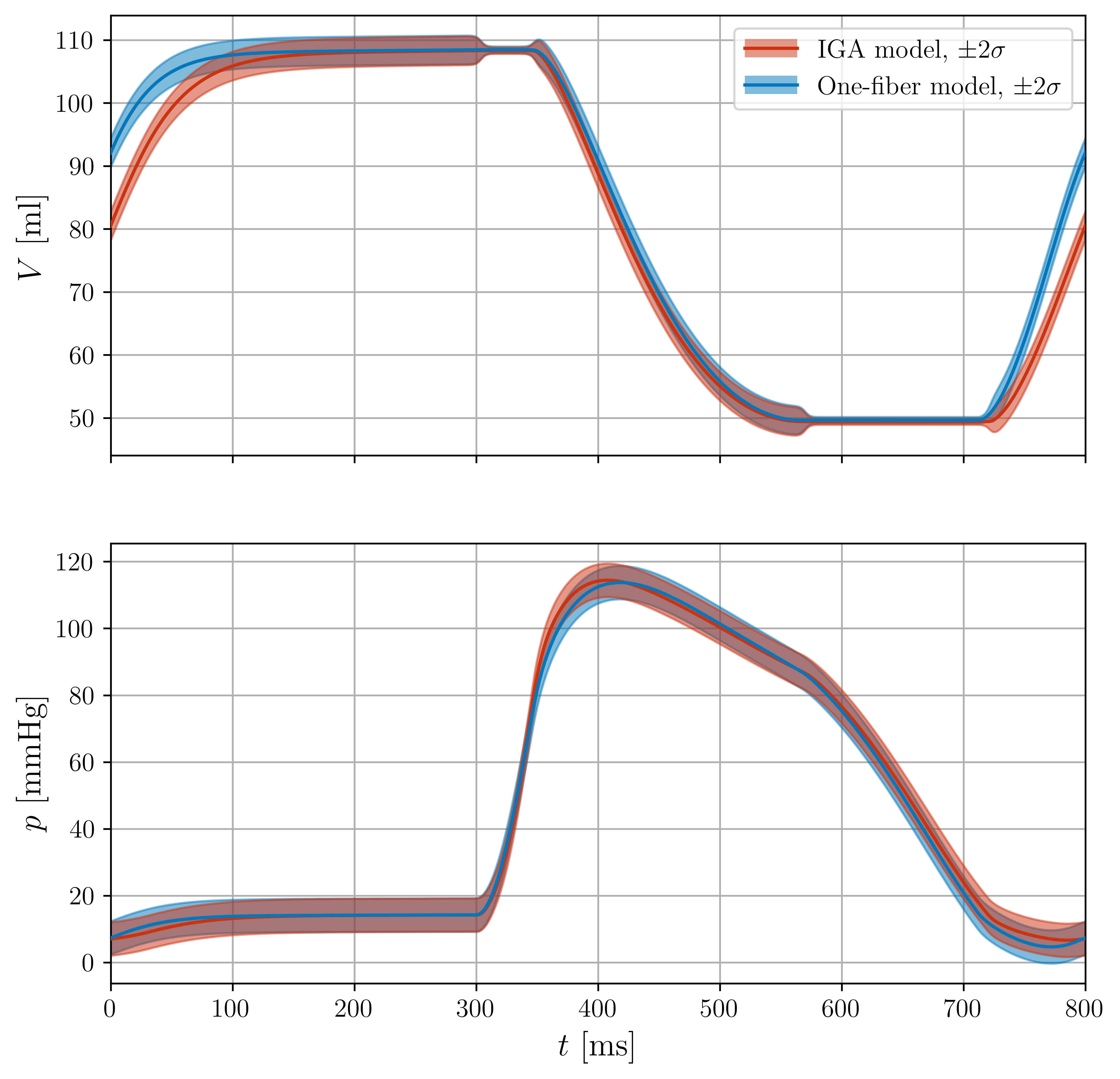}
                  \caption{}
                  \label{fig:BI2D_PVa}
     \end{subfigure}
     \hfill
     \begin{subfigure}[b]{0.48\textwidth}
         \centering
         \includegraphics[width=\textwidth]{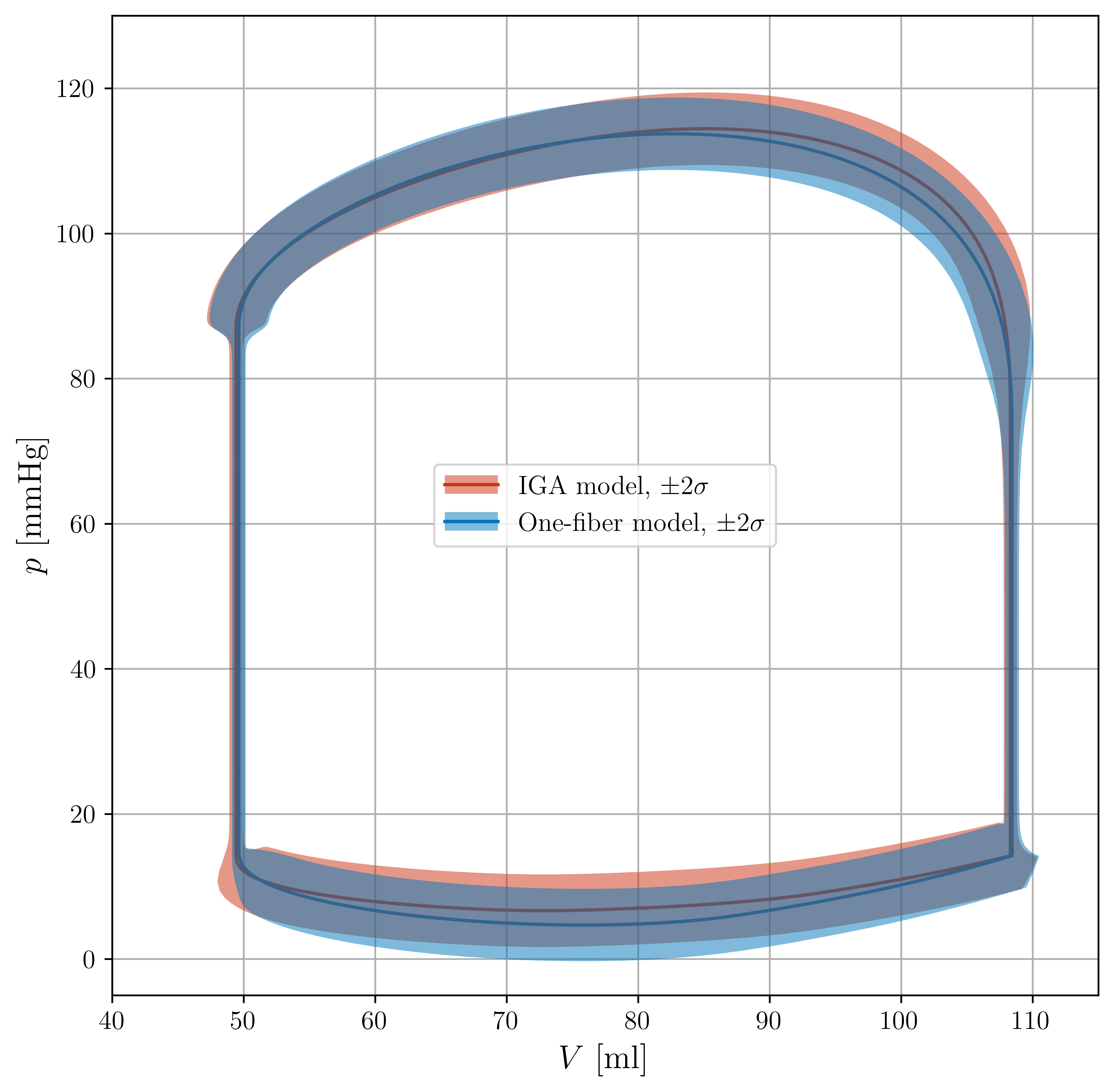}
                  \caption{}
                  \label{fig:BI2D_PVb}
     \end{subfigure}
        \caption{Hemodynamic model results for the $6^{\rm th}$ cardiac cycle of geometry $g_5$, obtained using the full-order IGA model and the calibrated generalized one-fiber model. The FOM results show the $95\%$ credible intervals corresponding to the assigned uncertainty in Section~\ref{ssec:BI2D}. The ROM results obtained from the Metropolis-Hastings algorithm show the $95\%$ credible intervals of the correction factor posteriors. Results are visualized as (a) pressure-volume traces and (b) pressure-volume loops.}
        \label{fig:BI2D_PV}
\end{figure*}

The reduced-order model correction factors, $\boldsymbol{\theta}^{\rm corr}$, are inferred from the FOM results as visualized in Figure~\ref{fig:Hemodynamics2D} using Bayesian inference. To attain these results, a total of $100,000$ adaptive MCMC samples with a reset at every $25,000$ samples is used to generate a proposal covariance matrix corresponding to a $0.234$ acceptance ratio. This optimized proposal covariance matrix is then used for the regular MCMC sampling, for which we run $25,000$ samples, excluding $5,000$ burn-in samples. The results of the random walk chain are visualized in Figure~\ref{fig:BI2Dchains} for the vertex geometries, $g_1$ ,$g_3$, $g_7$, and $g_9$, as shown in Figure~\ref{fig:LVvariations2D}. The Markov chains are observed to mix well and result in stationary posterior distributions \cite{lambert_students_2018}. Additional chains, not shown here, yield similar results, corroborating that the limit distributions represent the posteriors. A typical pressure-volume result following the Markov chain is visualized in Figure~\ref{fig:BI2D_PV} for geometry $g_5$, which displays the $95\%$ credible intervals. The results show a good fit between the IGA FOM and the calibrated ROM. The different noise levels assigned to the isovolumetric phases, discussed in Section~\ref{ssec:BI2D}, are clearly distinguishable and ensure that the inferred results capture these volumetric quantities accurately. Adversely, the ROM is unable to capture the filling phase accurately. This indicates that the current ROM lacks the physical complexity to describe all cardiac phases accurately. The observed differences are, however, well within clinically acceptable limits. Extension of the simplified-physics ROM to include additional physical phenomena is therefore not considered in this work.

\begin{figure*}[!t]
     \centering
     \begin{subfigure}[b]{0.48\textwidth}
         \centering
         \includegraphics[width=\textwidth]{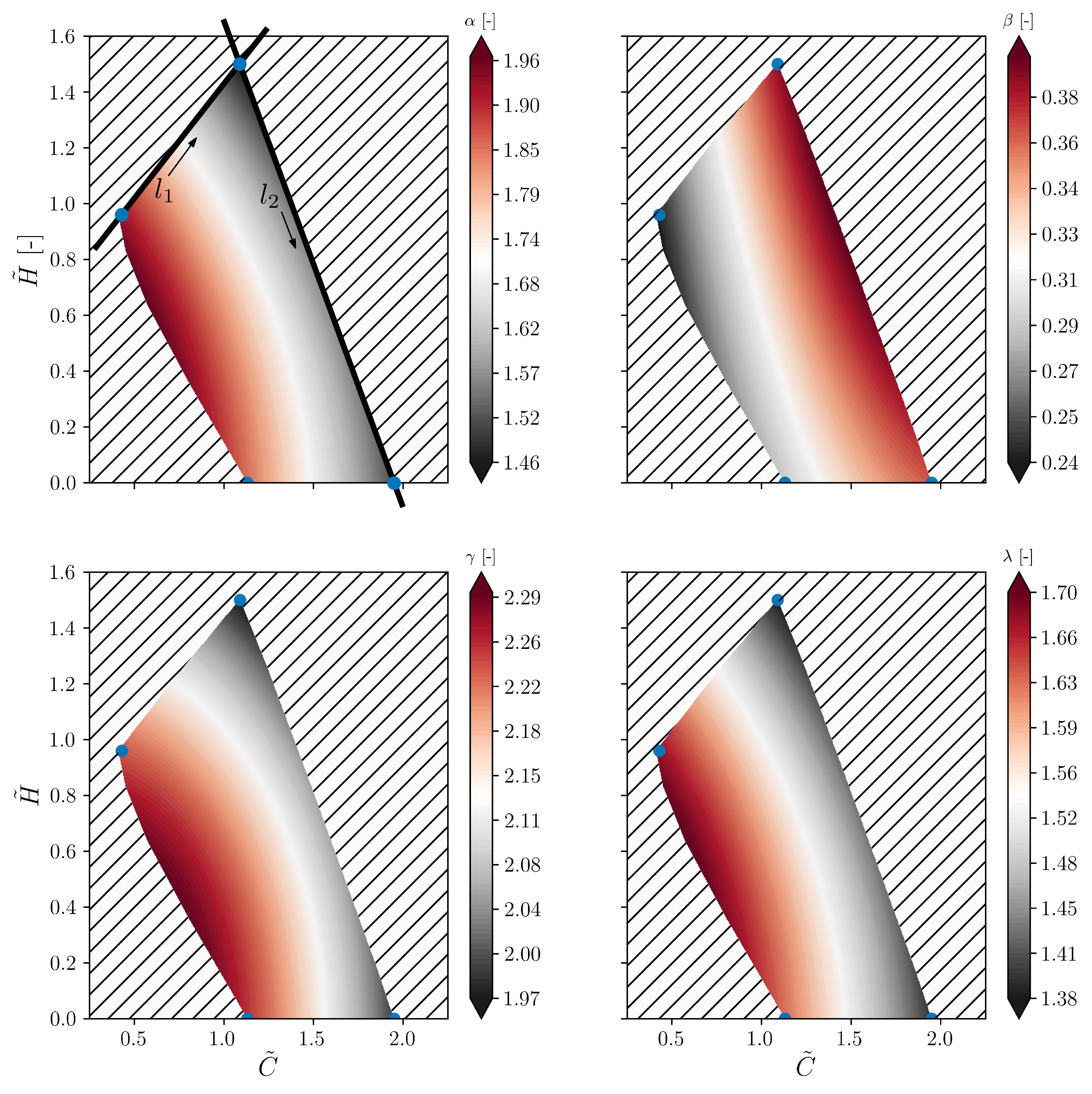}
                  \caption{}
                  \label{fig:GP2Dcornera}
     \end{subfigure}
     \hfill
     \begin{subfigure}[b]{0.48\textwidth}
         \centering
         \includegraphics[width=\textwidth]{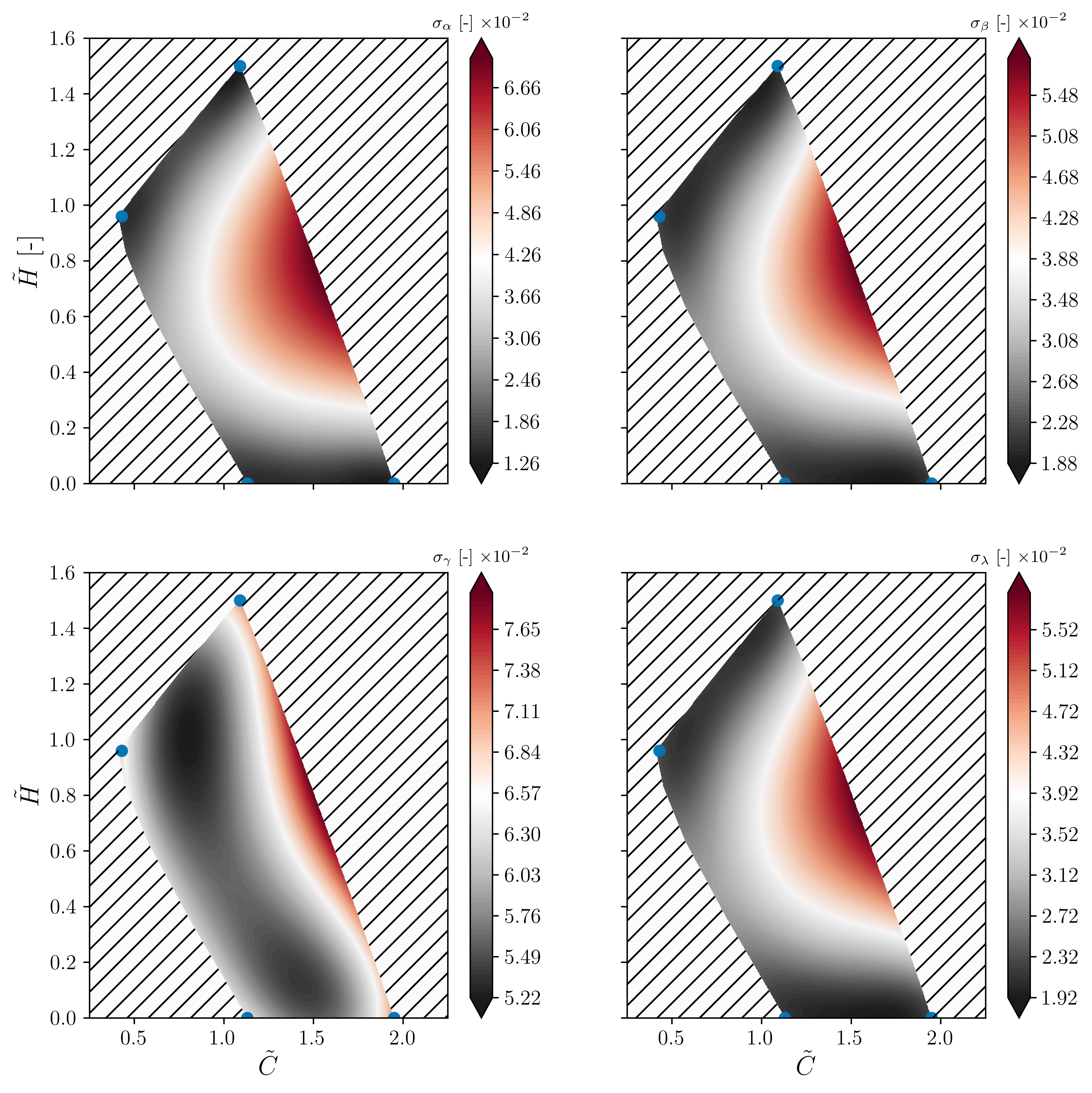}
                  \caption{}
                  \label{fig:GP2Dcornerb}
     \end{subfigure}
        \caption{Gaussian process results of the predicted correction factors, $\boldsymbol{\theta}^{\rm corr}$, based on the geometries that define the vertices of the constrained parameter space, $\{g_1,g_3,g_7,g_9\}$, plotted against the geometry parameters, $\tilde{C}$ and $\tilde{H}$. (a) Mean of the predicted correction factors. (b) Standard deviation of the predicted correction factors.}
        \label{fig:GP2Dcorner}
\end{figure*}

\begin{figure*}[!t]
     \centering
     \begin{subfigure}[b]{0.48\textwidth}
         \centering
         \includegraphics[width=\textwidth]{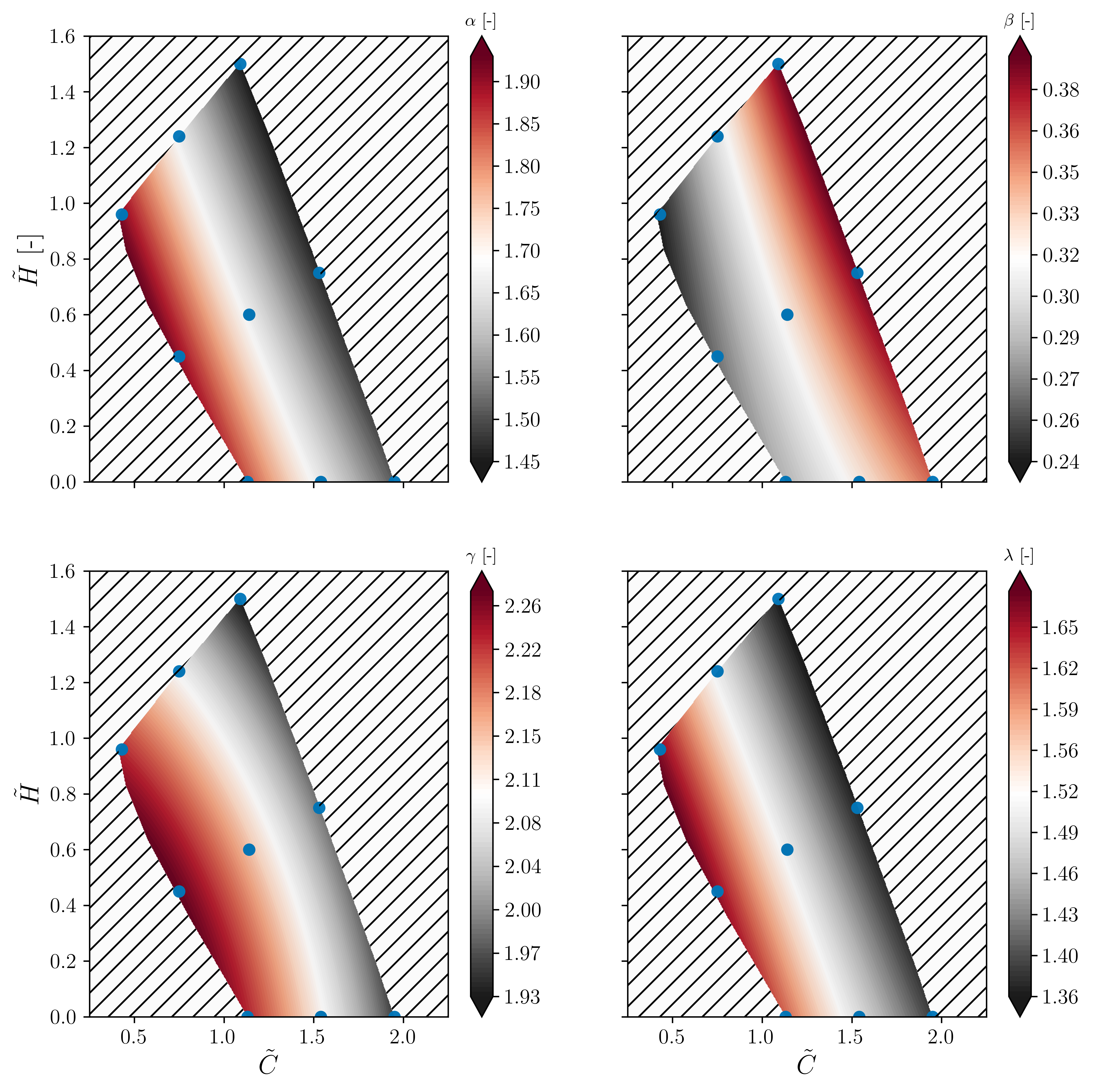}
                  \caption{}
                  \label{fig:GP2Dfulla}
     \end{subfigure}
     \hfill
     \begin{subfigure}[b]{0.48\textwidth}
         \centering
         \includegraphics[width=\textwidth]{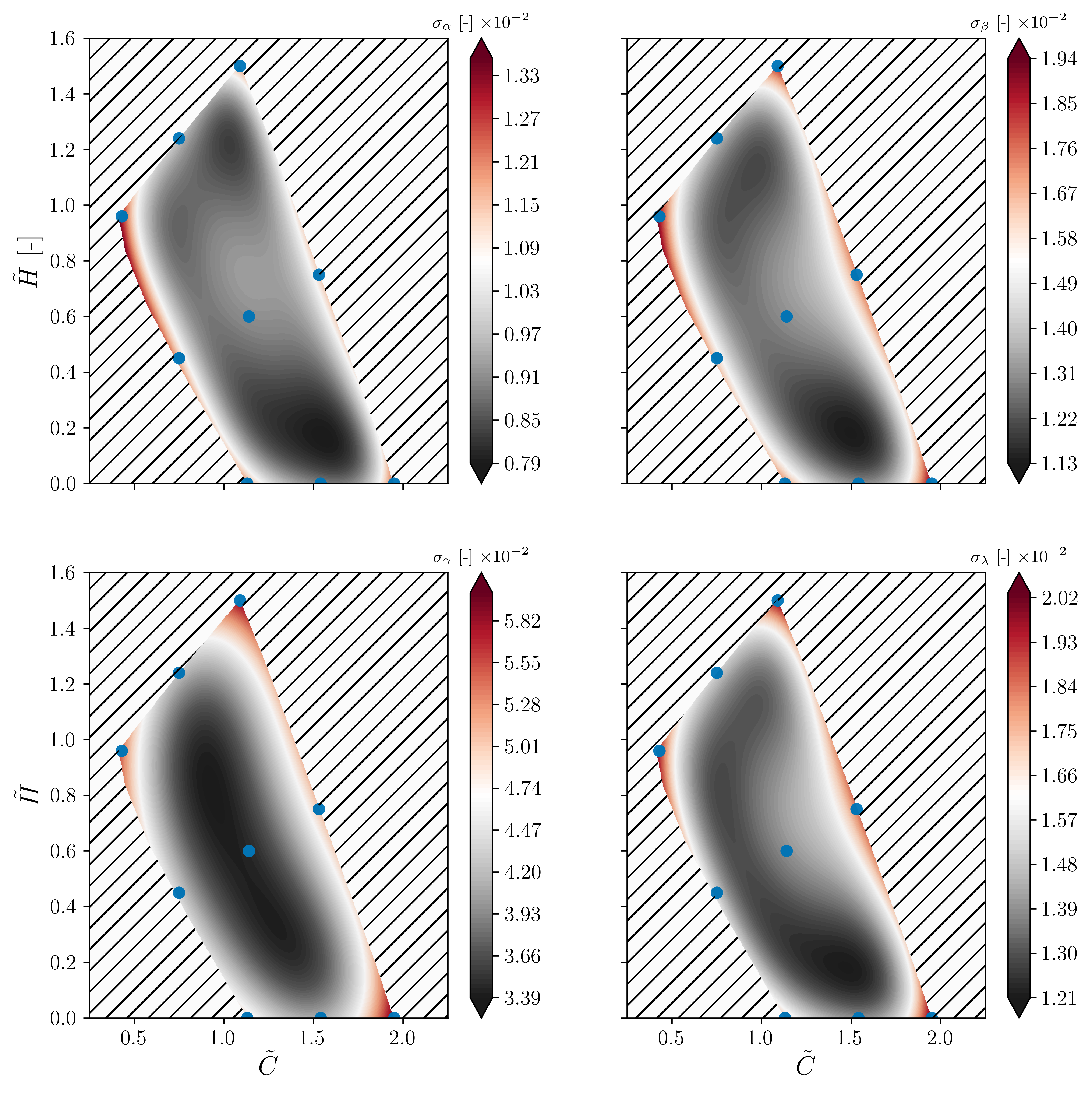}
                  \caption{}
                  \label{fig:GP2Dfullb}
     \end{subfigure}
        \caption{Gaussian process results of the predicted correction factors, $\boldsymbol{\theta}^{\rm corr}$, based on 
        all training geometries, $\{g_1,g_2,\ldots,g_9\}$, plotted against the geometry parameters, $\tilde{C}$ and $\tilde{H}$. (a) Mean of the predicted correction factors. (b) Standard deviation of the predicted correction factors.}
        \label{fig:GP2Dfull}
\end{figure*}

Following the Bayesian calibration of the ROM parameters, we construct the initial GP using the vertex geometries, $g_1,g_3,g_7,g_9$, which is visualized in Figure~\ref{fig:GP2Dcorner}. This figure shows the mean (Figure~\ref{fig:GP2Dcornera}) and standard deviation (Figure~\ref{fig:GP2Dcornerb}) of the predicted correction factors. The means exhibit a dependency on the data points similar to that observed for the ventricle diameter isocontours in Figure~\ref{fig:paramspace2D}. The $\alpha$, $\gamma$, and $\lambda$ means increase with the diameter, while the mean of $\beta$ decreases. This dependence is explained from the fact that ventricles with fixed cavity and wall volumes exhibit a stiffer behavior when the diameter increases (Figure~\ref{fig:Hemodynamics2Da}). On one hand, the stiffness factors, $\gamma$ and $\lambda$, are increased, while on the other hand, the slope of the linear generalized one-fiber model~\eqref{eq:fgenonefiber} is decreased, indicating a more restraint response (Section~\ref{sec:lowfidelity}). Upon adding additional data points to the GP, as shown in Figure~\ref{fig:GP2Dfulla}, this geometry dependence is maintained, while the uncertainty associated with the prediction is reduced, as conveyed by the standard deviations in Figure~\ref{fig:GP2Dfullb}. The observed dependence on geometry parameters other than the cavity and wall volumes is a crucial feature of our reduced-order modeling framework, as many standard one-fiber models (Section~\ref{sec:genrelations}) are fundamentally incapable of incorporating such parameters.

\begin{figure*}[!t]
     \centering
     \begin{subfigure}[b]{0.48\textwidth}
         \centering
         \includegraphics[width=\textwidth]{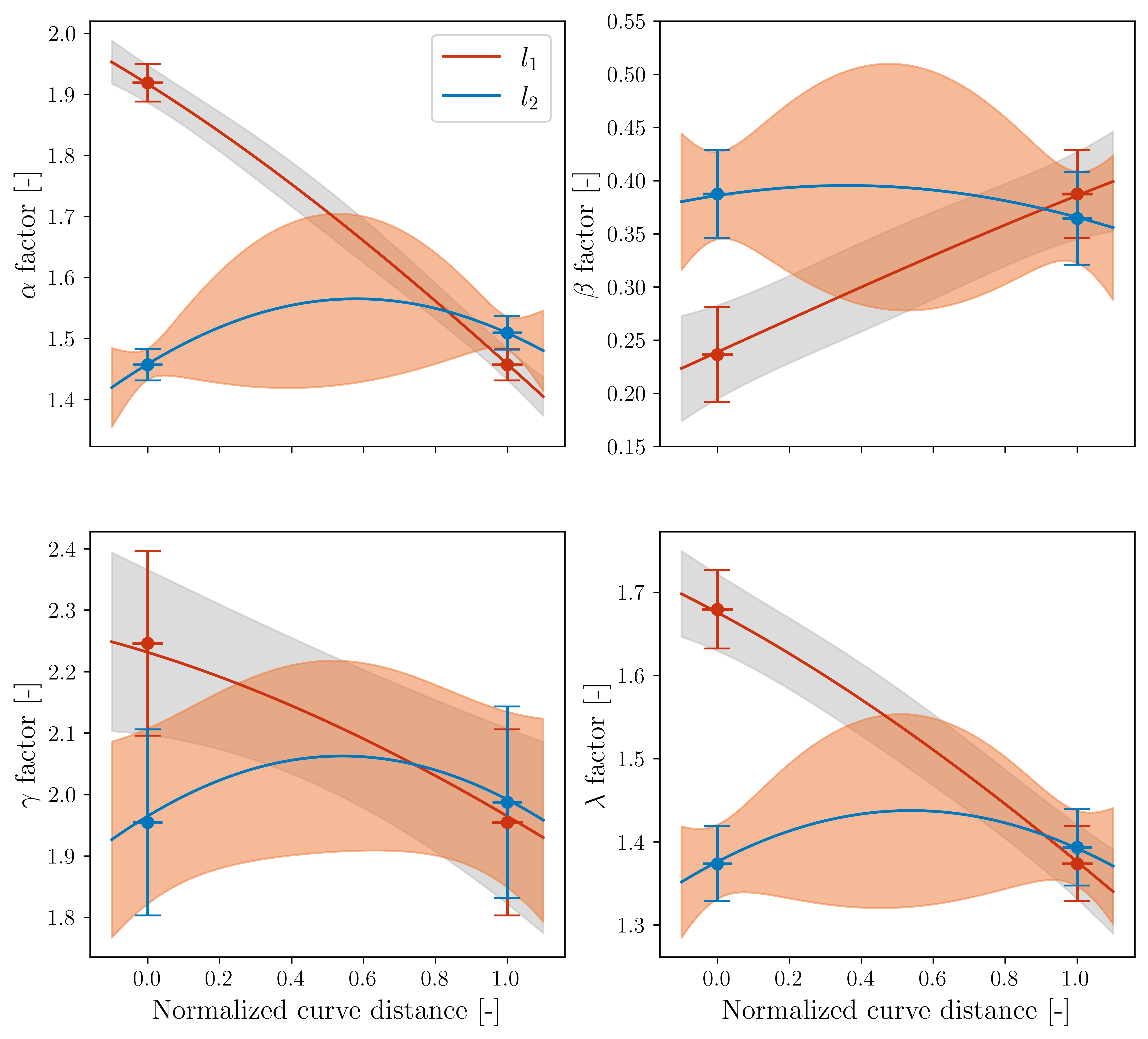}
                  \caption{}
                  \label{fig:GP2Dlinesa}
     \end{subfigure}
     \hfill
     \begin{subfigure}[b]{0.48\textwidth}
         \centering
         \includegraphics[width=\textwidth]{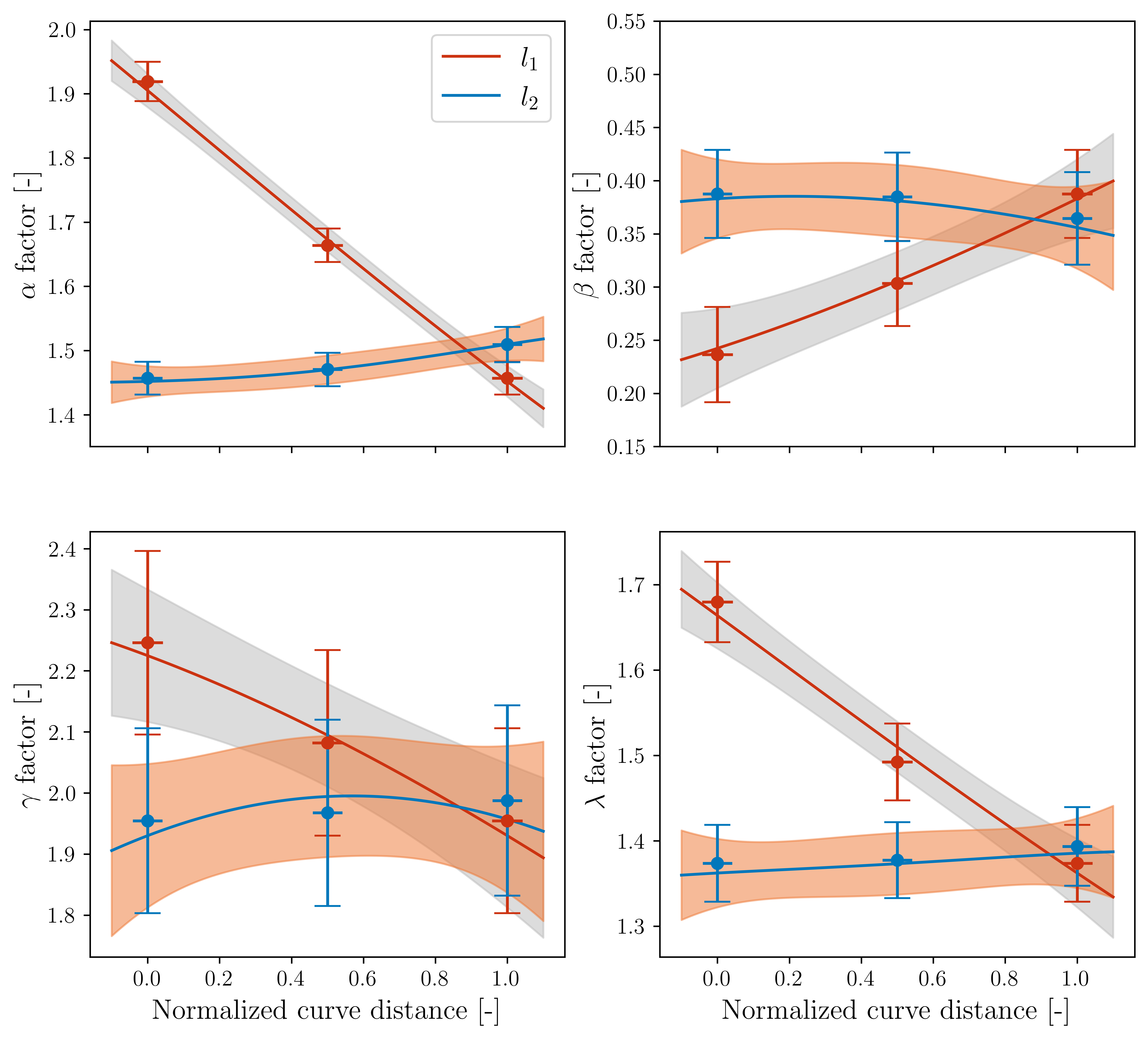}
                   \caption{}
                  \label{fig:GP2Dlinesb}
     \end{subfigure}
        \caption{Gaussian process for the correction factors plotted along the two characteristic lines, $l_1$ and $l_2$, in Figure~\ref{fig:GP2Dcornera}. The lines indicate the mean values of the correction factors, while the bands indicate the $95\%$ credible intervals. The data points and error bars are obtained from the Bayesian calibration. (a) GP constructed using the four corner vertices. (b) GP constructed using all nine geometries.}
        \label{fig:GP2Dlines}
\end{figure*}

To examine the local behavior of the GP prediction, we define two lines in Figure~\ref{fig:GP2Dcornera}, $l_1$ and $l_2$, along two boundaries of the prediction space. By evaluating the means and standard deviations along these two lines, we obtain the results as presented in Figure~\ref{fig:GP2Dlines}. When initializing the GP using only the boundary vertices, we observe that the GP matches the uncertainties associated with the data points. The GP prediction is observed to be more confident (lower credible interval) along $l_1$ compared to $l_2$, which can be explained by the length scale as defined in Equation~\eqref{eq:lengscales}. Given that the GP generally optimizes toward the upper bound of this length scale, and the physical length of $l_1$ is approximately twice as small as that of $l_2$, we obtain a higher confidence along $l_1$.

When we include the remaining data points, as shown in Figure~\ref{fig:GP2Dlinesb}, we observe a slightly different behavior. The GP prediction does no longer necessarily match the uncertainties associated with the data points and the prediction confidence increases. Both these observations are, again, related to the length scales. Provided that we obtain the same optimized length scales as in Figure~\ref{fig:GP2Dlinesa}, the additional data point in the middle effectively halves the physical distance between the data points. This ultimately improves the confidence, but also shifts the GP prediction slightly away from the data point mean values. This behavior is a direct consequence of the optimization of the log-marginal likelihood~\eqref{eq:gp_lml}. More specifically, the term that describes the model complexity is reduced in this optimization, \emph{i.e.}, the GP prefers linear predictions. This feature of GPs provides a natural way of combining uncertainties in data points with the likeliness of the resulting prediction. Ultimately, the GP prediction improves the data point uncertainties, as it combines the information of all data points.

To analyze the effects of the uncertainty propagation of the correction factors on the physical quantities of interest, \emph{i.e.}, the pressure and volume, we perform a forward analysis of the ROM. As an example, we use the results of $g_5$ and compare three results, \emph{viz.}: The Bayesian calibration result, the GP prediction when excluding $g_5$ (\emph{i.e.}, only considering the vertex geometries), and the GP prediction when including it as an additional training point. The forward analysis is run by performing a Monte-Carlo sampling of $10,000$ samples taken from the predicted multivariate normal distribution.

\begin{figure*}[!t]
     \centering
     \begin{subfigure}[b]{0.48\textwidth}
         \centering
         \includegraphics[width=\textwidth]{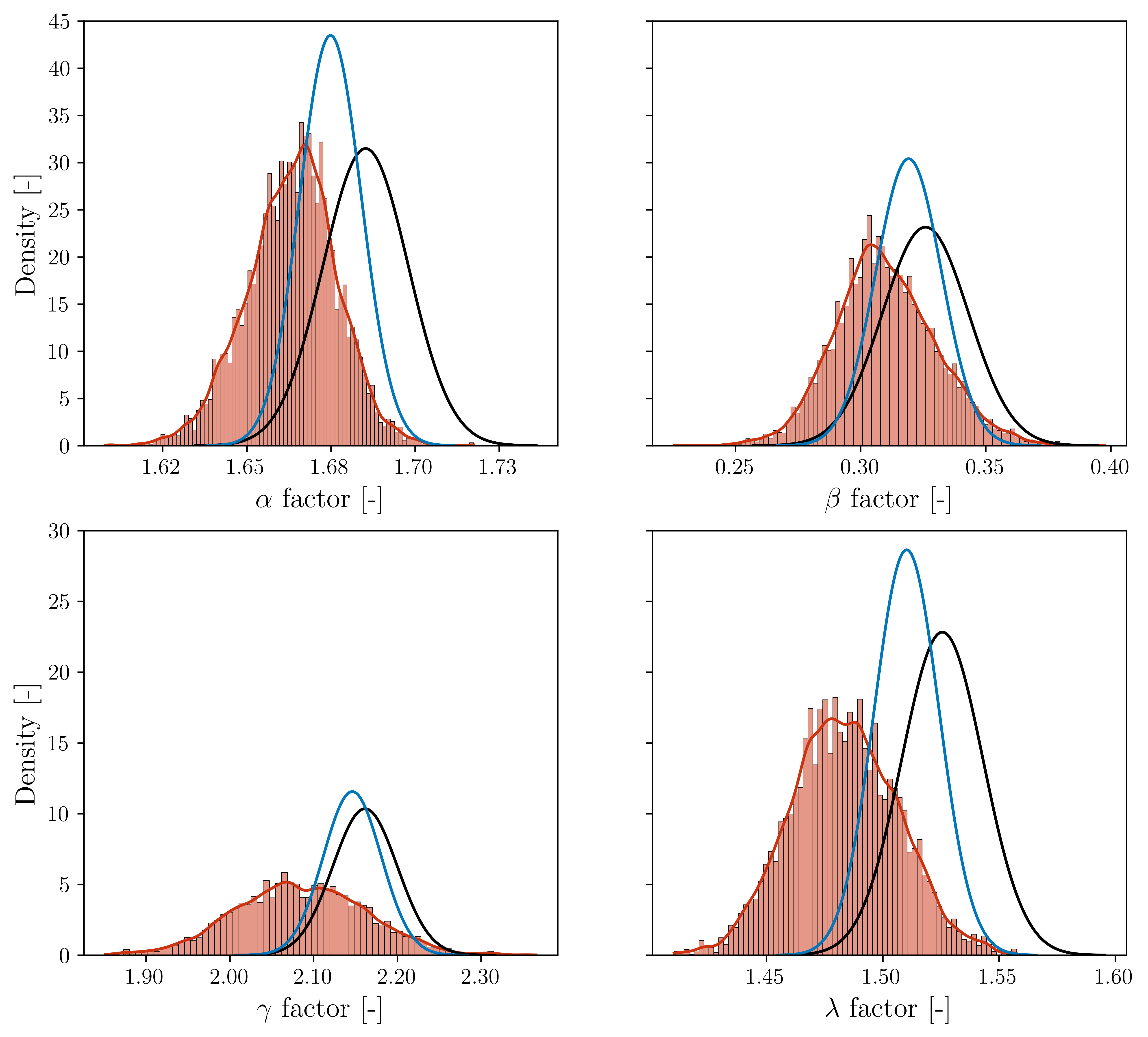}
                  \caption{}
                  \label{fig:GP2Dpredictiona}
     \end{subfigure}
     \hfill
     \begin{subfigure}[b]{0.44\textwidth}
         \centering
         \includegraphics[width=\textwidth]{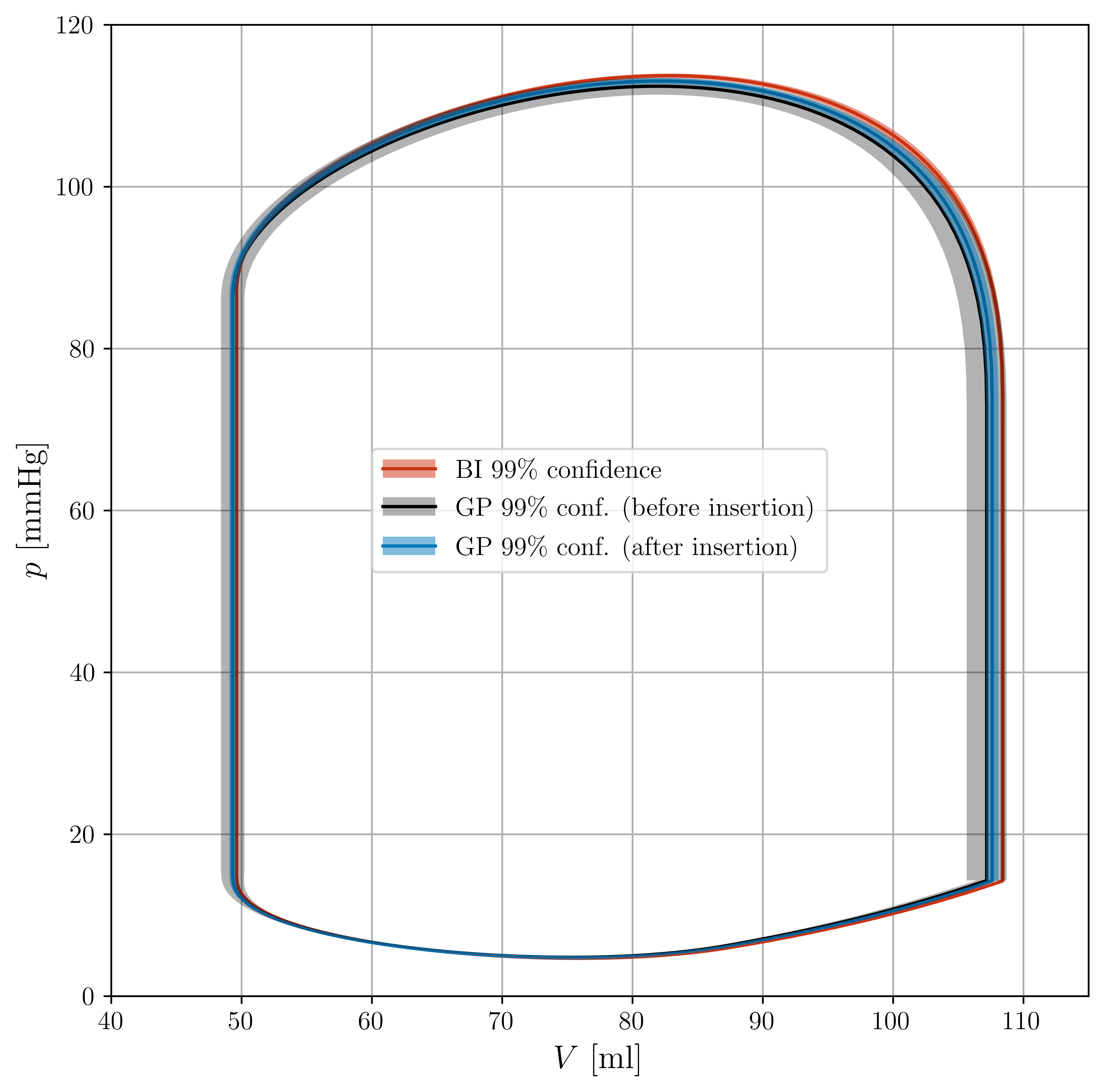}
                  \caption{}
                  \label{fig:GP2Dpredictionb}
     \end{subfigure}
        \caption{Comparison between the Bayesian inference result and the Gaussian process prediction of the correction factors for geometry $g_5$. (a) Posterior probability densities of the inferred correction factors (red) and the Gaussian process prediction based on the vertex training data (\emph{i.e.}, excluding $g_5$) (black) and after including $g_5$ as an additional point int he training set (blue). (b) Forward analysis of the ROM using the multivariate normal distributions of the correction factors, showing the 99\% credible intervals associated with the parametric uncertainty (\emph{i.e.}, excluding the noise model of Section~\ref{ssec:BIformulation}).}
        \label{fig:GP2Dprediction}
\end{figure*}

The results of the forward analysis are presented in Figure~\ref{fig:GP2Dprediction}. The initial GP prediction (before insertion) already yields an accurate prediction, in the sense that it shows a minimal change in pressure-volume response. After including the additional data point, the prediction is slightly improved, but a mismatch remains with the inferred result. This mismatch is, as mentioned in the discussion above, attributed to the model complexity minimized by the GP. Nonetheless, the effect on the pressure-volume curve is again limited. Furthermore, due to the correlation of the correction factors, the resulting pressure-volume credible interval associated with the parametric uncertainty (\emph{i.e.}, excluding the noise model of Section~\ref{ssec:BIformulation}) remains relatively small. Therefore, we can conclude that the trained GP for this example can accurately predict new geometries without the necessity to run the FOM.

\section{Framework application to scan-based data}\label{sec:ApplicationC}
In the previous section, the proposed reduced-order modeling framework was tested for an idealized ventricle with only two geometry parameters. This idealization would be questionable in clinical applications, as patient-specific scan geometries clearly have more detailed geometric features. The parametrized NURBS approach used for the idealized test case is not directly extensible to the case of feature-rich scan-based geometries, mandating the use of a different type of parametrization technique. As the number of geometry parameters impacts the complexity of the reduced-order modeling framework, it is important to employ a method that accurately represents scan-based geometries with a minimal number of geometry parameters. We propose to use the proper-orthogonal decomposition (POD) technique to parametrize scan-based geometries through a moderate number of modes.

In Section~\ref{ssec:modaldecomp} we introduce the modal decomposition technique in the context of scan-based geometries. We then discuss the generation of the FOM training data for the Gaussian process in Section~\ref{ssec:GP4D}. Application-specific aspects of the Bayesian calibration procedure are then discussed in Section~\ref{ssec:BI4D}, after which the results of our ROM framework for a scan-based analysis are presented in Section~\ref{ssec:Results4D}.

\subsection{Ventricle geometry parametrization through modal decomposition}\label{ssec:modaldecomp}
In our isogeometric FOM all ventricle geometries are three-dimensional NURBS volumes. We construct these volumes by lofting the NURBS surfaces for the endocardium and epicardium, which we represent by the column vector
\begin{equation}\label{eq:Xgeom}
    \mathbf{X} = \left[ \underbrace{x^{\mathrm{cps}}_1, y^{\mathrm{cps}}_1, z^{\mathrm{cps}}_1}_{\mathrm{endo}} ..., \underbrace{x^{\mathrm{cps}}_1, y^{\mathrm{cps}}_1, z^{\mathrm{cps}}_1}_{\mathrm{epi}}, ..., \underbrace{w^{\mathrm{cps}}_1, w^{\mathrm{cps}}_2, w^{\mathrm{cps}}_3}_{\mathrm{endo}} ,..., \underbrace{w^{\mathrm{cps}}_1, w^{\mathrm{cps}}_2, w^{\mathrm{cps}}_3}_{\mathrm{epi}}  \right]^{\mathrm{T}},
\end{equation}
where $(x,y,z)$ are the control point locations and $w$ the corresponding weight. All considered NURBS objects are topologically consistent, meaning that they are based on the same parameter domain and patch connectivity. The fitting algorithm proposed by Willems \emph{et al.}~\cite{willems_echocardiogram-based_2024} is used to convert scan data into an analysis-suitable NURBS object. The direct usage of the NURBS control points for the geometry parametrization in our ROM framework is impractical, as this would result in a high-dimensional input space for the Gaussian process mapping, the training of which would require a very large number of FOM simulations.

To reduce the number of geometry parameters we use a proper orthogonal decomposition \cite{liang_proper_2002} to approximate the control net \eqref{eq:Xgeom} as
\begin{equation}\label{eq:Xapprox}
    \mathbf{X}(\mathbf{c})  \approx \mathbf{X}_{\mathrm{ref}} + \overline{\mathbf{dU}} \, \mathbf{c},
\end{equation}
where the modal coefficients, $\mathbf{c} \in \mathbb{R}^{n^{\rm geom}}$, serve as the geometry parameters for the Gaussian process (Section~\ref{sec:GP}) and where $\mathbf{X}_{\mathrm{ref}}$ corresponds to the reference geometry as defined by Table~\ref{tab:refparam}. The columns of the matrix $\overline{\mathbf{dU}}$ are the dominant deformation modes corresponding to the modal coefficients.

\begin{table}[]
\caption{Physiological dimensions of the left ventricle at end-diastole. The ranges represent a $95\%$ credible interval.}\label{tab:EDphysvalues}
\centering
\begin{tabular}{lc}
\hline
                         & Range \\ \hline
LV diameter \cite{lang_recommendations_2015,galderisi_standardization_2017} & $4.2 - 5.8$ {[}cm{]}   \\
LV basal diameter \cite{anwar_true_2007,de_groot-de_laat_how_2019}  & $2.4-4.4$ {[}cm{]} \\
LV cavity volume \cite{lang_recommendations_2005,lang_recommendations_2015,galderisi_standardization_2017} & $62 - 150$ {[}ml{]} \\
LV wall volume \cite{lang_recommendations_2015,galderisi_standardization_2017} & $84 - 213$ {[}ml{]} 
\end{tabular}
\end{table}

\begin{figure*}[!t]
     \centering
     \begin{subfigure}[b]{0.48\textwidth}
         \centering
         \includegraphics[width=\textwidth]{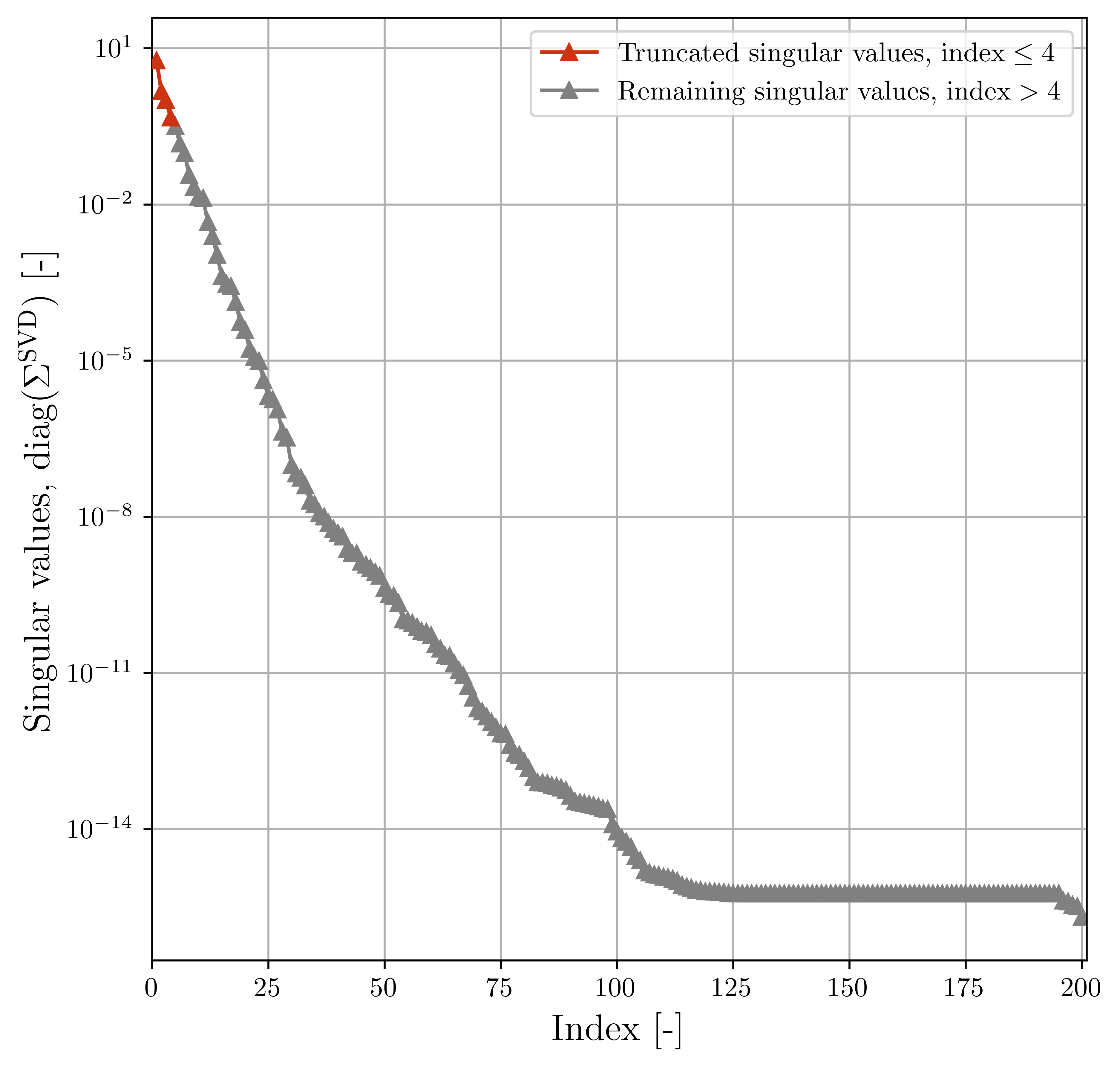}
                  \caption{}
                  \label{fig:GP4DPODa}
     \end{subfigure}
     \hfill
     \begin{subfigure}[b]{0.4\textwidth}
         \centering
         \includegraphics[width=\textwidth]{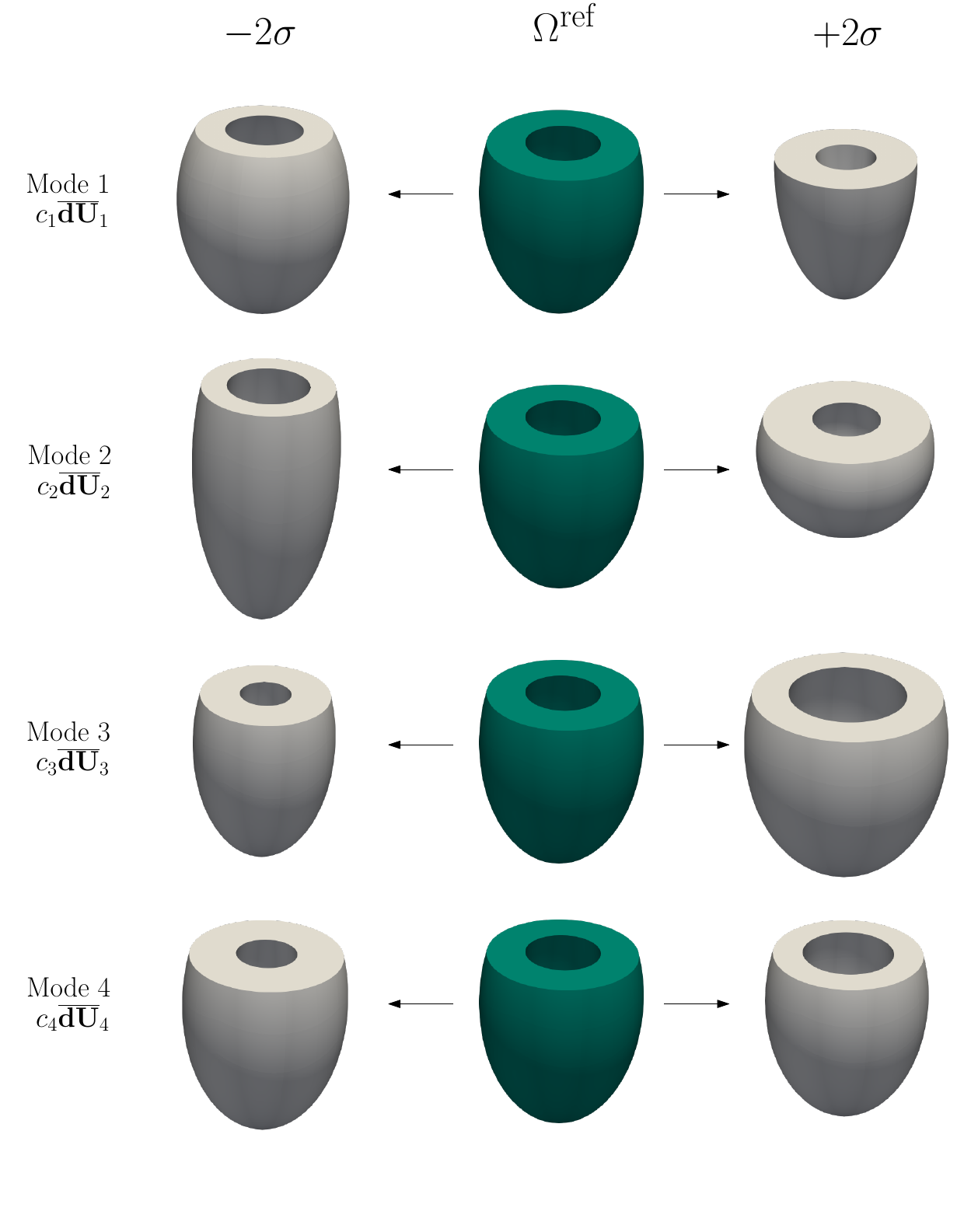}
                  \caption{}
                  \label{fig:GP4DPODb}
     \end{subfigure}
        \caption{Results of the singular value decomposition (SVD) based on synthetic population data with $n^{\rm pop}=200$ NURBS geometries. (a) The singular values resulting from the SVD show a rapidly decaying energy. (b) The $n^{\rm geom}=4$ dominant shape modes, with the ranges representing the $95$\% credible interval with respect to the population data.}
        \label{fig:GP4DPOD}
\end{figure*}

To attain the dominant shape modes, we perform a singular value decomposition on a set of population-based synthetic geometries. From the literature, we retrieve physiological ranges (Table~\ref{tab:EDphysvalues}) for the cavity volume, wall volume, basal diameter, and ventricle diameter, all evaluated at end-diastole. To generate synthetic population data, we perform a Monte-Carlo sampling using uncorrelated log-normal distributions corresponding to these ranges. The sampled quantities can then be used to determine the idealized geometry parameters, $\xi_{\mathrm{endo}}$, $\xi_{\mathrm{epi}}$, $H$, and $C$, \emph{cf.}~Equation~\eqref{eq:lvgeom}, from which NURBS ventricles can be constructed. The assumption that the geometry parameters in Table~\ref{tab:EDphysvalues} are uncorrelated is invalid, but necessary to make on account of usable correlation information being absent in the literature. The consequence of this assumption is that non-physiological ventricle geometries are sampled, \emph{e.g.}, geometries with very thin walls or with a very slender profile. We filter out such non-physiological geometries to attain a set of $n^{\rm pop}=200$ physiologically realistic ventricles,
\begin{equation}
\label{eq:populationset}
    \mathbf{dX} =  \begin{bmatrix}
\vdots & \vdots &  & \vdots \\
d\mathbf{X}_1 & d\mathbf{X}_2 & \cdots & d\mathbf{X}_{n^{\rm pop}} \\
\vdots & \vdots &  & \vdots \\
\end{bmatrix},
\end{equation}
expressed in terms of the deformation vectors
\begin{equation}\label{eq:differenceX}
    d\mathbf{X}_i = \mathbf{X}_i - \mathbf{X}_{\mathrm{ref}}.
\end{equation}
The dominant shape modes are obtained from the singular-value decomposition of the synthetic data
\begin{equation}\label{eq:SVD}
    \mathbf{dX} = \mathbf{dU} \ \boldsymbol{\Sigma}^{\mathrm{SVD}} \ \mathbf{V}^{\mathrm{T}},
\end{equation}
where $\mathbf{dU}$ is the matrix with all shape modes, $\boldsymbol{\Sigma}^{\mathrm{SVD}}$ is a matrix containing the singular values, and $\mathbf{V}$ contains the linear combinations corresponding to the data. Since the singular values -- visualized in Figure~\ref{fig:GP4DPODa} -- show a rapid decay in energy (representative of their weight in the decomposition \eqref{eq:SVD}), the population data can be approximated as
\begin{equation}\label{eq:trunc}
    \mathbf{dX} \approx \overline{\mathbf{dU}} \ \overline{\boldsymbol{\Sigma}}^{\mathrm{SVD}} \ \overline{\mathbf{V}}^{\mathrm{T}},
\end{equation}
where $\overline{\boldsymbol{\Sigma}}^{\mathrm{SVD}}$ is the $n^{\rm geom} \times n^{\rm geom}$ truncated singular value matrix, $\overline{\mathbf{dU}}$ is the matrix with the $n^{\rm geom}$ dominant shape modes, and $\overline{\mathbf{V}}^{\mathrm{T}}$ is the truncated linear combination matrix. Based on Figure~\ref{fig:GP4DPODa}, in the remainder of this work we consider $n^{\rm geom}=4$.

The truncated singular value decomposition \eqref{eq:trunc} can be expressed as
\begin{equation}
    \mathbf{dX} \approx \overline{\mathbf{dU}} \mathbf{C},
\end{equation}
where the modal coefficient matrix
\begin{equation}
    \mathbf{C} = \begin{bmatrix}
 \mathbf{c}_1 & \mathbf{c}_2 & \cdots & \mathbf{c}_{n^{\rm geom}} 
\end{bmatrix}= \overline{\boldsymbol{\Sigma}}^{\mathrm{SVD}} \overline{\mathbf{V}}^{\mathrm{T}},
\end{equation}
contains the $n^{\rm geom}$ modal coefficients for each geometry, \emph{i.e.}, $\mathbf{c}_i=[c_{i,1},c_{i,2},\ldots,c_{i,n^{\rm geom}}]$. Correspondingly, for an arbitrary geometry, $\mathbf{c}$, the difference vector is given by $d\mathbf{X} \approx \overline{\mathbf{dU}} \mathbf{C}$, which in turn yields the shape approximation~\eqref{eq:Xapprox}. Note that the modal coefficients $\mathbf{c}=\mathbf{0}$ refer to the reference geometry, $\mathbf{X}_{\mathrm{ref}}$. Furthermore, it is noted that the approximation \eqref{eq:Xapprox} in principle does not guarantee $G^1$-continuity across patch interfaces. For the geometries considered in this work -- which do not extend significantly beyond the $G^1$-continuous population data -- this effect is insignificant and has not been observed to affect the attained FOM results discernibly.

To apply our ROM framework to a new patient-specific scan-based NURBS geometry, $\Omega^{\mathrm{scan}} \ni \mathbf{x}^{\rm scan}$, the modal coefficients, $\mathbf{c}$, that approximate this geometry must be determined. We do this by minimizing the Euclidean distance between the scan and the geometry approximation \eqref{eq:Xapprox}:
\begin{equation}
    \label{eq:minimizationproblem}
     \mathbf{c} = \argmin_{ \tilde{\mathbf{c}} \, \subset \, \mathbb{R}^4}\left( \int \limits_{\Omega_{\mathrm{endo}}^{\mathrm{scan}}} \left\| \mathbf{x}^{\mathrm{scan}} - \mathbf{x}^{\rm modal}(\tilde{\mathbf{c}}) \right\|^2  \mathrm{d}\Omega_{\mathrm{endo}}^{\mathrm{scan}} + \int \limits_{\Omega_{\mathrm{epi}}^{\mathrm{scan}}} \left\| \mathbf{x}^{\mathrm{scan}} - \mathbf{x}^{\mathrm{modal}}(\tilde{\mathbf{c}}) \right\|^2  \mathrm{d}\Omega_{\mathrm{epi}}^{\mathrm{scan}}  \right).
\end{equation}

\begin{figure}
    \centering
    \includegraphics[width=0.75\linewidth]{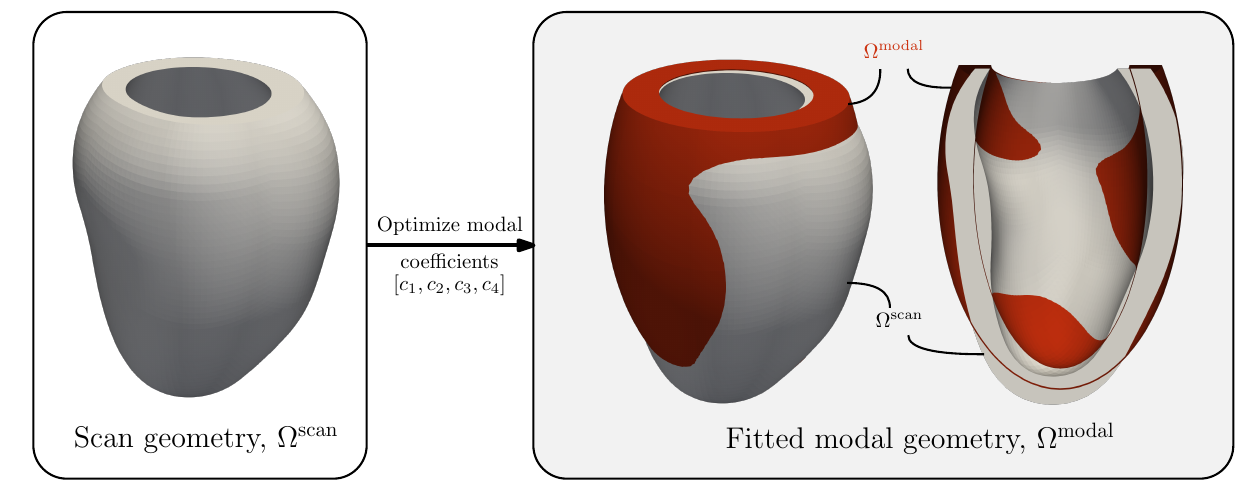}
    \caption{Graphical representation of a patient-specific NURBS geometry, $\Omega_{\mathrm{scan}}$, approximated by the dominant shape modes, $\Omega_{\mathrm{modal}}$. The approximate geometry is a result of the optimization of the modal coefficients, $\mathbf{c}$, such that the Euclidean distance between the two geometries is minimized.}
    \label{fig:modalfit}
\end{figure}

Figure~\ref{fig:modalfit} illustrates the minimization procedure for a patient-specific NURBS geometry similar to the one used in Ref.~\cite{willems_echocardiogram-based_2024}. This figure conveys that with only four shape modes, the scan-based ventricle can be matched closely. We note that the minimization problem \eqref{eq:minimizationproblem} does not incorporate  constraints for the cavity and wall volume. While the extension of the minimization problem with such constraints is relevant -- in the sense that these volumes are known to affect the hemodynamic response -- it is not considered to be essential in the context of the current work.

\subsection{Training data generation}\label{ssec:GP4D}
\begin{figure}
    \centering
    \includegraphics[width=0.75\linewidth]{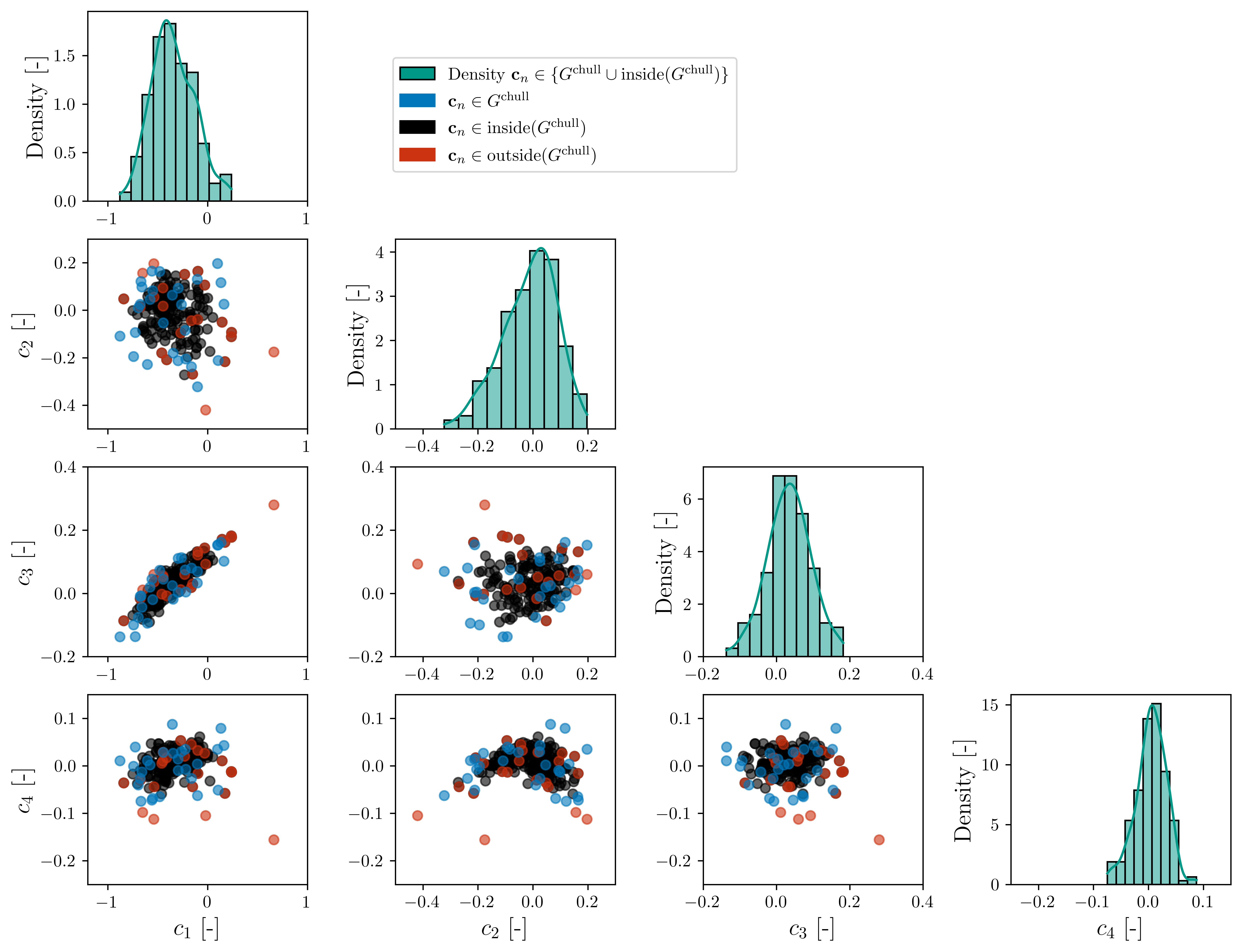}
    \caption{Corner plot of the modal coefficients, $\mathbf{c}=[c_1,c_2,c_3,c_4]$, resulting from the singular value decomposition (SVD). The initial SVD data set consists of $n^{\rm pop}=200$ geometries, of which each modal coefficient is plotted. A convex hull, $G^{\mathrm{chull}}$, with $27$ boundary points is determined such that approximately $90\%$ of the points are either on or inside the convex hull. The colors denote whether a synthetic geometry is either on the convex hull (blue), inside the convex hull (black), or outside and therefore neglected (red). Points located on the convex hull (blue) are used to initialize the Gaussian process, as future points are likely to be positioned within this hull, allowing for interpolation rather than extrapolation.}
    \label{fig:modalcoeff}
\end{figure}

To initialize the Gaussian process that converts the modal coefficients, $\mathbf{c}=[c_1,c_2,c_3,c_4]$, to correction factors for the generalized one-fiber model, $\boldsymbol{\theta}^{\rm corr}$, a training data set is required. Preferably, this initial set is as small as possible and allows for the interpolation of physiologically realistic geometries. In contrast to the two-dimensional constrained input space considered in Section~\ref{ssec:GP2D}, in the four-dimensional case considered here, the bounding vertices cannot be easily identified visually. To identify the initial training set of the Gaussian process in this higher-dimensional case, we assume that the $n^{\rm pop}=200$ synthetic population data set adequately samples the domain of physiologically realistic ventricles. We then use a convex hull algorithm to retrieve the boundary vertices of this domain. To limit the number of points in the training set (\emph{i.e.}, the number of FOM simulations), geometries with a low likelihood of occurrence are removed incrementally, until a convex hull, $G^{\mathrm{chull}}$, that contains approximately $90\%$ of the synthetic population is attained. This procedure to generate the initial training data is illustrated in Figure~\ref{fig:modalcoeff}, which displays the marginal distributions of the population data geometry parameters. The convex hull consists of $27$ boundary vertices, indicated in blue, for which the FOM is evaluated.

In contrast to the idealized test case in Section~\ref{ssec:BI2D}, the geometries in the training set are assumed to be registered at end-diastole. The cavity is therefore assumed to be loaded by an end-diastolic pressure of $12$~[mmHg]. To perform the isogeometric cardiac FOM simulations, we retrieve the stress-free configurations using the isogeometric generalized pre-stressing algorithm of Ref.~\cite{willems_echocardiogram-based_2024}. Once the pre-stress field is computed, we simulate $12$ consecutive cycles to obtain a hemodynamic steady state, from which the last cycle is used for the Bayesian calibration of the generalized one-fiber model.

\subsection{Bayesian calibration of the generalized one-fiber model}\label{ssec:BI4D}
To perform the Bayesian calibration of the correction factors, like the isogeometric FOM, the generalized one-fiber model is initialized at end-diastole and simulated for $12$ cycles while using the same reference cavity and wall volume. The prior distributions in Equation~\eqref{eq:2Dprior} are used, with the exception that $\sigma_{\beta}=0.05$. This more informed prior for the $\beta$-factor prevents it from moving toward zero, which is non-physiological~\cite{arts_epicardial_1982} as it would indicate that the fiber stress becomes independent of volume changes. A low value of the $\beta$-factor indicates a less homogeneous fiber stress distribution within the ventricle wall, following the explanation in Box~\ref{box:linearization}. This explanation is confirmed by the observation that the $\beta$-factor decreases when the FOM is considered with a fiber field that has not been optimized toward homogeneous strains (see \ref{app:fiberfield}). These observations either indicate that the current fiber-stress-pressure-volume relation used in the generalized one-fiber model has difficulties describing inhomogeneous fiber stresses, or that the IGA model exhibits significant heterogeneity. Either way, the choice to decrease the variance of the $\beta$-factor is reasonable when considering this effect. Additionally, due to the significant differences in pressure-volume results between the geometries in the training set, the FOM noise level is scaled according to hemodynamic landmarks, \emph{i.e.}, the maximum cavity pressure and stroke volume, $V^{\mathrm{stroke}}=V^{\mathrm{ED}} - V^{\mathrm{ES}}$. We set the different noise levels equal to $\sigma^{p}_i = 0.05 p^{\mathrm{max}}$  (independent of time), $\sigma^{V}_{\mathrm{max}} = 0.033 V^{\mathrm{stroke}}$, and $\sigma^{V}_{\mathrm{min}} = 0.017 V^{\mathrm{stroke}}$. The adaptive MCMC method with $125,000$ samples and a reset at every $25,000$ samples is used to generate the proposal covariance matrix that corresponds to a $0.234$ acceptance ratio. The resulting covariance matrix is then used for the regular MCMC, for which we run $50,000$ samples, excluding $10,000$ burn-in samples. Using the Bayesian calibration results for the 27 FOM simulations, the approach presented in Section~\ref{ssec:GP2D} is employed to initialize the Gaussian process.

\subsection{Patient-specific ventricle results}\label{ssec:Results4D}
The trained ROM framework is now analyzed in a clinical scenario, where an online prediction needs to be made for patients that are not incorporated in the training set. Below we discuss how to interpret these ROM predictions, specifically focusing on how to assess their trustworthiness.

\begin{figure*}[!t]
     \centering
     \begin{subfigure}[b]{0.50\textwidth}
         \centering
         \includegraphics[width=\textwidth]{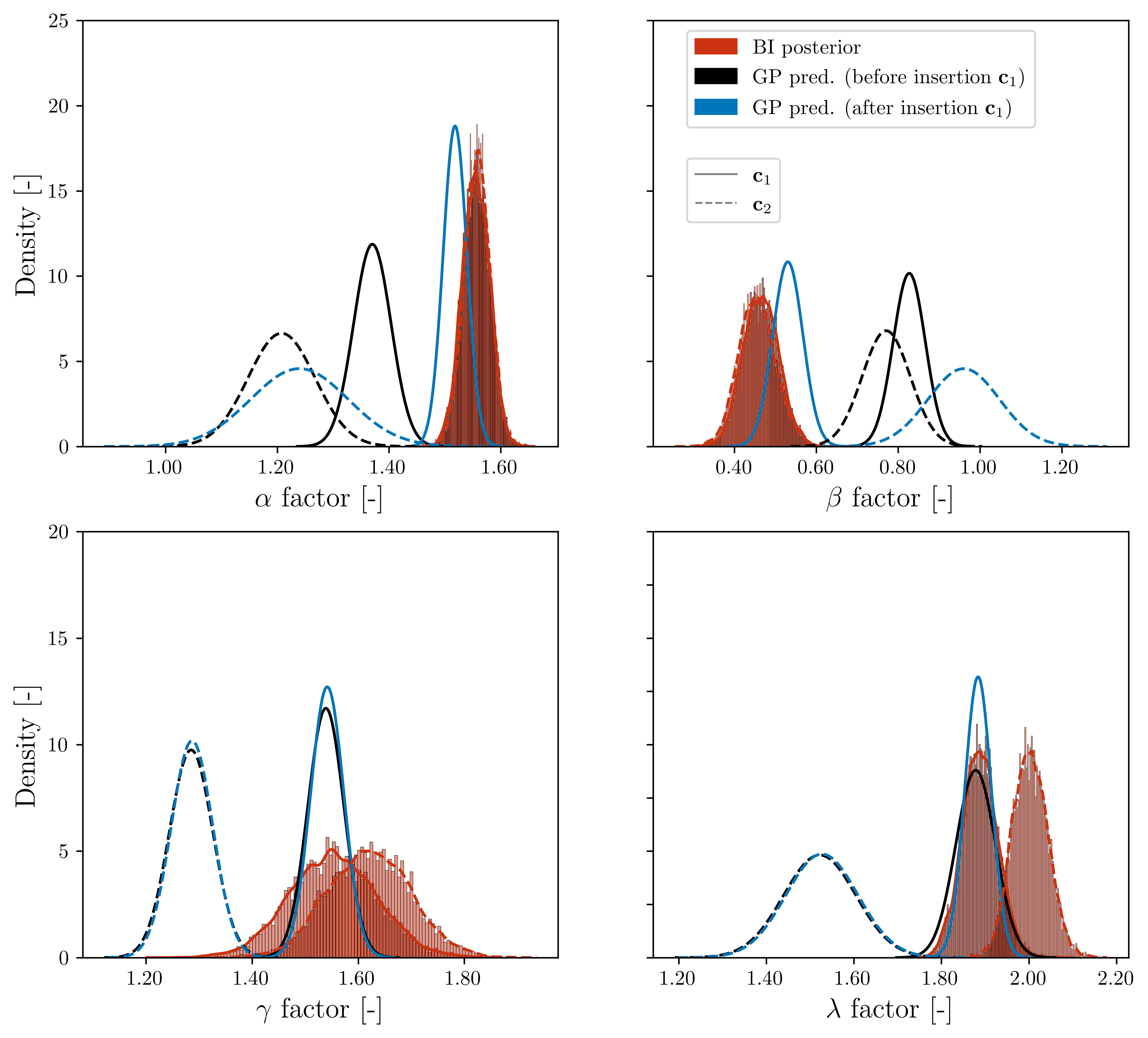}
                  \caption{}
                  \label{fig:GP4Dpatienta}
     \end{subfigure}
     \hfill
     \begin{subfigure}[b]{0.47\textwidth}
         \centering
         \includegraphics[width=\textwidth]{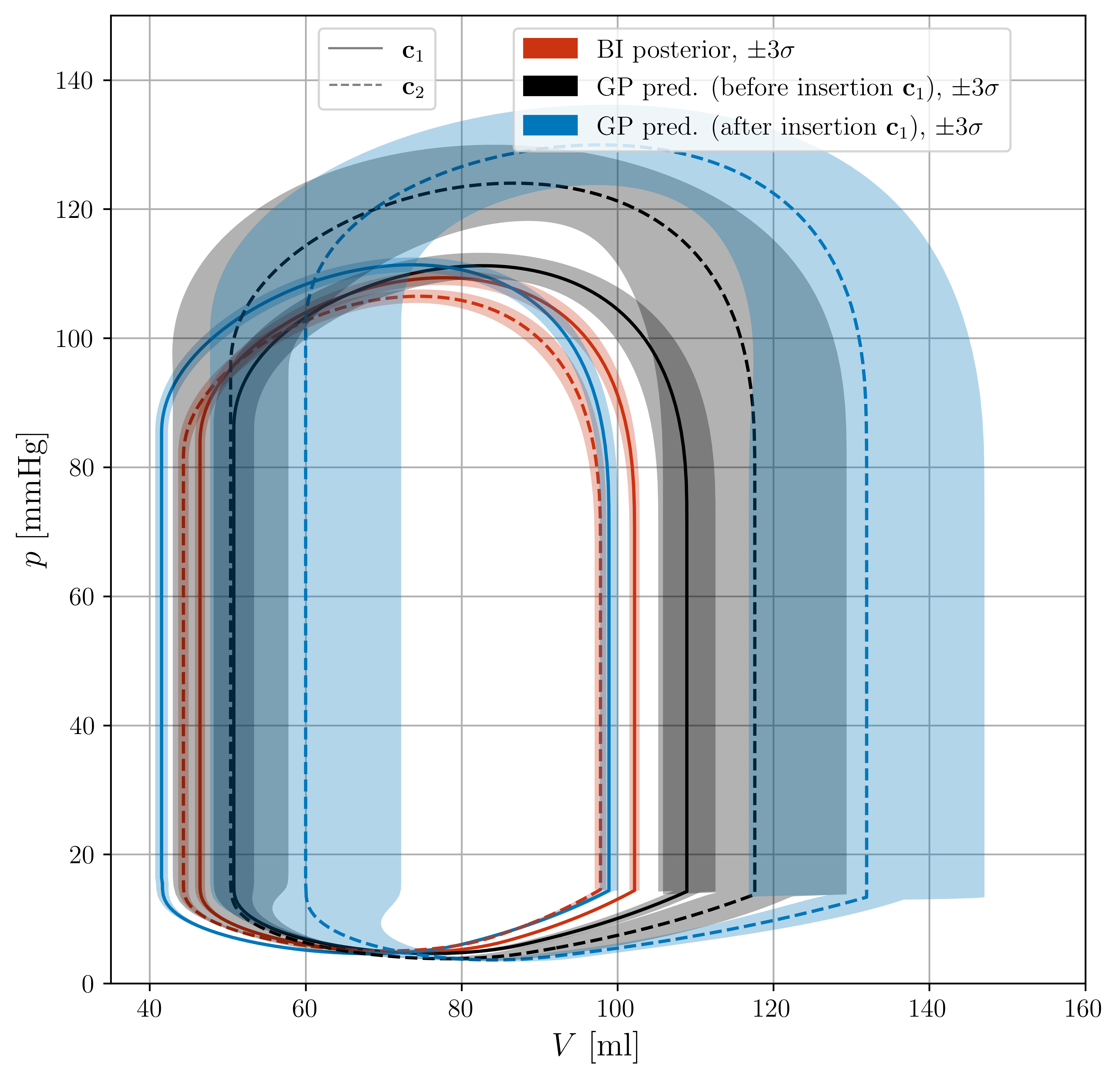}
                  \caption{}
                  \label{fig:GP4Dpatientb}
     \end{subfigure}
    \caption{Comparison between the ROM framework predictions and the FOM results for two patient-specific geometries, $\mathbf{c}_{1}$ (solid lines) and $\mathbf{c}_2$ (dashed lines). (a) Probability density curves of the inferred correction factors (red) and the Gaussian process prediction before including the inferred results of $\mathbf{c}_1$ as an additional GP training point (black) and after including it (blue). (b) Forward Monte-Carlo analysis of the ROM using the predicted multivariate normal distributions of the correction factors for both patient-specific geometries.}
        \label{fig:GP4Dpatient}
\end{figure*}

Consider two new patients, characterized by the modal coefficients, $\mathbf{c}_1$ and $\mathbf{c}_2$, obtained using Equation~\eqref{eq:minimizationproblem}. In the online stage, the Gaussian process produces the probability density curves for the generalized one-fiber model correction factors, shown in black in Figure~\ref{fig:GP4Dpatienta}. The corresponding pressure-volume curves, shown in black in Figure~\ref{fig:GP4Dpatientb}, follow from a Monte-Carlo analysis with $10,000$ samples, drawn from the multivariate normal distribution for the predicted correction factors. Note that a clinician only observes the pressure-volume responses in Figure~\ref{fig:GP4Dpatientb}, which reveal a relatively large uncertainty band, especially for the end-systolic, $V_{\mathrm{ES}}$, and end-diastolic, $V_{\mathrm{ED}}$, volumes. This uncertainty band represents only the correction factor uncertainty and excludes the Bayesian model noise, which is generally unavailable in the clinical setting.

The pressure-volume uncertainty band gives the clinician an initial indication of the predicted curve’s trustworthiness. Although the observed uncertainty band may be clinically acceptable, it is significantly broader than that of the FOM-calibrated result, which is shown in red and also excludes the model noise. Note that the FOM-calibrated result is not known in the online stage. The uncertainty level of this unknown result can be estimated, however, through the model noise level used for the Bayesian calibration, $\sigma^V_{\mathrm{min}}$. To assess the trustworthiness of the ROM prediction in the online stage, we therefore propose to compare the width of its uncertainty band to this model noise level. Based on insights gained from, among others, the results for the idealized test case in Section~\ref{sec:ApplicationS}, we propose that the width of the predicted uncertainty band should ideally be an order of magnitude smaller than the model noise level. If this is not the case, the clinician should be notified that additional FOM training data is necessary to improve the confidence of the online prediction. For the two specific cases considered in Figure~\ref{fig:GP4Dpatientb}, an end-diastolic volume $99\%$ credible interval of $V^{\mathrm{ED}}\pm3.5$[ml] and $V^{\mathrm{ED}}\pm12$[ml] is observed for $\mathbf{c}_1$ and $\mathbf{c}_2$, respectively. Given their respective stroke volumes of $V^{\mathrm{stroke}}\approx55$[ml] and $V^{\mathrm{stroke}}\approx67$[ml], a $99\%$ noise level of approximately $\pm2.7$[ml] and $\pm3.3$[ml] is expected, as discussed in Section~\ref{ssec:BI4D}. Since the predicted end-diastolic volume uncertainty is not an order of magnitude smaller than the noise level, this indicates a need for additional data points to be incorporated into the training set.

The addition of the FOM simulation corresponding to $\mathbf{c}_1$ to the training set illustrates an important effect encountered when considering Gaussian processes, \emph{viz}., predictions are likely to improve near the newly inserted data point, but may degrade in regions farther away from it. This effect is observed in Figure~\ref{fig:GP4Dpatienta}, where the blue curves show the results after incorporating the data of $\mathbf{c}_1$ into the Gaussian process. The solid blue line is closer to the inferred data than the original black line, conveying that the prediction for $\mathbf{c}_1$ indeed improves. However, the dashed blue line corresponding to $\mathbf{c}_2$ has moved farther away from the inferred data, indicating a decline in prediction accuracy. This result conveys that adding data points to the Gaussian process does not guarantee enhanced prediction quality across the entire prediction space. Fortunately, the observed decline in accuracy for $\mathbf{c}_2$ is accompanied with a widening of the uncertainty band in Figure~\ref{fig:GP4Dpatientb}, providing a visual indication of increased prediction uncertainty. We note that in the setting considered here, this effect is very pronounced, because an initial Gaussian process with only a limited number of data points is considered. As additional FOM data points are introduced in the training set, the Gaussian process' overall prediction quality will improve and this effect will become less pronounced.

\begin{figure*}[!t]
     \centering
     \begin{subfigure}[b]{0.48\textwidth}
         \centering
         \includegraphics[width=\textwidth]{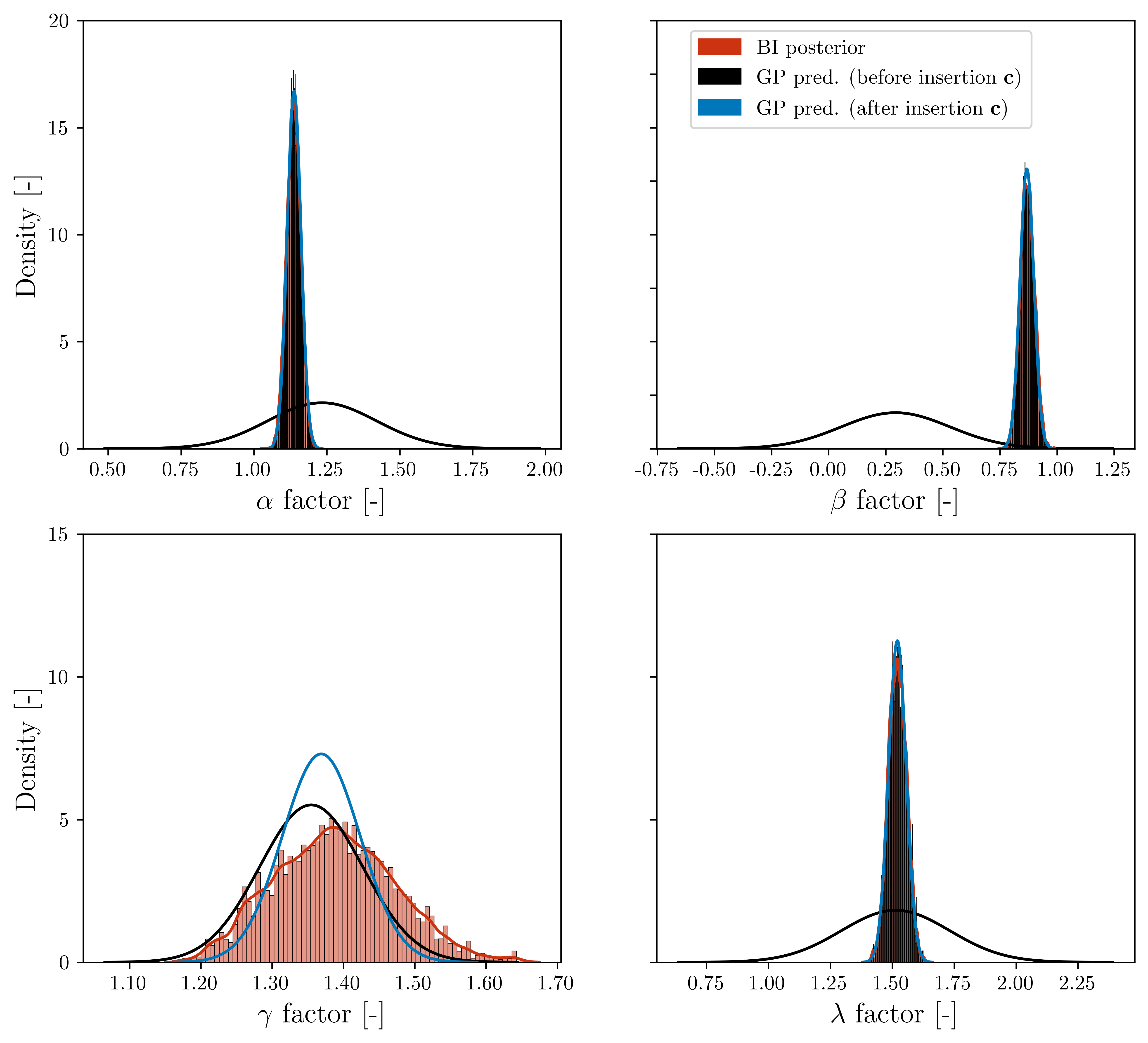}
                  \caption{}
                  \label{fig:GP4Dpatientspecifica}
     \end{subfigure}
     \hfill
     \begin{subfigure}[b]{0.48\textwidth}
         \centering
         \includegraphics[width=\textwidth]{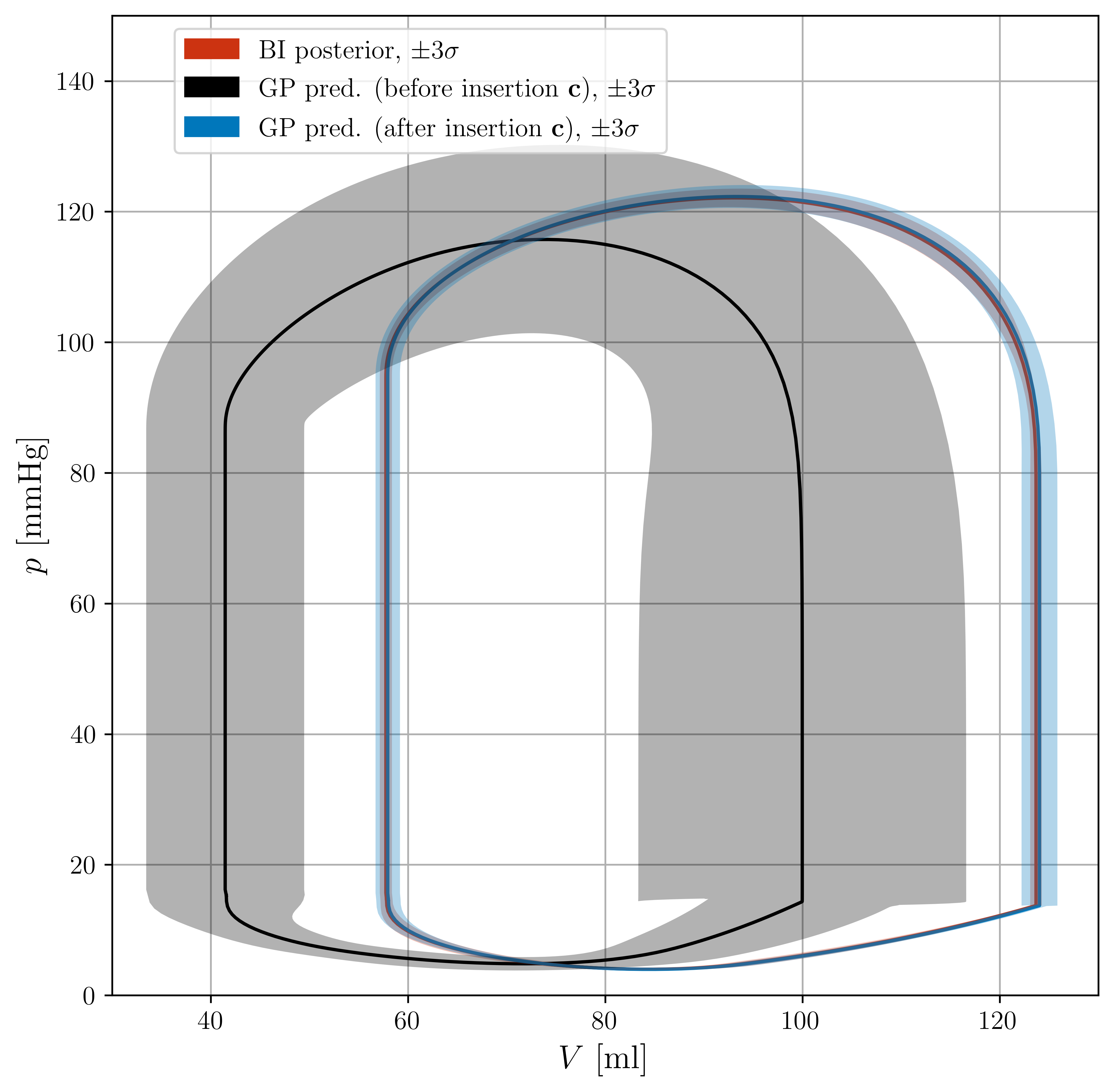}
                  \caption{}
                  \label{fig:GP4Dpatientspecificb}
     \end{subfigure}
        \caption{Comparison between the ROM framework prediction and the FOM results for the patient-specific scan geometry, $\Omega^{\rm scan}$, of Figure~\ref{fig:modalfit}. (a) Probability density curves of the inferred correction factors (red) and the Gaussian process prediction before including the inferred results of $\Omega^{\rm scan}$ as an additional GP training point (black) and after including it (blue). (b) Forward Monte-Carlo analysis of the ROM using the predicted multivariate normal distribution of the correction factors. Note that the pressure-volume curve (red) for the inferred result coincides with the mean prediction after adding the scan geometry as an additional point to the training set.}
        \label{fig:GP4Dpatientspecific}
\end{figure*}

To conclude our demonstration of the reduced-order modeling framework, we analyze its application to the patient-specific geometry, $\Omega^{\mathrm{modal}}$, in Figure~\ref{fig:modalfit}. This geometry was approximated by the modal coefficients $\mathbf{c}$ in Section~\ref{ssec:GP4D}. The online prediction based on a Gaussian process with 27 initial and 6 additional training simulations is shown in Figure~\ref{fig:GP4Dpatientspecific}. This prediction (black) shows a relatively high uncertainty for the correction factors (Figure~\ref{fig:GP4Dpatientspecifica}), which is attributed to the absence of sufficiently close (neighboring) data points. The predicted distribution for the $\beta$-factor is observed to be physiologically improbable, in the sense that it is significantly shifted toward zero (see Section~\ref{ssec:BI4D}). The corresponding pressure-volume curve in Figure~\ref{fig:GP4Dpatientspecificb} shows that the uncertainty in the correction factors is propagated through the generalized one-fiber ROM, resulting in a wide uncertainty band. This makes it clear that the provided prediction based on the Gaussian processes is not trustworthy. Following the criterion discussed above, the patient-specific data point is therefore added to the training set through a subsequent offline simulation. The updated prediction of the ROM framework, shown in blue in Figure~\ref{fig:GP4Dpatientspecific}, then results in a near-perfect match of the FOM. This illustrates the importance of incrementally populating the training set to gradually improve the online predictive capabilities of the ROM framework.

\section{Concluding remarks}\label{sec:ConcRec}
We have developed a reduced-order modeling (ROM) framework for the patient-specific mechanical analysis of the heart. The design of this ROM framework follows three principles: \emph{(i)} The end-user should be able to consider the ROM as a black box replacement of the full-order model (FOM) on which it is based, meaning that the input and output spaces are identical. \emph{(ii)} The ROM framework should be modular, facilitating the incorporation of established models and methods, thereby promoting the adaptation of the framework for different applications, as well as its integration into existing workflows. \emph{(iii)} The ROM framework should allow for the quantification of uncertainties, which are inherent to the clinical setting in which it is to be used. Through the incorporation of uncertainties, the loss in accuracy can be made explicit, in the sense that it is reflected by the uncertainty bounds. Below we draw conclusions regarding the most prominent aspects of the developed framework in relation to these design principles.

The main challenge with respect to maintaining the input space of the FOM pertains to the scan data. While scan data can be introduced in the considered continuum mechanics FOM through a NURBS fitting algorithm, the geometry assumptions made to derive simplified-physics models generally make them insensitive to local shape variations. In our framework, we solve this problem by calibrating a generalized one-fiber model to the FOM using Bayesian inference. The essential modeling step is to introduce geometry correction factors in the one-fiber model, consistent with the assumptions made for its derivation. Similarly, correction factors are introduced in the generalized one-fiber model to relate its constitutive model parameters to those of the FOM. In total, the generalized one-fiber model has four correction factors, which are calibrated to the pressure-volume curves of the FOM using a Markov chain sampler. It is demonstrated that through this approach the online ROM evaluations are made sensitive to local shape variations. The specific modeling choices made for the generalized one-fiber model correction factors can limit the sensitivity of the ROM. For example, the considered generalized one-fiber model cannot be expected to correct for material heterogeneity. Although this has not been limiting in the scope of this work, further generalization of the one-fiber model can extend its range of applicability.

To make online predictions, \emph{i.e.}, without the need to solve the FOM, a Gaussian process trained on synthetic population data is used to interpolate the correction factors for a geometry that is not in the training set. Since the required size of the FOM training set is strongly influenced by the input space of the Gaussian process, it is crucial to parametrize the geometry with as few parameters as possible. In this work, we propose to achieve this through a proper orthogonal decomposition technique, based on shape modes constructed using synthetically generated ventricles. We have found that, with only four shape modes, scan-based ventricle shapes can be matched accurately. While the synthetically generated ventricles are based on population data, the usage of a large number of patient scans to construct the shape modes can further improve the quality of the low-dimensional geometry approximation.

With respect to modularity, the designed ROM framework is rather abstract, in the sense that different choices can be made for most of the framework's components (blocks in Figure~\ref{fig:LFmodel}). This holds for the generalized one-fiber model, for the Gaussian process mapping, for the Bayesian calibration, and for the proper orthogonal decomposition. A clear argumentation has been provided for all our choices, but we emphasize that these are context-dependent. The ROM framework is adaptable to different contexts, not necessarily restricted to cardio-mechanical modeling. In such different contexts, it may very well be that alternative choices for some of the components are preferable. For example, the Gaussian process is ideally suited for the considered case with a moderate number of data points in the training set. However, in a setting where training data is abundant, the use of, \emph{e.g.}, a neural network may be preferable.

To enable the quantification of uncertainties, the components of the framework are required to propagate their random inputs. This is a prominent argument for the consideration of Gaussian processes in combination with Bayesian calibration of the correction factors on the FOM training data, as these techniques are inherently probabilistic. For the Gaussian process component, we have made the practical choice to construct a vector-valued field from multiple scalar-valued fields, as the latter approach allows for the usage of standard software implementations. However, an extension of the framework with a vector-valued correlated Gaussian process is desirable.

To translate the uncertainties in the generalized one-fiber correction factors to probabilistic pressure-volume curves, the generalized one-fiber model is evaluated using Monte Carlo sampling. On account of its computational efficiency, this forward uncertainty quantification step can be performed efficiently. The presented framework demonstrations illustrate how the obtained uncertainty bounds give insight into the trustworthiness of the ROM results. We believe that this is a valuable feature of the proposed framework, as it can indicate whether an online prediction should be considered with extra care. This indicator can also serve as a criterion for further populating the training set. In terms of probabilistic modeling, it would be interesting to apply the ROM framework to (forward and/or inverse) uncertainty quantification, incorporating variability in all model parameters (circulatory system parameters, stiffness parameters, \emph{etc.}).

The characteristics of the developed framework make it well-suited for clinical use. However, for actual implementation in a clinical setting, further clinical validation and integration into existing workflows are necessary. An important step to be made in the further development is the construction of a high-quality training set for a well-defined clinical problem setting.

\section*{Acknowledgment}
This publication is part of the COMBAT-VT project (no.~17983) of the research program High Tech Systems and Materials which is partly financed by the Dutch Research Council (NWO). Additionally, this work was performed within the IMPULS framework under the Picasso project (no.~TKI HTSM/20.0022) of the Eindhoven MedTech Innovation Center (e/MTIC, incorporating Eindhoven University of Technology, Philips Research, and the Catharina Hospital), including a PPS-supplement from the Dutch Ministry of Economic Affairs and Climate Policy. We would also like to thank the European Union’s Horizon 2020 research and innovation program for the financial support under grant agreement no.~101017578 (SIMCor) and the ECSEL Joint Undertaking (JU) under grant agreement no.~101007319 (AI-TWILIGHT). We acknowledge the team of Nutils~\cite{van_zwieten_nutils_2022} for their support regarding the numerical implementation of the cardiac model.

%% Appendices
\appendix

\section{Fiber orientation effect} \label{app:fiberfield}
A key component of the isogeometric FOM is the definition of the fiber field, ${\mathbf{e}}^{\mathrm{f0}}$ (Figure~\ref{fig:igafoma}). Retrieving the fiber field is a complex task and often prone to errors involving the segmentation procedure. The common approach is to use a rule-based method, in which the fiber field adheres to population-based measurements. For the IGA FOM, we have considered two methods in Ref.~\cite{willems_isogeometric_2024}: the method proposed by Rossi~\emph{et al.}~\cite{quarteroni_integrated_2017} and the one of Bovendeerd~\emph{et al.} \cite{bovendeerd_dependence_1992}. The former can be applied to any ventricular shape, while the latter is only defined for the idealized ventricle as shown in Figure~\ref{fig:LVideal3D}. The fiber field proposed by Bovendeerd~\emph{et al.} is optimized for homogeneous fiber strain during ejection. Compared to the method of Rossi \emph{et al.}, this optimized field is in better agreement with the homogeneous fiber stress assumption used to derive the one-fiber model (Section~\ref{sec:genrelations}). To clarify the role of the FOM fiber field on the reduced-order modeling framework, we here show the effect of the fiber field on the IGA FOM results and the resulting inferred values of the generalized one-fiber model.

\begin{figure*}[!t]
     \centering
     \begin{subfigure}[b]{0.48\textwidth}
         \centering
         \includegraphics[width=\textwidth]{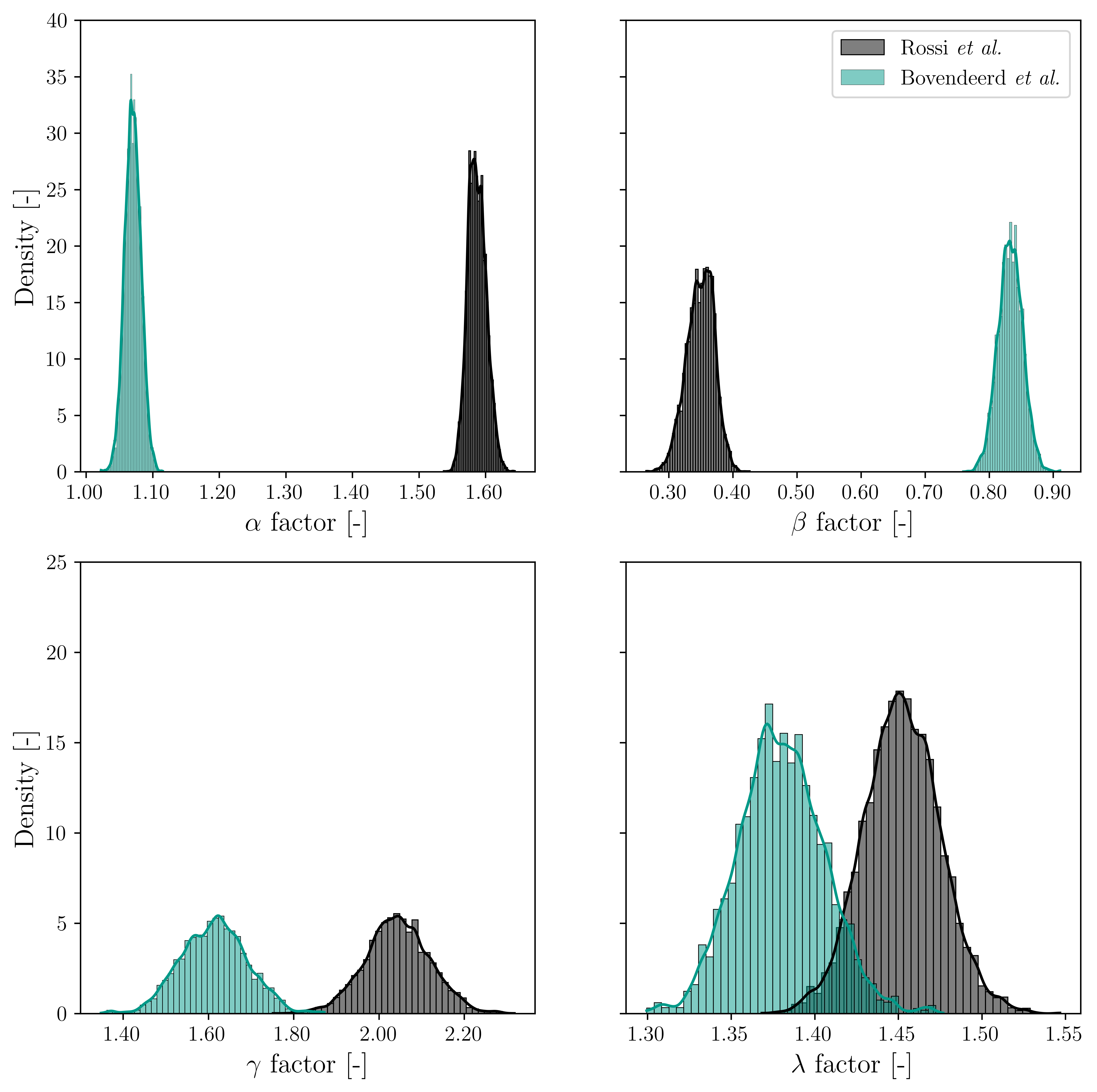}
                  \caption{}
                  \label{fig:BIFibera}
     \end{subfigure}
     \hfill
     \begin{subfigure}[b]{0.48\textwidth}
         \centering
         \includegraphics[width=\textwidth]{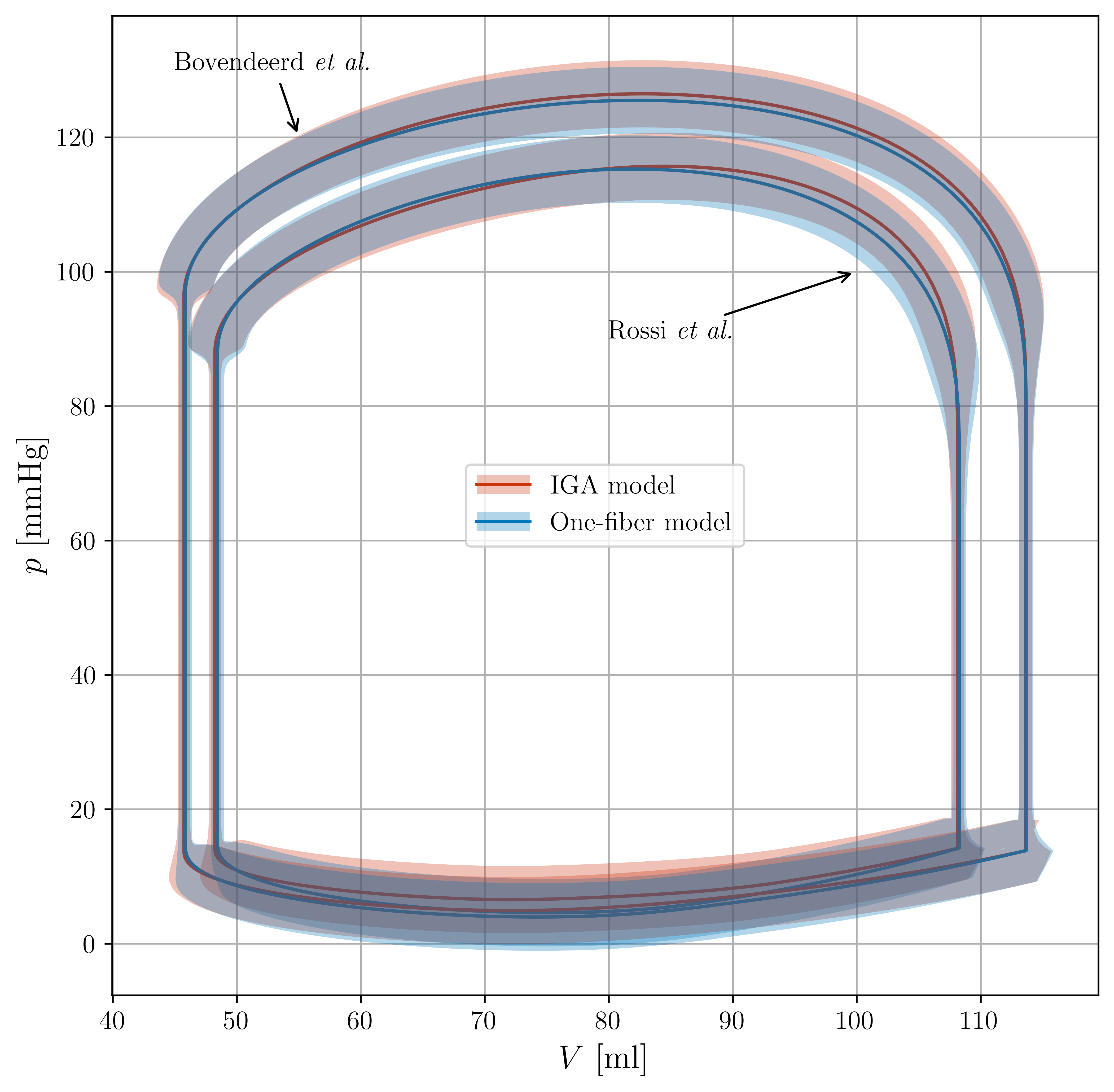}
                  \caption{}
                  \label{fig:BIFiberb}
     \end{subfigure}
        \caption{Influence of the fiber field definition in the isogeometric FOM on the inferred generalized one-fiber model, comparing the methods of  Rossi~\emph{et al.} and Bovendeerd~\emph{et al.}. (a) Posterior distributions of the inferred correction factors of the generalized one-fiber model based on the FOM results. (b) Pressure-volume response of the FOM and the corresponding Bayesian inference results for the generalized one-fiber model.}
        \label{fig:BIFiber}
\end{figure*}

The results of the isogeometric FOM are computed using the model outlined in Section~\ref{sec:fommodel} and simulated up to $6$ cardiac cycles. The results of the $6^{\mathrm{th}}$ cycle are visualized in Figure~\ref{fig:BIFiber}, including the inferred correction factors. The fiber fields result in a distinct difference when observing the pressure-volume curves in Figure~\ref{fig:BIFiberb}. The optimized field of Bovendeerd~\emph{et al.} exhibits, as expected, an overall bigger cardiac output in terms of, \emph{e.g.}, total pump work, ejection fraction, and peak pressure. The fiber field of Rossi~\emph{et al.} shows a more stiffened response as a result of increased strain heterogeneity.

Bayesian inference is performed using similar settings as described in Section~\ref{ssec:BI2D}. From the posterior distributions (Figure~\ref{fig:BIFibera}) it is observed that all correction factors, $\boldsymbol{\theta}^{\mathrm{corr}}$, are closer to unity for the Bovendeerd~\emph{et al.} model. The $\alpha$ and $\beta$ factors are particularly close to unity for this model, indicating a close alignment with the original one-fiber model (Section~\ref{sec:genrelations}). From these results, we can infer that the parameters derived from the FOM results based on the fiber field of Bovendeerd \emph{et al.} show a better agreement with the original one-fiber model. This observation indicates that deviation of the inferred correction factors from unity can, at least partly, be interpreted to be a result of stress and strain heterogeneity.

\section{Estimating the covariance matrix using Gaussian processes}
\begin{figure}[!t]
\centering
\includegraphics[width=0.8\textwidth]{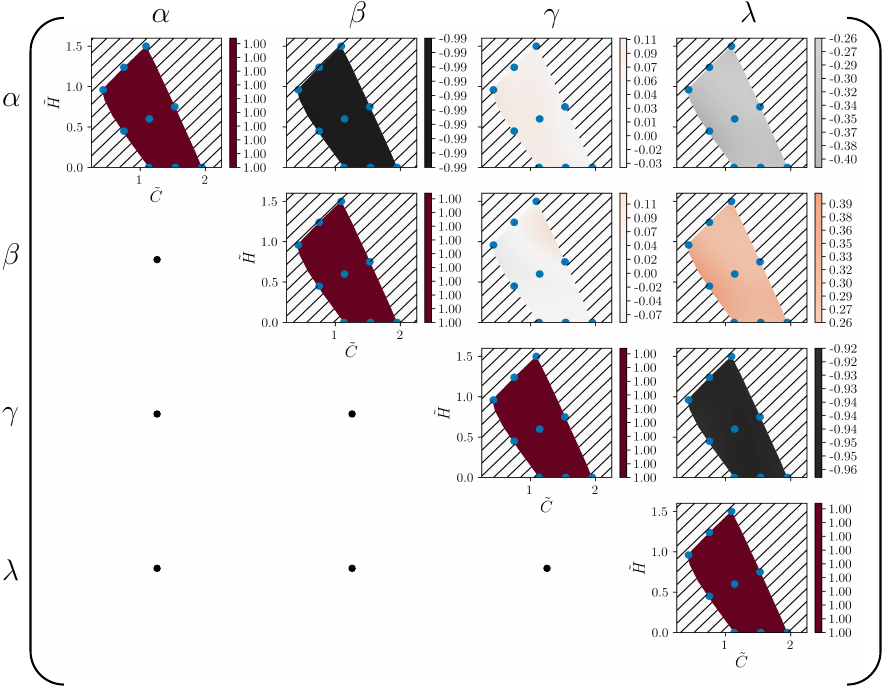}
\caption{Graphical illustration of the symmetric correlation matrix, showcasing the diagonal and upper triangular part. Each unique entry is predicted by a separate Gaussian process with zero noise. Data point values are based on the correlation of the Bayesian inference Markov chain.} \label{fig:Cormatrix}
\end{figure}

As discussed in Section~\ref{sec:GPcorrection}, we reconstruct the correlated multivariate normal distribution for a predicted geometry point by modeling the correlation matrix with individual Gaussian processes (Algorithm~\ref{alg:GP}). This construction starts with calculating the correlation matrix for each training simulation using Bayesian inference. These matrix entries are then used to train separate Gaussian processes, with zero noise applied. Figure~\ref{fig:Cormatrix} provides a graphical representation of the correlation matrix for the two-dimensional test case in Section~\ref{sec:ApplicationS}. The diagonal elements, by definition, equal unity, while the off-diagonal entries reveal the inter-dependencies between the correction factors. The results show that the correlation coefficients vary moderately across the data points. This moderate variation does affect the prediction, however, as applying a constant correlation across all points would lead to slightly different prediction outcomes.

In addition to its predictive role, Figure~\ref{fig:Cormatrix} highlights the correlation among the correction factors, $\boldsymbol{\theta}^{\mathrm{corr}}$, derived from Bayesian inference. For instance, we observe a negative correlation between $\alpha$ and $\beta$, as well as between $\gamma$ and $\lambda$. This relationship is expected, as these pairs of factors exert similar influences on the results. Specifically, both $\gamma$ and $\lambda$ directly affect the effective stiffness. During the Metropolis-Hastings sampling procedure, an increase in one of these factors typically leads to a compensatory decrease in the other.

\section{Time-dependent noise level}\label{app:noise}
To quantify the noise magnitude associated with the model data during Bayesian inference, Equation~\eqref{eq:covariancematrices}, we define a time-dependent noise level, $\sigma(t)$, which varies over the course of a single cardiac cycle. This variation is based on the clinical relevance of certain phases in the cycle, as specific segments are considered more diagnostically significant. For instance, the ejection fraction -- a common biomarker for assessing the cardiac function -- depends solely on the cavity volume evaluated at two key instances within the cycle.

The cardiac cycle comprises four distinct phases: the filling phase, the isovolumetric contraction phase, the ejection phase, and the isovolumetric relaxation phase. To accurately determine the ejection fraction, both the isovolumetric phases should be captured appropriately. To ensure this, we decrease the noise level for these phases relative to the other cardiac phases. This essentially assigns more weight or importance to these parts of the data in the Bayesian inference procedure. 

\begin{figure}[!t]
\centering
\includegraphics[width=0.5\textwidth]{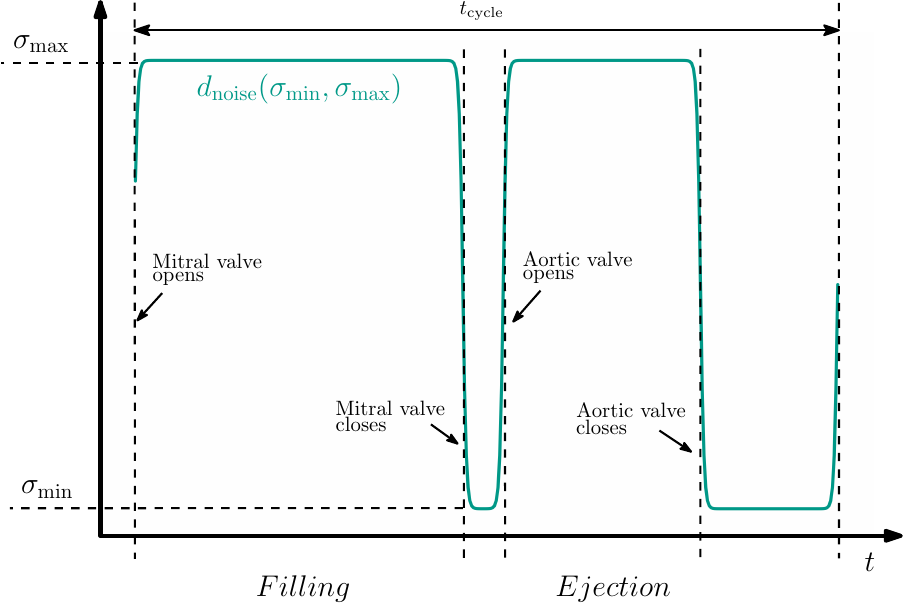}
\caption{Schematic representation of a time-dependent noise level function, $d_{\mathrm{noise}}(\sigma_{\mathrm{min}}, \sigma_{\mathrm{max}})$, which varies depending on the location within the cardiac cycle. The lower bound, $\sigma_{\mathrm{min}}$, relates to the isovolumetric phases, while the upper bound, $\sigma_{\mathrm{max}}$, corresponds to the filling and ejection phases.} \label{fig:stdnoise}
\end{figure}

A function $d_{\mathrm{noise}}(\sigma_{\mathrm{min}}, \sigma_{\mathrm{max}})$ is introduced to gradually vary the noise levels between $\sigma_{\mathrm{min}}$ and $\sigma_{\mathrm{max}}$ according to the cardiac phases. This transition is achieved through a combination of hyperbolic tangent functions, with the resulting curve shown in Figure~\ref{fig:stdnoise}. We note that, for the Bayesian inference, this time-dependent noise level is applied solely to the volumetric component of the likelihood function in Equation~\eqref{eq:covtotal}. Specifically, the values of $\sigma^{V}_i$ in Equation~\eqref{eq:covariancematrices} are defined such that
\begin{equation}\label{eq:timednoise}
    \sigma^{V}_i = d_{\mathrm{noise},i}(\sigma^{V}_{\mathrm{min}}, \sigma^{V}_{\mathrm{max}}),
\end{equation}
where the index $i$ denotes the time index. The values of $\sigma^{V}_{\mathrm{min}}$ and $\sigma^{V}_{\mathrm{max}}$ are problem-specific parameters, which are specified in Sections~\ref{sec:ApplicationS} and \ref{sec:ApplicationC}.

%% Bibliography


\begin{thebibliography}{10}
\expandafter\ifx\csname url\endcsname\relax
  \def\url#1{\texttt{#1}}\fi
\expandafter\ifx\csname urlprefix\endcsname\relax\def\urlprefix{URL }\fi
\expandafter\ifx\csname href\endcsname\relax
  \def\href#1#2{#2} \def\path#1{#1}\fi

\bibitem{trayanova_cardiac_2011}
N.~A. Trayanova, J.~J. Rice,
  \href{https://www.frontiersin.org/journals/physiology/articles/10.3389/fphys.2011.00043/full}{Cardiac
  {Electromechanical} {Models}: {From} {Cell} to {Organ}}, Frontiers in
  Physiology 2, publisher: Frontiers (Aug. 2011).
\newblock \href {https://doi.org/10.3389/fphys.2011.00043}
  {\path{doi:10.3389/fphys.2011.00043}}.

\bibitem{sun_computational_2014}
W.~Sun, C.~Martin, T.~Pham,
  \href{https://www.annualreviews.org/content/journals/10.1146/annurev-bioeng-071813-104517}{Computational
  {Modeling} of {Cardiac} {Valve} {Function} and {Intervention}}, Annual Review
  of Biomedical Engineering 16~(Volume 16, 2014) (2014) 53--76, publisher:
  Annual Reviews.
\newblock \href {https://doi.org/10.1146/annurev-bioeng-071813-104517}
  {\path{doi:10.1146/annurev-bioeng-071813-104517}}.

\bibitem{niederer_computational_2019}
S.~A. Niederer, J.~Lumens, N.~A. Trayanova,
  \href{https://www.nature.com/articles/s41569-018-0104-y}{Computational models
  in cardiology}, Nature Reviews Cardiology 16~(2) (2019) 100--111, number: 2
  Publisher: Nature Publishing Group.
\newblock \href {https://doi.org/10.1038/s41569-018-0104-y}
  {\path{doi:10.1038/s41569-018-0104-y}}.

\bibitem{sung_whole-heart_2021}
E.~Sung, S.~Etoz, Y.~Zhang, N.~A. Trayanova, Whole-heart ventricular arrhythmia
  modeling moving forward: {Mechanistic} insights and translational
  applications, Biophysics Reviews 2~(3) (2021) 031304.
\newblock \href {https://doi.org/10.1063/5.0058050}
  {\path{doi:10.1063/5.0058050}}.

\bibitem{de_lepper_evidence-based_2022}
A.~G.~W. de~Lepper, C.~M.~A. Buck, M.~van~‘t Veer, W.~Huberts, F.~N. van~de
  Vosse, L.~R.~C. Dekker,
  \href{https://royalsocietypublishing.org/doi/10.1098/rsif.2022.0317}{From
  evidence-based medicine to digital twin technology for predicting ventricular
  tachycardia in ischaemic cardiomyopathy}, Journal of The Royal Society
  Interface 19~(194) (2022) 20220317, publisher: Royal Society.
\newblock \href {https://doi.org/10.1098/rsif.2022.0317}
  {\path{doi:10.1098/rsif.2022.0317}}.

\bibitem{aguado-sierra_patient-specific_2011}
J.~Aguado-Sierra, A.~Krishnamurthy, C.~Villongco, J.~Chuang, E.~Howard, M.~J.
  Gonzales, J.~Omens, D.~E. Krummen, S.~Narayan, R.~C.~P. Kerckhoffs, A.~D.
  McCulloch, Patient-specific modeling of dyssynchronous heart failure: a case
  study, Progress in Biophysics and Molecular Biology 107~(1) (2011) 147--155.
\newblock \href {https://doi.org/10.1016/j.pbiomolbio.2011.06.014}
  {\path{doi:10.1016/j.pbiomolbio.2011.06.014}}.

\bibitem{kuijpers_modeling_2012}
N.~H.~L. Kuijpers, E.~Hermeling, P.~H.~M. Bovendeerd, T.~Delhaas, F.~W.
  Prinzen, \href{https://doi.org/10.1007/s12265-012-9346-y}{Modeling {Cardiac}
  {Electromechanics} and {Mechanoelectrical} {Coupling} in {Dyssynchronous} and
  {Failing} {Hearts}}, Journal of Cardiovascular Translational Research 5~(2)
  (2012) 159--169.
\newblock \href {https://doi.org/10.1007/s12265-012-9346-y}
  {\path{doi:10.1007/s12265-012-9346-y}}.

\bibitem{arevalo_arrhythmia_2016}
H.~J. Arevalo, F.~Vadakkumpadan, E.~Guallar, A.~Jebb, P.~Malamas, K.~C. Wu,
  N.~A. Trayanova,
  \href{https://www.nature.com/articles/ncomms11437}{Arrhythmia risk
  stratification of patients after myocardial infarction using personalized
  heart models}, Nature Communications 7~(1) (2016) 11437, number: 1 Publisher:
  Nature Publishing Group.
\newblock \href {https://doi.org/10.1038/ncomms11437}
  {\path{doi:10.1038/ncomms11437}}.

\bibitem{willems_isogeometric_2024}
R.~Willems, K.~L. P.~M. Janssens, P.~H.~M. Bovendeerd, C.~V. Verhoosel,
  O.~van~der Sluis, \href{https://doi.org/10.1007/s00466-023-02376-x}{An
  isogeometric analysis framework for ventricular cardiac mechanics},
  Computational Mechanics 73~(3) (2024) 465--506.
\newblock \href {https://doi.org/10.1007/s00466-023-02376-x}
  {\path{doi:10.1007/s00466-023-02376-x}}.

\bibitem{willems_echocardiogram-based_2024}
R.~Willems, L.~Verberne, O.~van~der Sluis, C.~V. Verhoosel,
  \href{https://www.sciencedirect.com/science/article/pii/S0045782524002147}{Echocardiogram-based
  ventricular isogeometric cardiac analysis using multi-patch fitted {NURBS}},
  Computer Methods in Applied Mechanics and Engineering 425 (2024) 116958.
\newblock \href {https://doi.org/10.1016/j.cma.2024.116958}
  {\path{doi:10.1016/j.cma.2024.116958}}.

\bibitem{willems_isogeometric-mechanics-driven_2023}
R.~Willems, E.~Kruithof, K.~L. P.~M. Janssens, M.~J.~M. Cluitmans, O.~van~der
  Sluis, P.~H.~M. Bovendeerd, C.~V. Verhoosel,
  Isogeometric-{Mechanics}-{Driven} {Electrophysiology} {Simulations}
  of {Ventricular} {Tachycardia}, in: O.~Bernard, P.~Clarysse, N.~Duchateau,
  J.~Ohayon, M.~Viallon (Eds.), Functional {Imaging} and {Modeling} of the
  {Heart}, Lecture {Notes} in {Computer} {Science}, Springer Nature
  Switzerland, Cham, 2023, pp. 97--106.
\newblock \href {https://doi.org/10.1007/978-3-031-35302-4\_10}
  {\path{doi:10.1007/978-3-031-35302-4\_10}}.

\bibitem{manzoni_reduced_2018}
A.~Manzoni, D.~Bonomi, A.~Quarteroni,
  \href{https://doi.org/10.1007/978-3-319-96649-6_6}{Reduced {Order} {Modeling}
  for {Cardiac} {Electrophysiology} and {Mechanics}: {New} {Methodologies},
  {Challenges} and {Perspectives}}, in: D.~Boffi, L.~F. Pavarino, G.~Rozza,
  S.~Scacchi, C.~Vergara (Eds.), Mathematical and {Numerical} {Modeling} of the
  {Cardiovascular} {System} and {Applications}, Springer International
  Publishing, Cham, 2018, pp. 115--166.
\newblock \href {https://doi.org/10.1007/978-3-319-96649-6_6}
  {\path{doi:10.1007/978-3-319-96649-6_6}}.

\bibitem{peherstorfer_survey_2018}
B.~Peherstorfer, K.~Willcox, M.~Gunzburger,
  \href{https://epubs.siam.org/doi/10.1137/16M1082469}{Survey of
  {Multifidelity} {Methods} in {Uncertainty} {Propagation}, {Inference}, and
  {Optimization}}, SIAM Review 60~(3) (2018) 550--591, publisher: Society for
  Industrial and Applied Mathematics.
\newblock \href {https://doi.org/10.1137/16M1082469}
  {\path{doi:10.1137/16M1082469}}.

\bibitem{liang_proper_2002}
Y.~C. Liang, H.~P. Lee, S.~P. Lim, W.~Z. Lin, K.~H. Lee, C.~G. Wu,
  \href{https://www.sciencedirect.com/science/article/pii/S0022460X01940416}{{PROPER}
  {ORTHOGONAL} {DECOMPOSITION} {AND} {ITS} {APPLICATIONS}—{PART} {I}:
  {THEORY}}, Journal of Sound and Vibration 252~(3) (2002) 527--544.
\newblock \href {https://doi.org/10.1006/jsvi.2001.4041}
  {\path{doi:10.1006/jsvi.2001.4041}}.

\bibitem{quarteroni_reduced_2016}
A.~Quarteroni, A.~Manzoni, F.~Negri,
  \href{http://link.springer.com/10.1007/978-3-319-15431-2}{Reduced {Basis}
  {Methods} for {Partial} {Differential} {Equations}}, Vol.~92 of {UNITEXT},
  Springer International Publishing, Cham, 2016.
\newblock \href {https://doi.org/10.1007/978-3-319-15431-2}
  {\path{doi:10.1007/978-3-319-15431-2}}.

\bibitem{boulakia_reduced-order_2012}
M.~Boulakia, E.~Schenone, J.-F. Gerbeau,
  \href{https://onlinelibrary.wiley.com/doi/abs/10.1002/cnm.2465}{Reduced-order
  modeling for cardiac electrophysiology. {Application} to parameter
  identification}, International Journal for Numerical Methods in Biomedical
  Engineering 28~(6-7) (2012) 727--744.
\newblock \href {https://doi.org/10.1002/cnm.2465}
  {\path{doi:10.1002/cnm.2465}}.

\bibitem{rama_towards_2020}
R.~R. Rama, S.~Skatulla,
  \href{https://www.sciencedirect.com/science/article/pii/S0045794917306727}{Towards
  real-time modelling of passive and active behaviour of the human heart using
  {PODI}-based model reduction}, Computers \& Structures 232 (2020) 105897.
\newblock \href {https://doi.org/10.1016/j.compstruc.2018.01.002}
  {\path{doi:10.1016/j.compstruc.2018.01.002}}.

\bibitem{gurney_introduction_2017}
K.~Gurney, An {Introduction} to {Neural} {Networks}, CRC Press, London, 2017.
\newblock \href {https://doi.org/10.1201/9781315273570}
  {\path{doi:10.1201/9781315273570}}.

\bibitem{williams_gaussian_2006}
C.~K. Williams, C.~E. Rasmussen, Gaussian processes for machine learning,
  Vol.~2, MA: MIT press, Cambridge, 2006.

\bibitem{raissi_physics-informed_2019}
M.~Raissi, P.~Perdikaris, G.~E. Karniadakis,
  \href{https://www.sciencedirect.com/science/article/pii/S0021999118307125}{Physics-informed
  neural networks: {A} deep learning framework for solving forward and inverse
  problems involving nonlinear partial differential equations}, Journal of
  Computational Physics 378 (2019) 686--707.
\newblock \href {https://doi.org/10.1016/j.jcp.2018.10.045}
  {\path{doi:10.1016/j.jcp.2018.10.045}}.

\bibitem{arts_relation_1991}
T.~Arts, P.~H. Bovendeerd, F.~W. Prinzen, R.~S. Reneman,
  \href{https://www.sciencedirect.com/science/article/pii/S0006349591822019}{Relation
  between left ventricular cavity pressure and volume and systolic fiber stress
  and strain in the wall}, Biophysical Journal 59~(1) (1991) 93--102.
\newblock \href {https://doi.org/10.1016/S0006-3495(91)82201-9}
  {\path{doi:10.1016/S0006-3495(91)82201-9}}.

\bibitem{arts_adaptation_2005}
T.~Arts, T.~Delhaas, P.~Bovendeerd, X.~Verbeek, F.~W. Prinzen,
  \href{https://journals.physiology.org/doi/full/10.1152/ajpheart.00444.2004}{Adaptation
  to mechanical load determines shape and properties of heart and circulation:
  the {CircAdapt} model}, American Journal of Physiology-Heart and Circulatory
  Physiology 288~(4) (2005) H1943--H1954, publisher: American Physiological
  Society.
\newblock \href {https://doi.org/10.1152/ajpheart.00444.2004}
  {\path{doi:10.1152/ajpheart.00444.2004}}.

\bibitem{arts_model_1979}
T.~Arts, R.~S. Reneman, P.~C. Veenstra,
  \href{https://doi.org/10.1007/BF02364118}{A model of the mechanics of the
  left ventricle}, Annals of Biomedical Engineering 7~(3) (1979) 299--318.
\newblock \href {https://doi.org/10.1007/BF02364118}
  {\path{doi:10.1007/BF02364118}}.

\bibitem{hill_heat_1997}
A.~V. Hill,
  \href{https://royalsocietypublishing.org/doi/10.1098/rspb.1938.0050}{The heat
  of shortening and the dynamic constants of muscle}, Proceedings of the Royal
  Society of London. Series B - Biological Sciences 126~(843) (1997) 136--195,
  publisher: Royal Society.
\newblock \href {https://doi.org/10.1098/rspb.1938.0050}
  {\path{doi:10.1098/rspb.1938.0050}}.

\bibitem{lumens_three-wall_2009}
J.~Lumens, T.~Delhaas, B.~Kirn, T.~Arts,
  \href{https://doi.org/10.1007/s10439-009-9774-2}{Three-{Wall} {Segment}
  ({TriSeg}) {Model} {Describing} {Mechanics} and {Hemodynamics} of
  {Ventricular} {Interaction}}, Annals of Biomedical Engineering 37~(11) (2009)
  2234--2255.
\newblock \href {https://doi.org/10.1007/s10439-009-9774-2}
  {\path{doi:10.1007/s10439-009-9774-2}}.

\bibitem{bovendeerd_modeling_2012}
P.~H.~M. Bovendeerd,
  \href{https://www.sciencedirect.com/science/article/pii/S0021929011007111}{Modeling
  of cardiac growth and remodeling of myofiber orientation}, Journal of
  Biomechanics 45~(5) (2012) 872--881.
\newblock \href {https://doi.org/10.1016/j.jbiomech.2011.11.029}
  {\path{doi:10.1016/j.jbiomech.2011.11.029}}.

\bibitem{huntjens_influence_2014}
P.~R. Huntjens, J.~Walmsley, S.~Ploux, P.~Bordachar, F.~W. Prinzen, T.~Delhaas,
  J.~Lumens, \href{https://doi.org/10.1093/europace/euu231}{Influence of left
  ventricular lead position relative to scar location on response to cardiac
  resynchronization therapy: a model study}, EP Europace 16~(suppl\_4) (2014)
  iv62--iv68.
\newblock \href {https://doi.org/10.1093/europace/euu231}
  {\path{doi:10.1093/europace/euu231}}.

\bibitem{bovendeerd_dependence_2006}
P.~H.~M. Bovendeerd, P.~Borsje, T.~Arts, F.~N. van De~Vosse,
  \href{https://www.ncbi.nlm.nih.gov/pmc/articles/PMC1705493/}{Dependence of
  {Intramyocardial} {Pressure} and {Coronary} {Flow} on {Ventricular} {Loading}
  and {Contractility}: {A} {Model} {Study}}, Annals of Biomedical Engineering
  34~(12) (2006) 1833--1845.
\newblock \href {https://doi.org/10.1007/s10439-006-9189-2}
  {\path{doi:10.1007/s10439-006-9189-2}}.

\bibitem{arts_epicardial_1982}
T.~Arts, P.~C. Veenstra, R.~S. Reneman,
  \href{https://journals.physiology.org/doi/abs/10.1152/ajpheart.1982.243.3.H379}{Epicardial
  deformation and left ventricular wall mechanisms during ejection in the dog},
  American Journal of Physiology-Heart and Circulatory Physiology 243~(3)
  (1982) H379--H390, publisher: American Physiological Society.
\newblock \href {https://doi.org/10.1152/ajpheart.1982.243.3.H379}
  {\path{doi:10.1152/ajpheart.1982.243.3.H379}}.

\bibitem{arts_dynamics_1989}
T.~Arts, R.~S. Reneman,
  \href{https://www.sciencedirect.com/science/article/pii/0021929089900936}{Dynamics
  of left ventricular wall and mitral valve mechanics—{A} model study},
  Journal of Biomechanics 22~(3) (1989) 261--271.
\newblock \href {https://doi.org/10.1016/0021-9290(89)90093-6}
  {\path{doi:10.1016/0021-9290(89)90093-6}}.

\bibitem{golestaneh_how_2024}
P.~Golestaneh, M.~Taheri, J.~Lederer,
  \href{http://arxiv.org/abs/2405.16696}{How many samples are needed to train a
  deep neural network?} (May 2024).
\newblock \href {https://doi.org/10.48550/arXiv.2405.16696}
  {\path{doi:10.48550/arXiv.2405.16696}}.

\bibitem{wendland_scattered_2004}
H.~Wendland,
  \href{https://www.cambridge.org/core/books/scattered-data-approximation/980EEC9DBC4CAA711D089187818135E3}{Scattered
  {Data} {Approximation}}, Cambridge {Monographs} on {Applied} and
  {Computational} {Mathematics}, Cambridge University Press, Cambridge, 2004.
\newblock \href {https://doi.org/10.1017/CBO9780511617539}
  {\path{doi:10.1017/CBO9780511617539}}.

\bibitem{bonilla_multi-task_2007}
E.~V. Bonilla, K.~Chai, C.~Williams,
  \href{https://proceedings.neurips.cc/paper_files/paper/2007/hash/66368270ffd51418ec58bd793f2d9b1b-Abstract.html}{Multi-task
  {Gaussian} {Process} {Prediction}}, in: Advances in {Neural} {Information}
  {Processing} {Systems}, Vol.~20, Curran Associates, Inc., 2007.

\bibitem{lambert_students_2018}
B.~Lambert, A {Student}'s {Guide} to {Bayesian} {Statistics}, 1st Edition, SAGE
  Publications Ltd, 2018.

\bibitem{biehler_towards_2015}
J.~Biehler, M.~W. Gee, W.~A. Wall,
  \href{https://doi.org/10.1007/s10237-014-0618-0}{Towards efficient
  uncertainty quantification in complex and large-scale biomechanical problems
  based on a {Bayesian} multi-fidelity scheme}, Biomechanics and Modeling in
  Mechanobiology 14~(3) (2015) 489--513.
\newblock \href {https://doi.org/10.1007/s10237-014-0618-0}
  {\path{doi:10.1007/s10237-014-0618-0}}.

\bibitem{agostinelli_weighted_2012}
C.~Agostinelli, L.~Greco, Weighted likelihood in {Bayesian} inference, in: 46th
  {Scientific} {Meeting} of the {Italian} {Statistical} {Society}, 2012, pp.
  746--757.

\bibitem{brooks_handbook_2011}
S.~Brooks, A.~Gelman, G.~Jones, X.-L. Meng (Eds.), Handbook of {Markov} {Chain}
  {Monte} {Carlo}, Chapman and Hall/CRC, New York, 2011.
\newblock \href {https://doi.org/10.1201/b10905} {\path{doi:10.1201/b10905}}.

\bibitem{haario_adaptive_2001}
H.~Haario, E.~Saksman, J.~Tamminen,
  \href{https://projecteuclid.org/journals/bernoulli/volume-7/issue-2/An-adaptive-Metropolis-algorithm/bj/1080222083.full}{An
  adaptive {Metropolis} algorithm}, Bernoulli 7~(2) (2001) 223--242, publisher:
  Bernoulli Society for Mathematical Statistics and Probability.

\bibitem{bovendeerd_determinants_2009}
P.~H.~M. Bovendeerd, W.~Kroon, T.~Delhaas, Determinants of left ventricular
  shear strain, American Journal of Physiology. Heart and Circulatory
  Physiology 297~(3) (2009) H1058--1068.
\newblock \href {https://doi.org/10.1152/ajpheart.01334.2008}
  {\path{doi:10.1152/ajpheart.01334.2008}}.

\bibitem{lang_recommendations_2015}
R.~M. Lang, L.~P. Badano, V.~Mor-Avi, J.~Afilalo, A.~Armstrong, L.~Ernande,
  F.~A. Flachskampf, E.~Foster, S.~A. Goldstein, T.~Kuznetsova, P.~Lancellotti,
  D.~Muraru, M.~H. Picard, E.~R. Rietzschel, L.~Rudski, K.~T. Spencer,
  W.~Tsang, J.-U. Voigt,
  \href{https://doi.org/10.1093/ehjci/jev014}{Recommendations for {Cardiac}
  {Chamber} {Quantification} by {Echocardiography} in {Adults}: {An} {Update}
  from the {American} {Society} of {Echocardiography} and the {European}
  {Association} of {Cardiovascular} {Imaging}}, European Heart Journal -
  Cardiovascular Imaging 16~(3) (2015) 233--271.
\newblock \href {https://doi.org/10.1093/ehjci/jev014}
  {\path{doi:10.1093/ehjci/jev014}}.

\bibitem{galderisi_standardization_2017}
M.~Galderisi, B.~Cosyns, T.~Edvardsen, N.~Cardim, V.~Delgado, G.~Di~Salvo,
  E.~Donal, L.~E. Sade, L.~Ernande, M.~Garbi, J.~Grapsa, A.~Hagendorff,
  O.~Kamp, J.~Magne, C.~Santoro, A.~Stefanidis, P.~Lancellotti, B.~Popescu,
  G.~Habib, {Reviewers: This document was reviewed by members of the
  2016–2018 EACVI Scientific Documents Committee},
  \href{https://doi.org/10.1093/ehjci/jex244}{Standardization of adult
  transthoracic echocardiography reporting in agreement with recent chamber
  quantification, diastolic function, and heart valve disease recommendations:
  an expert consensus document of the {European} {Association} of
  {Cardiovascular} {Imaging}}, European Heart Journal - Cardiovascular Imaging
  18~(12) (2017) 1301--1310.
\newblock \href {https://doi.org/10.1093/ehjci/jex244}
  {\path{doi:10.1093/ehjci/jex244}}.

\bibitem{lang_recommendations_2005}
R.~M. Lang, M.~Bierig, R.~B. Devereux, F.~A. Flachskampf, E.~Foster, P.~A.
  Pellikka, M.~H. Picard, M.~J. Roman, J.~Seward, J.~S. Shanewise, S.~D.
  Solomon, K.~T. Spencer, M.~St~John~Sutton, W.~J. Stewart,
  \href{https://www.sciencedirect.com/science/article/pii/S0894731705009831}{Recommendations
  for {Chamber} {Quantification}: {A} {Report} from the {American} {Society} of
  {Echocardiography}’s {Guidelines} and {Standards} {Committee} and the
  {Chamber} {Quantification} {Writing} {Group}, {Developed} in {Conjunction}
  with the {European} {Association} of {Echocardiography}, a {Branch} of the
  {European} {Society} of {Cardiology}}, Journal of the American Society of
  Echocardiography 18~(12) (2005) 1440--1463.
\newblock \href {https://doi.org/10.1016/j.echo.2005.10.005}
  {\path{doi:10.1016/j.echo.2005.10.005}}.

\bibitem{nielsen_accuracy_2010}
J.~C. Nielsen, I.~D. Lytrivi, H.~H. Ko, J.~Yau, P.~Bhatla, I.~A. Parness,
  S.~Srivastava,
  \href{https://onlinelibrary.wiley.com/doi/abs/10.1111/j.1540-8175.2009.01120.x}{The
  {Accuracy} of {Echocardiographic} {Assessment} of {Left} {Ventricular} {Size}
  in {Children} by the 5/6 {Area} × {Length} ({Bullet}) {Method}},
  Echocardiography 27~(6) (2010) 691--695.
\newblock \href {https://doi.org/10.1111/j.1540-8175.2009.01120.x}
  {\path{doi:10.1111/j.1540-8175.2009.01120.x}}.

\bibitem{anwar_true_2007}
A.~M. Anwar, O.~I.~I. Soliman, F.~J. ten Cate, A.~Nemes, J.~S. McGhie, B.~J.
  Krenning, R.-J. van Geuns, T.~W. Galema, M.~L. Geleijnse,
  \href{https://doi.org/10.1007/s10554-006-9181-9}{True mitral annulus diameter
  is underestimated by two-dimensional echocardiography as evidenced by
  real-time three-dimensional echocardiography and magnetic resonance imaging},
  The International Journal of Cardiovascular Imaging 23~(5) (2007) 541--547.
\newblock \href {https://doi.org/10.1007/s10554-006-9181-9}
  {\path{doi:10.1007/s10554-006-9181-9}}.

\bibitem{de_groot-de_laat_how_2019}
L.~de~Groot-de Laat, B.~Ren, J.~S. McGhie, E.~J. Wiegers-Groeneweg, O.~I.
  Soliman, A.~J. Bogers, M.~L. Geleijnse, How to {Measure} {Mitral} {Annulus}
  {Size} with {Two}-{Dimensional} {Transthoracic} {Echocardiography}., J Heart
  Vasc Dis 1~((1)) (2019) 100005.

\bibitem{van_zwieten_nutils_2022}
J.~S.~B. van Zwieten, G.~J. van Zwieten, W.~Hoitinga,
  \href{10.5281/zenodo.6006701}{Nutils} (Jan. 2022).

\bibitem{quarteroni_integrated_2017}
A.~Quarteroni, T.~Lassila, S.~Rossi, R.~Ruiz-Baier,
  \href{https://www.sciencedirect.com/science/article/pii/S0045782516304662}{Integrated
  {Heart}—{Coupling} multiscale and multiphysics models for the simulation of
  the cardiac function}, Computer Methods in Applied Mechanics and Engineering
  314 (2017) 345--407.
\newblock \href {https://doi.org/10.1016/j.cma.2016.05.031}
  {\path{doi:10.1016/j.cma.2016.05.031}}.

\bibitem{bovendeerd_dependence_1992}
P.~H.~M. Bovendeerd, T.~Arts, J.~M. Huyghe, D.~H. van Campen, R.~S. Reneman,
  \href{https://www.sciencedirect.com/science/article/pii/002192909290069D}{Dependence
  of local left ventricular wall mechanics on myocardial fiber orientation: {A}
  model study}, Journal of Biomechanics 25~(10) (1992) 1129--1140.
\newblock \href {https://doi.org/10.1016/0021-9290(92)90069-D}
  {\path{doi:10.1016/0021-9290(92)90069-D}}.

\end{thebibliography}
\end{document}